\documentclass[fleqn,usenatbib]{mnras}

\usepackage{newtxtext,newtxmath}

\usepackage[T1]{fontenc}
\usepackage{ae,aecompl}

\usepackage{graphicx}	
\usepackage{amsmath}	
\usepackage{multicol}
\usepackage{subfig}
\usepackage{array}
\usepackage{xcolor}
\usepackage[switch,columnwise]{lineno}

\newcommand{\PreserveBackslash}[1]{\let\temp=\\#1\let\\=\temp}
\newcolumntype{C}[1]{>{\PreserveBackslash\centering}p{#1}}

\title[Multi-wavelength Variability of PKS~0027-426]{Multi-wavelength Optical and NIR Variability Analysis of the Blazar PKS~0027-426}

\author[E.~Guise~et~al.]{E.~Guise,$^{1}$\thanks{E-mail: ella.guise@soton.ac.uk}
S.~F.~H{\"o}nig,$^{1}$
T.~Almeyda,$^{1,2}$
K.~Horne,$^{3}$
M.~Kishimoto,$^{4}$
M.~Aguena,$^{5}$
S.~Allam,$^{6}$
\newauthor
F.~Andrade-Oliveira,$^{5,7}$
J.~Asorey,$^{8}$
M.~Banerji,$^{9,10}$
E.~Bertin,$^{11,12}$
B.~Boulderstone,$^{1}$
D.~Brooks,$^{13}$
\newauthor
D.~L.~Burke,$^{14,15}$
A.~Carnero~Rosell,$^{5,16,17}$
D.~Carollo,$^{18}$
M.~Carrasco~Kind,$^{19,20}$
J.~Carretero,$^{21}$
\newauthor
M.~Costanzi,$^{22,23,24}$
L.~N.~da Costa,$^{5,25}$
T.~M.~Davis,$^{26}$
J.~De~Vicente,$^{8}$
P.~Doel,$^{13}$
S.~Everett,$^{27}$
\newauthor
I.~Ferrero,$^{28}$
B.~Flaugher,$^{6}$
J.~Frieman,$^{6,29}$
P.~Gandhi,$^{1}$
M.~Goad,$^{30}$ 
D.~Gruen,$^{31}$
R.~A.~Gruendl,$^{19,20}$
\newauthor
J.~Gschwend,$^{5,25}$
G.~Gutierrez,$^{6}$
S.~R.~Hinton,$^{26}$
D.~L.~Hollowood,$^{27}$
K.~Honscheid,$^{32,33}$
D.~J.~James,$^{34}$
\newauthor
M.~A.~C.~Johnson,$^{1,35,36}$
K.~Kuehn,$^{37,38}$
G.~F.~Lewis,$^{39}$
C.~Lidman,$^{40,41}$
M.~Lima,$^{5,42}$
M.~A.~G.~Maia,$^{5,25}$
\newauthor
U.~Malik,$^{41}$
F.~Menanteau,$^{19,20}$
R.~Miquel,$^{21,43}$
R.~Morgan,$^{44}$
R.~L.~C.~Ogando,$^{5,25}$
A.~Palmese,$^{6,29}$
\newauthor
F.~Paz-Chinch\'{o}n,$^{9,19}$
M.~E.~S.~Pereira,$^{45}$
A.~Pieres,$^{5,25}$
A.~A.~Plazas~Malag\'on,$^{46}$
E.~Sanchez,$^{8}$
V.~Scarpine,$^{6}$
\newauthor
S.~Serrano,$^{47,48}$
I.~Sevilla-Noarbe,$^{8}$
N.~Seymour,$^{49}$
M.~Smith,$^{1}$
M.~Soares-Santos,$^{45}$
E.~Suchyta,$^{50}$
\newauthor
M.~E.~C.~Swanson,$^{19}$
G.~Tarle,$^{45}$
C.~To,$^{14,15,51}$
and B.~E.~Tucker $^{41}$
\\
~
\\
\textit{\Large Affiliations are listed at end of paper}
}
\date{Accepted XXX. Received YYY; in original form ZZZ}

\pubyear{2021}

\begin{document}
\label{firstpage}
\pagerange{\pageref{firstpage}--\pageref{lastpage}}
\maketitle

\begin{abstract}


We present multi-wavelength spectral and temporal variability analysis of \mbox{PKS~0027-426} using optical \emph{griz} observations from DES (Dark Energy Survey) between 2013-2018 and VOILETTE (VEILS Optical Light curves of Extragalactic TransienT Events) between 2018-2019 and near infrared (NIR) \emph{JKs} observations from VEILS (VISTA Extragalactic Infrared Legacy Survey) between 2017-2019. Multiple methods of cross-correlation of each combination of light curve provides measurements of possible lags between optical-optical, optical-NIR, and NIR-NIR emission, for each observation season and for the entire observational period. Inter-band time lag measurements consistently suggest either simultaneous emission or delays between emission regions on timescales smaller than the cadences of observations. The colour-magnitude relation between each combination of filters was also studied to determine the spectral behaviour of PKS~0027-426. Our results demonstrate complex colour behaviour that changes between bluer when brighter (BWB), stable~when brighter (SWB) and redder when brighter (RWB) trends over different timescales and using different combinations of optical filters. Additional analysis of the optical spectra is performed to provide further understanding of this complex spectral behaviour. 
\end{abstract}




\section{Introduction}
\label{Sect:Intro}

Blazars are the most variable subclass of active galactic nuclei (AGN), whose radiation is considered to be dominated by a bright, relativistic jet less than 10$^{\circ}$ from the line of sight (e.g. \citealt{UrryandPadovani1995}). They can be divided into two subclasses based on their spectra; Flat Spectrum Radio Quasars (FSRQs) and BL Lacertae objects (BL Lacs), as the spectra of FSRQs contain strong, broad emission lines, whilst BL Lacs are characterized by a relatively featureless optical continuum.

The emission from blazars is strongly variable over the entire electromagnetic spectrum, and is composed of both thermal and non-thermal contributions which originate in different components of the AGN. Their spectral energy distributions (SEDs) contain two characteristic bumps; one at low energies which covers the range from radio to UV and a higher energy bump which is located in the X-rays to gamma-rays (e.g. \citealt{Fossati1998}). The dominating emission processes corresponding to these bumps are considered to be synchrotron radiation from the relativistic electrons in the jet at lower energies (e.g. \citealt{Urry1982}) and at higher energies can be described by either leptonic models where the bump is due to inverse Compton scattering of the low-energy emission (e.g. \citealt{Bottcher2007}), or hadronic models in which the bump is due to emission from relativistic protons (e.g. \citealt{Mucke2001}). In the optical and near infrared (NIR), additional contributions are expected from thermal emission from the accretion disk and torus. 

Multi-wavelength variability studies of blazars provide further information on these emission processes; for example, analysis of temporal variability can be used to infer their location within the AGN using correlations between the radiation from the wavelength ranges corresponding to these processes. Additionally, spectral variability studies give insight into how the contributions from the thermal and non-thermal emission vary with respect to each other, as the ratio of thermal and non-thermal emission changes with the variations in the flux, and can result in changes to the spectral shape and colour of the blazar (e.g. \citealt{Gu2011}). 

Several studies of the correlations between the flux variations from different wavelength ranges of blazars commonly show that they are strongly correlated with short lags between the light curves on timescales $<$ 1 day or with no significant lag determined on the order of days \citep[e.g.][]{DAmmando2013, Zhang2013, Kaur2018}. This implies that the dominant source of the emission in the different wavelength ranges is co-temporal in the blazar, and possibly co-spatial. However, some studies have also reported lags on the order of 10-100 days between optical and NIR light curves in blazars (e.g. \citealt{Li2018}, \citealt{Safna2020}), which could imply that the sources of the emission are located with a distance between them, or \cite{Li2018} suggest that if the emission is produced by shocks in the jet, the higher energy emission could emerge closer to the front of the shock than the lower energy emission, resulting in a delay.

Studies of the spectral variation of blazars have shown three main colour behaviours; bluer when brighter (BWB), redder when brighter (RWB) or achromatic/stable when brighter (SWB). These colour trends are often explained as a result of variations between the different emission processes that contribute to the overall emission, for example a RWB trend could indicate that a red component, such as synchrotron emission from the relativistic jet, is more quickly varying than the bluer component, such as the thermal emission from the accretion disk, and vice versa for a BWB trend (e.g. \citealt{Fiorucci2004}, \citealt{Bonning2012}, \citealt{Agarwal2019}). Alternatively, BWB trends have also been explained by processes associated with the relativistic jet only; for example \cite{Fiorucci2004} describe a one component synchrotron model in which the more intense the energy release, the higher the particle's energy. A shock-in-jet model has also be used to describe the BWB trend as accelerated electrons at the front of the shock lose energy while propagating away, and because of synchrotron cooling the higher frequency electrons lose energy faster, thus making the high frequency bands more variable (e.g. \citealt{Kirk1998}, \citealt{Agarwal2019}). The RWB trend is most frequently observed with FSRQs, and similarly the BWB trend is most commonly observed in BL Lacs (e.g. \citealt{Gu2006}, \citealt{Bonning2012}, \citealt{Meng2018}), however this is not always the case, as some studies find the reverse or find SWB trends (e.g. \citealt{Gu2011}, \citealt{Zhang2015}, \citealt{Mao2016}). Furthermore, while many studies find these simple colour behaviours, some find that the colour trends can be complex; for example, \cite{Isler2017} showed that the B-J colour behaviour of the FSRQ 3C~279 varied on different timescales, and over different periods during the 7 years of observation. Specifically, the average colour trend of the entire 7 years is BWB, however the colour variability is shown to deviate for individual observation seasons, such as from achromatic or a SWB trend between May and August 2008, RWB between September 2009 and April 2010 and BWB between February and August 2011. Furthermore, the colour trend of some blazars has be shown to change at a certain magnitude; for example, \cite{Zhang2015} found that several sources showed RWB trends in the low flux state and then kept a SWB trend or a BWB trend in the high flux states. 

\begin{figure}
    \centering
    \includegraphics[width=\columnwidth]{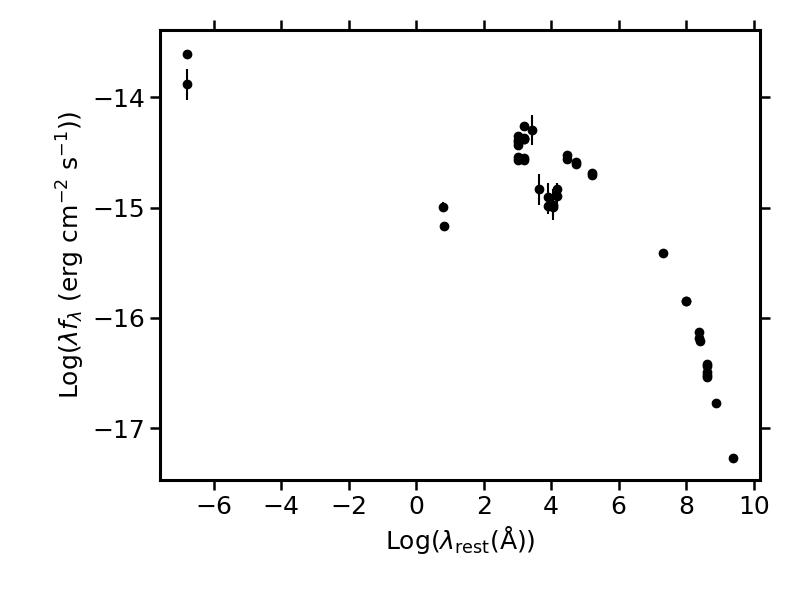}
    \caption{The spectral energy distribution of PKS~0027-426 using data from NASA/IPAC Extragalactic Database.}
    \label{fig:NED_SED}
\end{figure}

PKS~0027-426\footnote{RA = 00h30m17.584s, DEC = -42d24m46.02s (J2000)} is classified as a FSRQ  with z = 0.495 (\citealt{Hook2002}) . It has been observed in the optical \emph{griz} bands with DES (Dark Energy Survey) from 2013-2018 and VOILETTE (VEILS OptIcal Light curves of Extragalactic TransienT Events) from 2018 onwards, with concurrent observations in the NIR \emph{J} and \emph{Ks} bands with VEILS (VISTA Extragalactic Infrared Legacy Survey) from 2017 onward. PKS~0027-426 was found to be the most variable AGN detected in the VEILS fields thus far. The SED of PKS~0027-426 is displayed in Figure~\ref{fig:NED_SED} and is made using data from NASA/IPAC Extragalactic Database (NED). The lower energy peak can be seen in the wavelength range corresponding to log($\lambda_\text{rest}$(\AA)) $\sim$ 0-10, and a dip is present at $\sim$ 4 which corresponds to the optical-NIR wavelength range. 

In this paper, we analyse the temporal and spectral variability of PKS~0027-426 using optical \emph{griz} observations from DES and VOILETTE between 2013-2019 and NIR \emph{JKs} observations from VEILS between 2017-2019. The structure is as follows; in Section~\ref{Sect:Data} we describe the observations and data reduction. In Sections \ref{Sect:Time} and \ref{Sect:Colour} we present the temporal and spectral variability analysis of PKS~0027-426 respectively. In Section~\ref{Sect:Discuss} we discuss the results and provide further analysis of the spectra to explain the colour behaviour observed. Finally, in Section~\ref{Sect:Concl} we present a summary of the conclusions.

\section{Observations and Data Reduction}
\label{Sect:Data}

\begin{figure*}
    \begin{minipage}{\textwidth}
    	\includegraphics[width=0.98\columnwidth]{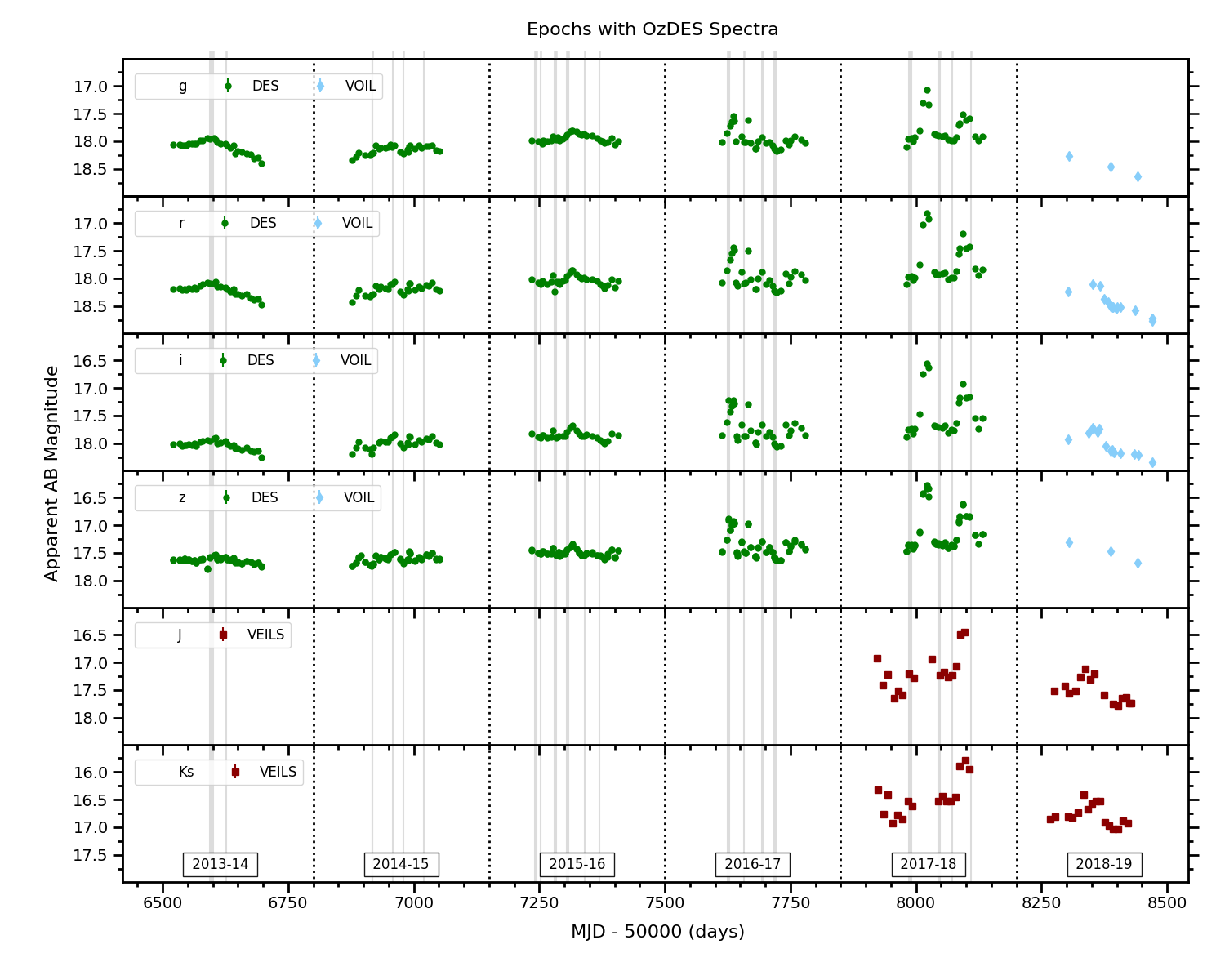}
    \end{minipage}
    \caption{Light curves of PKS~0027-426 in the optical \emph{griz} bands and NIR \emph{JKs} bands. The optical light curves contain a combination of DES (green circles) in the seasons starting in 2013-2017 and VOILETTE (blue diamonds) in the season starting in 2018 and the NIR observations are from VEILS (red squares) in seasons starting in 2017-2018 \label{fig:lc}. Each observation season is separated by the dotted lines and the epochs corresponding to OzDES observations are shown with the grey lines.}
\end{figure*}

\subsection{Overview of Optical and NIR Surveys}

\subsubsection{DES-SN}

DES was a 5-year survey that observed using DECam on the 4m Blanco telescope at the Cerro Tololo Inter-American Observatory (CTIO) in the \emph{grizY} bands between 2013-2018 (\citealt{DESFlaugher2015}). It consisted of two programmes; a wide-area survey that covered 5000 square degrees, in which each region was observed 10 times in each of the filters over the course of the survey, and a time-domain survey (DES-SN \citep{Kessler2015}) that covered a smaller region of 27 square degrees, but was observed repeatedly and regularly. The observed 27 square degrees of the DES-SN programme was divided between 4 fields; the Chandra Deep Field South (CDFS), Elias South (ES), Sloan Digital Sky Survey (SDSS) Stripe 82 field and the XMM deep field, each of which were observed with $\sim$ 6 month observation seasons per year, with $\sim$ 6 day cadences.

\subsubsection{VEILS}

VEILS is a current ESO Public Survey which repeatedly targets 9 square degrees of sky in the \emph{JKs} bands starting in 2017 using VIRCAM on the Visible and Infrared Survey Telescope for Astronomy (VISTA) telescope at the Paranal Observatory (\citealt{VISTAEmerson2006}). It was designed to observe regions that are covered by DES (3 square degrees in each of the CDFS, ES and XMM fields), also with $\sim$~6 month observation seasons and cadences of 10-14 days, to provide concurrent optical and NIR observations which will allow for multi-wavelength time domain studies of AGN. For example, one of the primary science goals of VEILS is to measure the time lags between the accretion disk variability and the response from the hot dust in the surrounding torus in a process referred to as dust reverberation mapping.

\subsubsection{VOILETTE}

From 2018 onwards, the optical \emph{griz} band observations continued with VOILETTE, which uses OmegaCAM on the 2.6m VLT Survey Telescope (VST) at the Paranal Observatory (\citealt{VSTKiujken2002}). VOILETTE was designed as the optical counterpart to VEILS, and as such covered approximately the same region of sky with planned cadences of $\sim$ 6-10 days.

\subsection{Data Reduction and Calibration}

The reduction of the data from DES included correcting for cross-talk and non-linear pixel response, as well as subtraction of bias and sky frames, bad pixel masking and flat fielding as explained by \cite{Morganson2018}. The raw data from VOILETTE and VEILS were similarly reduced by bad pixel masking, flat fielding and subtraction of bias, dark current and sky frames. Aperture photometry was then performed using fixed aperture sizes in each survey on PKS~0027-426, and also on nearby objects in the same detector that were variable by less than 0.5 dex over the entire observation period. These non-varying objects (listed in Appendix~\ref{ap:ref_stars}), whose magnitudes were calibrated using the DES photometric catalogue \citep{Abbott2018} for the optical observations and the 2MASS catalogue \citep{Skrutskie2006} for the NIR observations, were used to correct the observed counts of \mbox{PKS~0027-426} for the nightly effects such as the seeing or change in atmospheric conditions. The corrected counts for \mbox{PKS 0027-426} in each filter were converted into apparent AB magnitudes, creating the light curves displayed in Figure~\ref{fig:lc}.

\subsection{Light Curves of PKS~0027-426}

The top 4 panels of Figure~\ref{fig:lc} show the optical light curves of PKS~0027-426 over 6 years in the \emph{griz} bands, with each season~separated by the dotted lines and with overall magnitude variations (brightest - dimmest magnitude) of $|\Delta$\emph{g}| = 1.56 $\pm$ 0.06 mags, $|\Delta$\emph{r}| = 1.95 $\pm$ 0.07 mags, $|\Delta$\emph{i}| = 1.79 $\pm$ 0.06 mags and $|\Delta$\emph{z}| = 1.52 $\pm$ 0.06 mags. The cadences of the observations in the \emph{griz} bands are 6.3 $\pm$ 4.9 nights, 6.5 $\pm$ 5.3 nights, 6.6 $\pm$ 4.7 nights and 6.2 $\pm$ 4.8 nights respectively over the entire observational period, and Appendix \ref{ap:cadences} displays the average cadence for each observation season. The variability in the optical bands is relatively low in the seasons starting 2013-2015, but increases in the seasons starting in 2016-2018. The 2017~season displays the largest variability in each optical filter, and contains its peak magnitude in the first of 2 flares that are separated by approximately 75 days. The 2018~season contains only 3 observations in the \emph{g} and \emph{z} filters, however displays a decreasing brightness in all optical filters. The lower 2 panels of Figure~\ref{fig:lc} show the NIR light curves with 2 years of observations in the \emph{J} and \emph{Ks} bands respectively, with magnitude variations of $|\Delta$J| = 1.32 $\pm$ 0.01 mags and $|\Delta$Ks| = 1.24 $\pm$ 0.02 mags and cadences of 11.3 $\pm$ 5.9 nights and 11.2 $\pm$ 8.1 nights. Similarly to the optical, the 2017~season displays the largest variability in both NIR filters, however, it also contains a gap in the observations between $\sim$ 58000 and 58050 MJD, which corresponds to the epochs containing the first and brightest peak in the optical. The brightness of the 2018~season is also shown to decrease in the NIR.

\subsubsection{Light Curve Variability with the Amplitude Variability Parameter}

To characterise the variability of PKS~0027-426 in each filter and in each season, the amplitude variability parameter, A, was calculated using Equation~\ref{eq:PerAmp} \citep{HeidtWagner1996}.

\begin{equation}
    \label{eq:PerAmp}
    A = \sqrt{(A_\text{max}-A_\text{min})^2-2\sigma^2} 
\end{equation}
Where $A_\text{max}$ and $A_\text{min}$ are the maximum and minimum apparent magnitudes, and $\sigma$ is the average measurement error.

Table~\ref{tab:perampvar} shows the amplitude variation of each filter in each year. The variation in the \emph{griz} light curves is relatively small in the seasons starting in 2013-2015 with amplitude variability parameters of $\sim$~0.25~-~0.5, but in the 2016 and 2017 seasons they increase to $\sim$~0.6~-~0.9 and $>$~1 respectively, and in the 2018 season decreases back to $\sim$~0.6~-~0.7. The \emph{J} and \emph{Ks} bands in both the 2017 and 2018 seasons are shown to vary similarly to the optical.

\begin{table}
	\begin{minipage}{\columnwidth}
        \centering
        \footnotesize{
        \caption{The amplitude variation of the light curves in different filters in each observation season, calculated using Equation~\ref{eq:PerAmp}. The 2013-2016~seasons contain only data from the \emph{griz} bands from DES, the 2017~season also contains the \emph{JKs} bands from VEILS, and the 2018~season only has the \emph{r} and \emph{i} bands from VOILETTE and the \emph{J} and \emph{Ks} bands from VEILS due to limited \emph{g} and \emph{z} data. \label{tab:perampvar}}        
        \begin{tabular}{C{0.12\textwidth}C{0.08\textwidth}C{0.08\textwidth}C{0.08\textwidth}C{0.08\textwidth}C{0.08\textwidth}C{0.08\textwidth}}
            \hline
                Year Starting & \emph{g} (mag) & \emph{r} (mag) & \emph{i} (mag) & \emph{z} (mag) & \emph{J} (mag) & \emph{Ks} (mag) \\
            \hline
               2013 & 0.46 & 041 & 0.35 & 0.26 & - & - \\
               2014 & 0.28 & 0.36 & 0.36 & 0.27 & - & - \\
               2015 & 0.26 & 0.39 & 0.32 & 0.27 & - & - \\
               2016 & 0.63 & 0.81 & 0.85 & 0.76 & - & - \\
               2017 & 1.03 & 1.29 & 1.33 & 1.20 & 1.13 & 1.19 \\
               2018 & - & 0.66 & 0.61 & - & 0.62 & 0.66 \\
            \hline
        \end{tabular}}
	\end{minipage}
\end{table}

\begin{figure}
    \centering
    \includegraphics[width=\columnwidth]{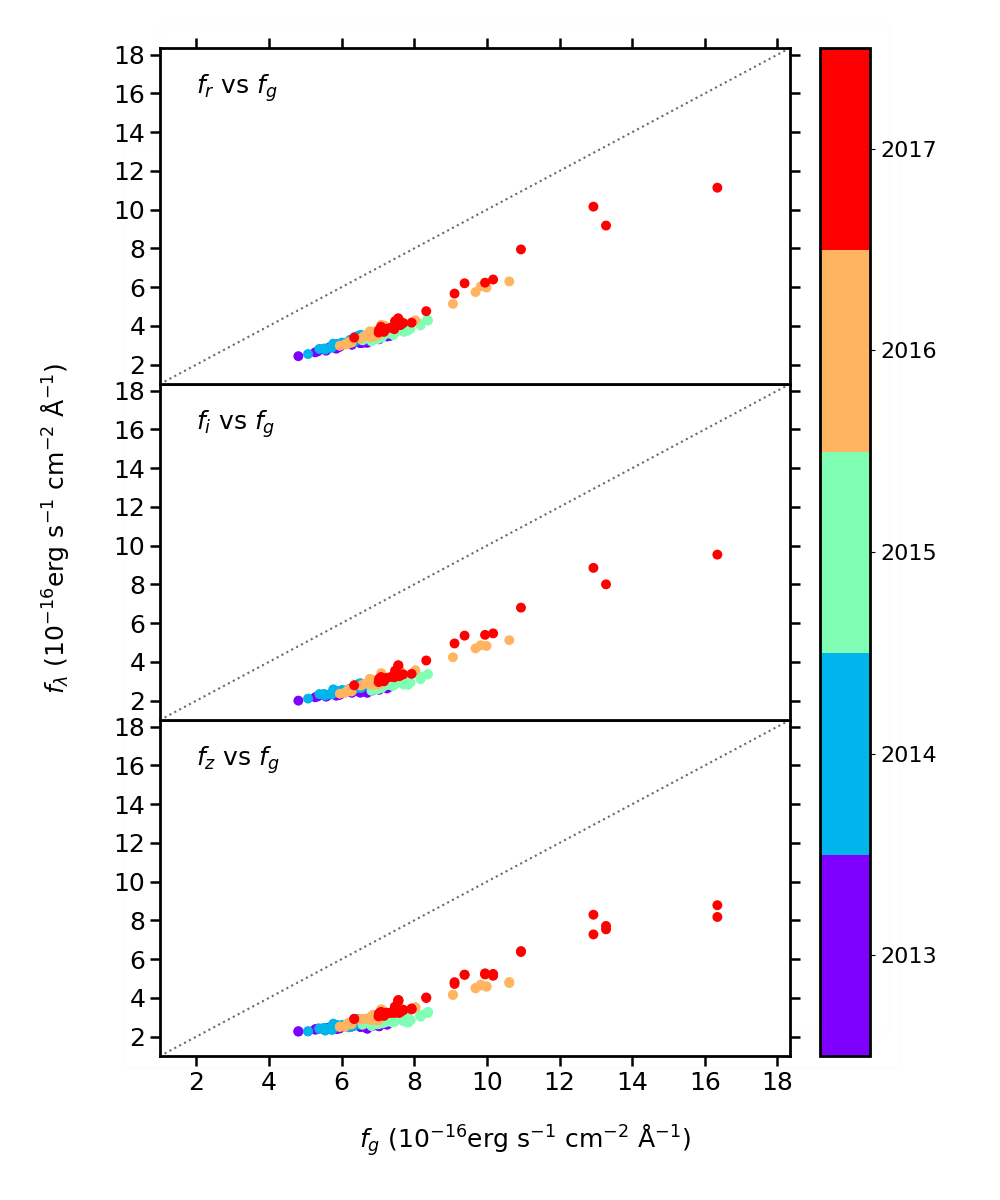}
    \caption{Flux variations in each DES \emph{r}, \emph{i}, \emph{z} filter compared to the DES \emph{g} band, where the data points are coloured according to observation season.}
    \label{fig:flux_flux}
\end{figure}

\subsubsection{DES Light Curve Variability with Flux-Flux Plots}

As PKS~0027-426 was often observed on the same night with each DES filter, the flux variations in each optical filter relative to another could be further analysed using flux-flux plots for the seasons starting 2013-2017, for example, Figure~\ref{fig:flux_flux} displays the \emph{r}, \emph{i} and \emph{z} band fluxes compared to the \emph{g} band fluxes. The flux in each filter is shown to increase as the \emph{g} band flux increases, and though the relationships aren't necessarily linear overall, Figure~\ref{fig:gz_flux_flux} in Appendix~\ref{ap:flux-flux} shows that the relationship in the individual seasons of the g-z flux-flux plots are approximately linear. Furthermore, Figure~\ref{fig:flux_flux_nir} displays the flux-flux plots of the 2017 season including VEILS observations, but as PKS~\mbox{0027-426} was not observed simultaneously in the NIR, the light curves were interpolated.

\subsection{Spectroscopy}

\begin{figure}
    \centering
    \includegraphics[width=\columnwidth]{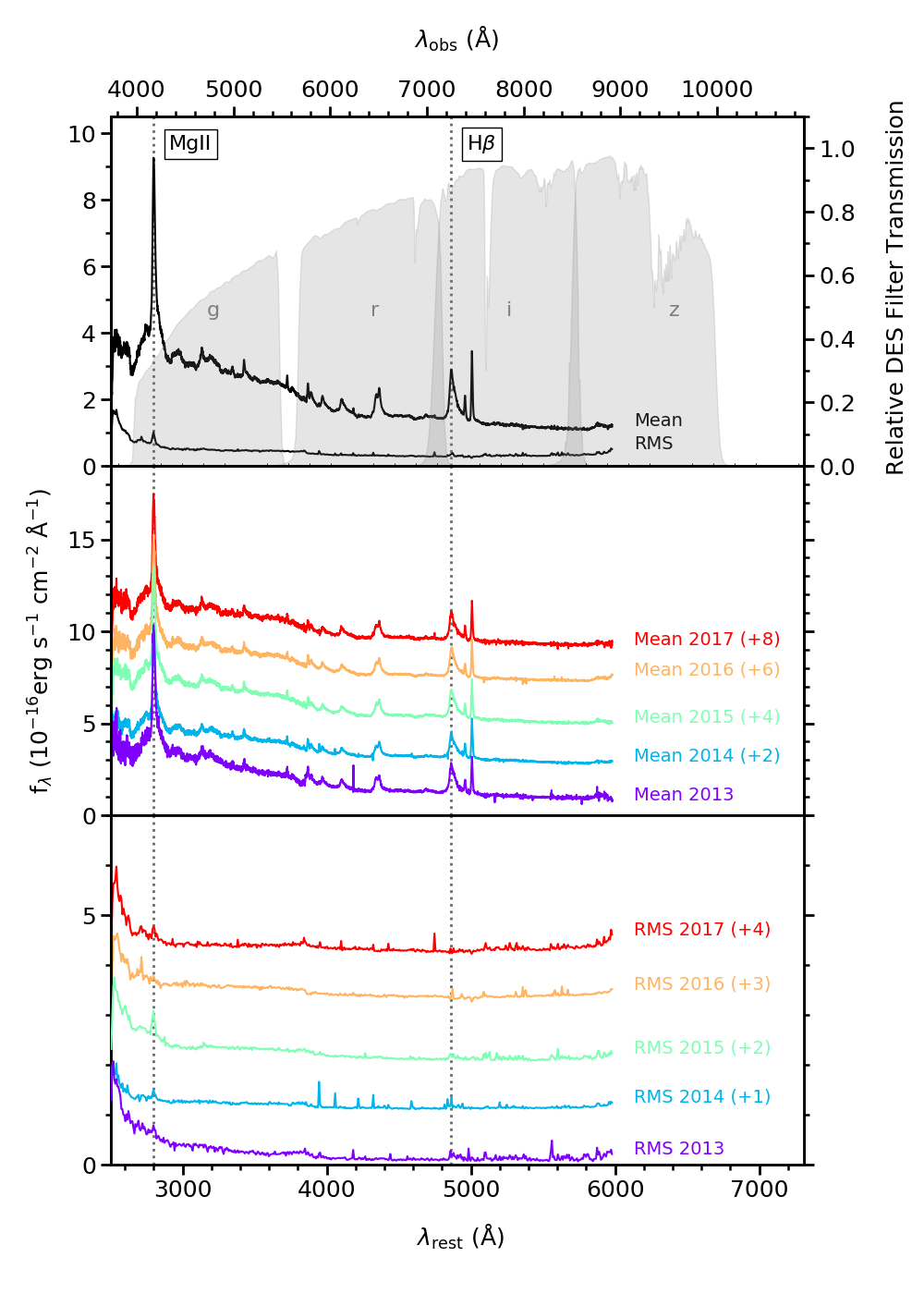}
    \caption{\textit{Top Panel}: Overall mean spectra of PKS~0027-426 using 92 observations from OzDES over 37 epochs between 2013-2018, with some of the relevant emission lines labelled. The filter transmission curves for DES are overlaid to demonstrate which filter each emission line lies in. \textit{Middle panel}: the mean spectra for each individual observation season, labelled with the starting year of observations. \textit{Lower panel}: The Smoothed RMS spectra for each individual observation season, labelled with the starting year of observations.}
    \label{fig:coadded_spectra}
\end{figure}

Optical spectra of PKS~0027-426 were obtained on 37 epochs between 2013 to 2018 by OzDES (Australian spectroscopic Dark Energy Survey), the spectroscopic follow up survey for DES. OzDES uses the 3.9m Anglo-Australian Telescope (AAT) (\citealt{Smith2004}) at Siding Spring Observatory in Australia, along with the AAOmega spectragraph with the Two Degree Field (2dF) 400 multi object fibre positioning system (\citealt{Lewis2002}), which covers the wavelength range of 3700-8800\AA \ with a spectral resolution of 1400-1700 \citep{Lidman2020}.

The spectra were flux calibrated using the photometry from DES, as described by \cite{Hoormann2019}, to remove any differences from each epoch due to factors including the image quality, airmass, transparency, and accuracy of the fibre placement. Figure~\ref{fig:coadded_spectra} displays the mean and smoothed RMS spectra of PKS~0027-426, for the entire observational period and the individual observations seasons, with most relevant emission lines labelled. 

An excess in the red wing can be seen in some of the broad emission lines (BELs) in Figure \ref{fig:coadded_spectra}, which is a phenomenon observed in many radio loud quasars. \cite{Punsly2020} studied the red asymmetry in the BELs of radio loud quasars and found that the blazars with the most redward asymmetric BELs had a low Eddington rate, a strong jet relative to the accretion flow bolometric luminosity and a polar line of sight. 

The \mbox{Mg II} and H$\beta$ lines in the spectra were used to obtain an estimate for the virial mass of the SMBH of \mbox{PKS 0027-426}, $M_\text{BH}$, using Equation~\ref{eq:smbh_mass}.

\begin{equation}
\label{eq:smbh_mass}
    \text{log} \bigg(\frac{M_\text{BH}}{M_{\odot}}\bigg) = a + b \text{log} \bigg(\frac{\lambda L_{\lambda}}{10^{44} \text{erg s}^{-1}} \bigg) + 2 \text{log} \bigg(\frac{\text{FWHM}}{\text{kms}^{-1}}\bigg)
\end{equation}
Where the coefficients a and b for the \mbox{Mg II} line are 0.74 and 0.62 respectively \citep{Shen2011}, and for the H$\beta$ line are 0.91 and 0.50 respectively \citep{Vestergaard2006}, $\lambda L_{\lambda}$ is the monochromatic luminosity at 3000 $\dot{A}$ and 5100 $\dot{A}$ for the \mbox{Mg II} and H$\beta$ lines respectively, which are calculated from the \emph{gri} magnitudes as described by \cite{Kozlowski2015}, and the FWHM is the FWHM of the emission line. This was calculated for each OzDES spectra, and the mean of Log($M_\text{BH}$) was found to be \mbox{8.16 $\pm$ 0.08 $M_{\odot}$}  for the \mbox{Mg II} line and \mbox{8.06 $\pm$ 0.28 $M_{\odot}$} for the H$\beta$ line, which are consistent within their 1$\sigma$ uncertainties.

\section{Temporal Variability}
\label{Sect:Time}

Temporal variability studies of blazars typically report lags between the optical and NIR emission on timescales < 1 day \citep[e.g.][]{DAmmando2013, Zhang2013, Kaur2018}, which implies that the dominant emission regions are co-spatial and could be expected to be due to the synchrotron emission from the relativistic jet. This expected lag is smaller than can be detected with the cadences of observations from DES, VOILETTE and VEILS, as these surveys were designed to detect dust reverberation lags in AGN. However, temporal variability analysis is used here to explore whether any larger lags, possibly one corresponding to a delay between the thermal emission in the optical and NIR, could be detected as well, as other studies have reported significant lags between the optical and NIR in blazars on the order of 10-100 days (e.g. \citealt{Li2018},  \citealt{Safna2020}).

In order to quantitatively study the temporal variability between the optical and NIR light curves from DES, VOILETTE and VEILS, the cross-correlation function (CCF) was computed (e.g. \citealt{Peterson1993}). The CCF requires continuous light curves, therefore they were interpolated using the structure function (SF), which is a measure of the fractional change in flux for observations that are separated by a given time interval, $\tau$, (e.g. \citealt{Suganuma2006}, \citealt{Emmanoulopoulos2010}), to simulate data points where there were no observations. The first order structure function, SF($\tau$), is defined in Equation \ref{eq:SF} \citep{Suganuma2006}:

\begin{equation}
    \label{eq:SF}
    SF(\tau) = \frac{1}{N(\tau)} \sum_{i<j} [f(t_i)-f(t_j)]^2
\end{equation}
Where $f(t)$ is the flux at time $t$ and the sum is over all pairs for which $t_j - t_i = \tau$ and $N(\tau)$ is the number of pairs. 

Interpolating the light curves introduces uncertainties however, as it creates large portions of simulated data in the $\sim$ 6 month gaps between the observation seasons. Therefore, to reduce the effect of the simulated data in the observation gaps, two methods of interpolated cross-correlation function (ICCF) were compared. The first method is the standard ICCF (S-ICCF), which utilises as much of the observed data as possible by cross correlating both light curves which were interpolated with 1 day cadences. This method was computationally inexpensive, and generally worked well when cross correlating individual observation seasons, however it treated the interpolated data between the observing seasons of the entire light curve equally with the data which could decrease its reliability. 

The second method was used to limit the impact of the simulated data between the observations, by only interpolating one light curve and extracting the epochs that matched the other filter's observations plus the range of possible lags being tested to get simultaneous light curves. This method is more reliable when using light curves that contain multiple observing seasons as it includes less of the interpolation in the gaps between observations. However, it did not utilise the entirety of the observations available as only the epochs of one of the light curves were used at a time. This method was used twice, alternatively interpolating each filter, which will be referred to as the modified ICCF (M-ICCF) and reverse modified ICCF (RM-ICCF) respectively.

\subsection{Cross-Correlation Results}

\begin{figure*}
    \centering
    \begin{minipage}{0.32\textwidth}
    \subfloat[figure][CCFs and ACFs of the entire \emph{r} and \emph{i} light \\ curves. \label{fig:ri_CCFs_all}]{
    \begin{minipage}{\textwidth}
        \includegraphics[width=\textwidth]{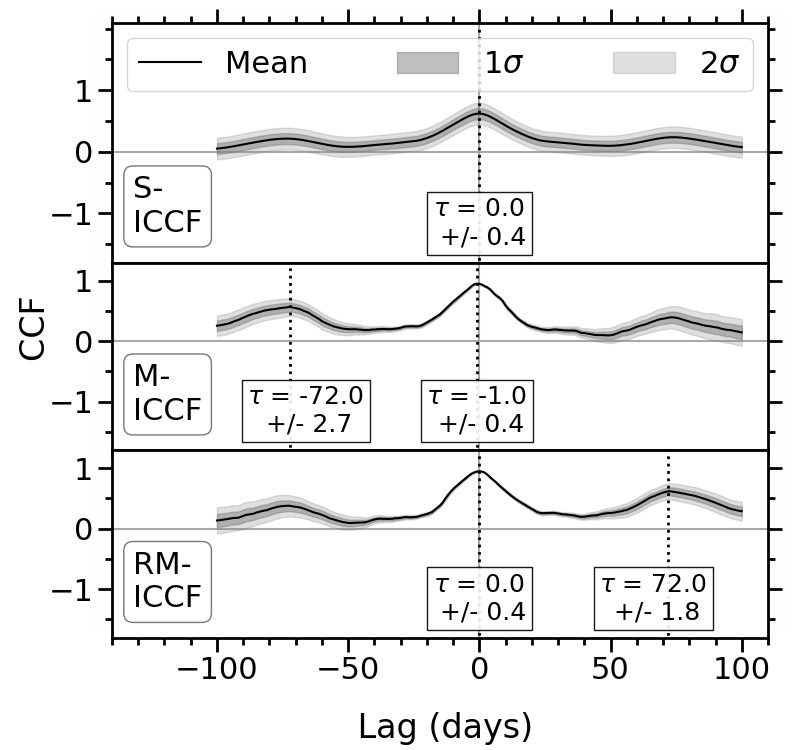} \\
        \includegraphics[width=\textwidth]{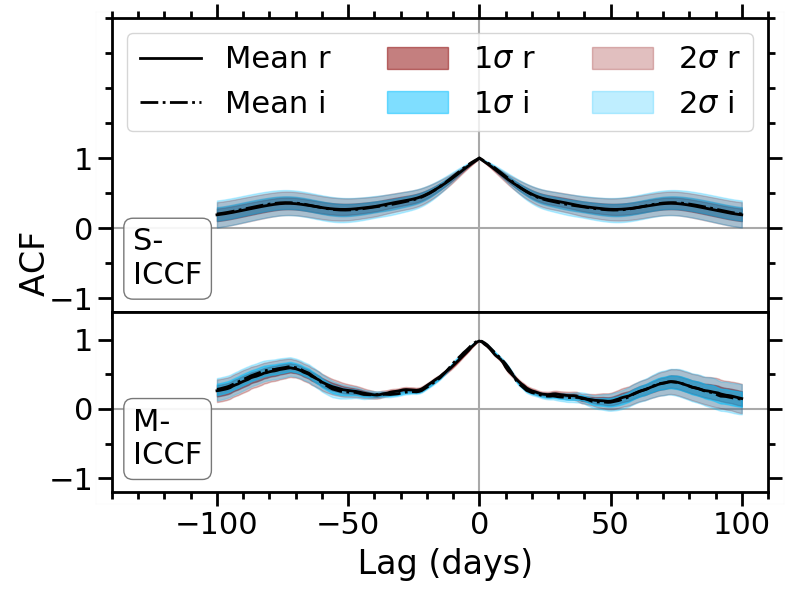}
    \end{minipage}} \\
    \subfloat[][CCFs and ACFs of the 2017 \emph{r} and \emph{i} light \\ curves. \label{fig:ri_CCF_2017}]{
    \begin{minipage}{\textwidth}
        \includegraphics[width=\textwidth]{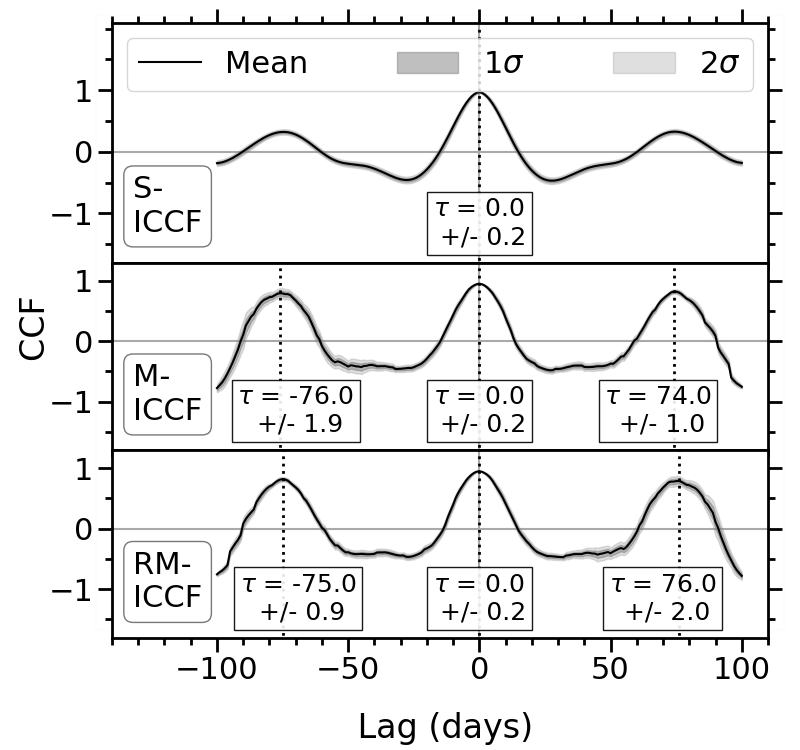} \\
        \includegraphics[width=\textwidth]{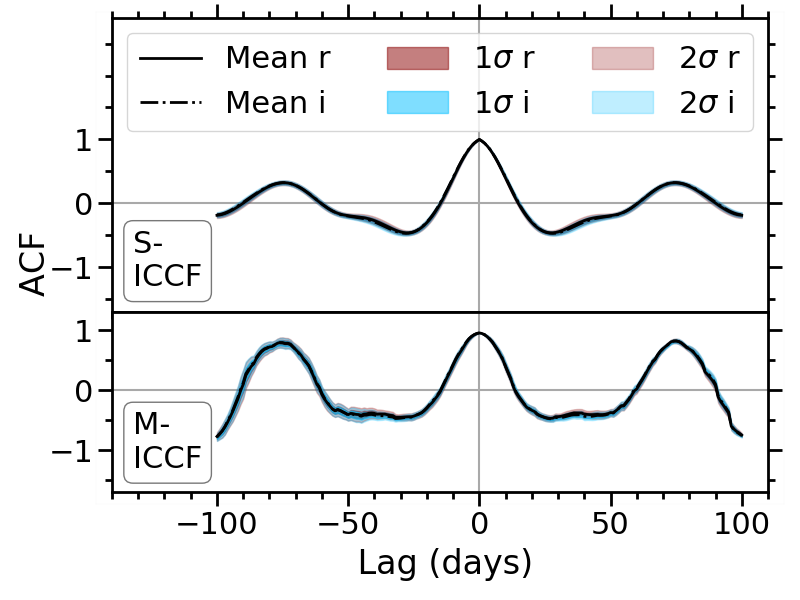}
    \end{minipage}} 
    \captionof{figure}{\textit{Top Panels}: Mean CCFs of the \emph{r} and \emph{i} light curves in the season starting 2017 and in the entire observation period between 2013 and 2019. \textit{Lower panels}: The corresponding ACFs. Here, the \mbox{M-ICCF} method refers to the interpolated \emph{r} band, and the \mbox{RM-ICCF} method refers to the interpolated \emph{i} band. \label{fig:ri_CCFs_years}}
    \end{minipage}
    \hspace{0.2cm}
    \begin{minipage}{0.32\textwidth}
            \subfloat[][CCFs and ACFs of the entire \emph{J} and \emph{Ks} light curves. \label{fig:JKs_CCF_all}]{
            \begin{minipage}{\textwidth}
                \includegraphics[width=\textwidth]{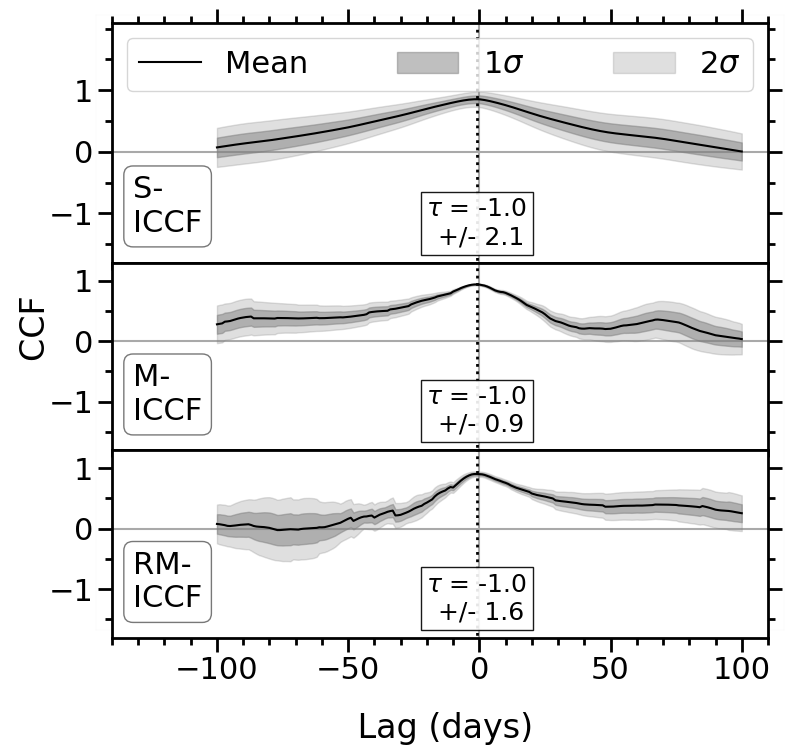} \\
                \includegraphics[width=\textwidth]{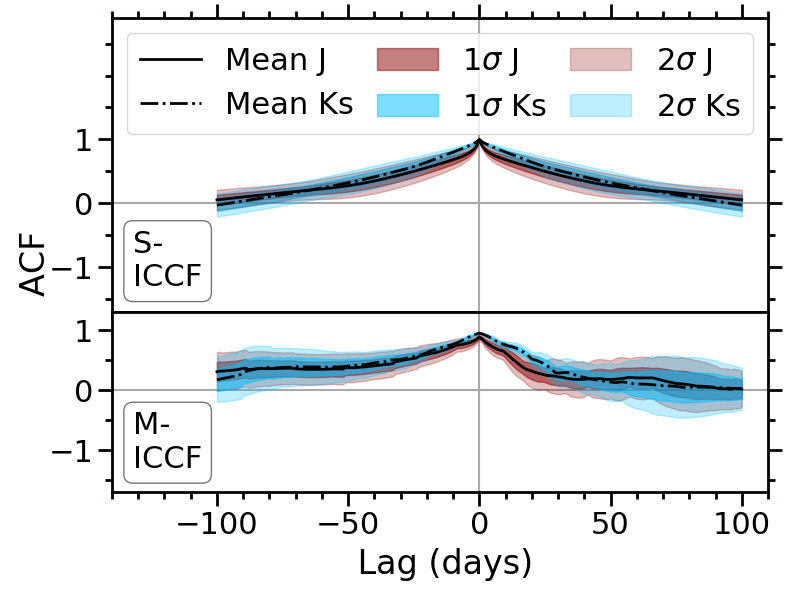}
            \end{minipage}} \\
            \subfloat[][CCFs and ACFs of the 2017 \emph{J} and \emph{Ks} light curves. \label{fig:JKs_CCF_2017}]{
            \begin{minipage}{\textwidth}
                \includegraphics[width=\textwidth]{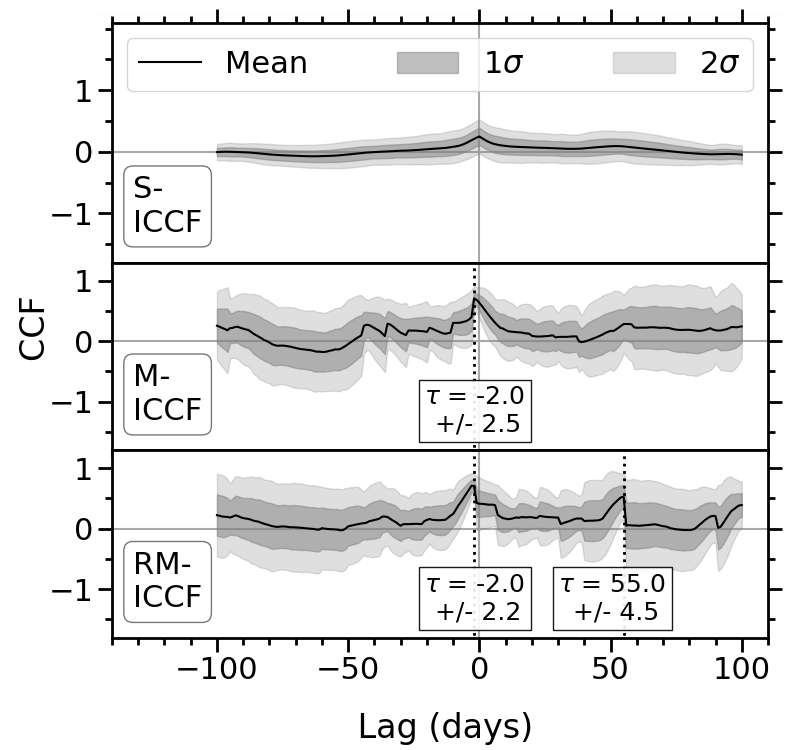} \\
                \includegraphics[width=\textwidth]{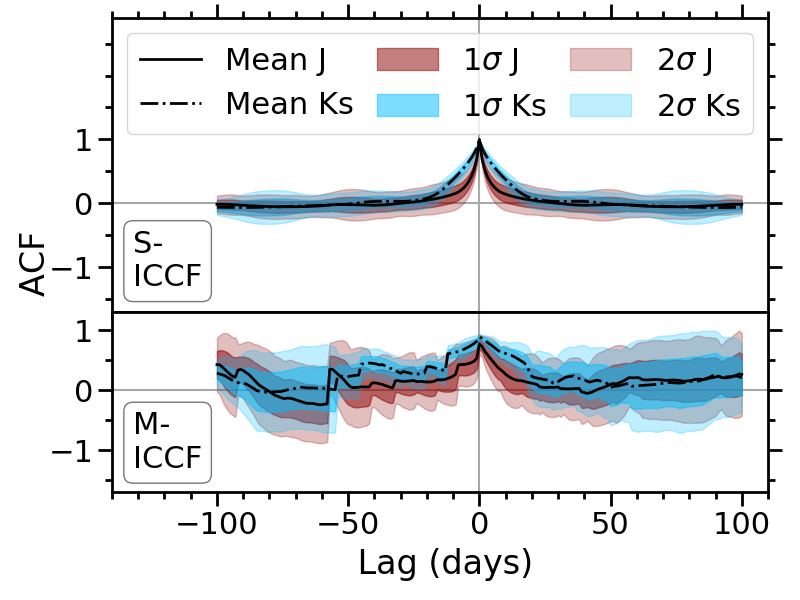}
            \end{minipage}} 
        \caption{\small \textit{Top Panels}: Mean CCFs of \emph{J} and \emph{Ks} light curves in the season starting 2017 and in the entire observation period between 2017 and 2019. \textit{Lower Panels}: The corresponding ACFs. Here, the \mbox{M-ICCF} method refers to the interpolated \emph{J} band and the \mbox{RM-ICCF} method refers to the interpolated \emph{Ks} band. \label{fig:JKs_CCF}}
    \end{minipage}
    \hspace{0.2cm}
    \begin{minipage}{0.32\textwidth}
        \centering
        \subfloat[][CCFs and ACFs of the combined 2017 and 2018 \emph{r} and \emph{Ks} light curves. \label{fig:rKs_CCF_all}]{
            \begin{minipage}{\textwidth}
                \includegraphics[width=\textwidth]{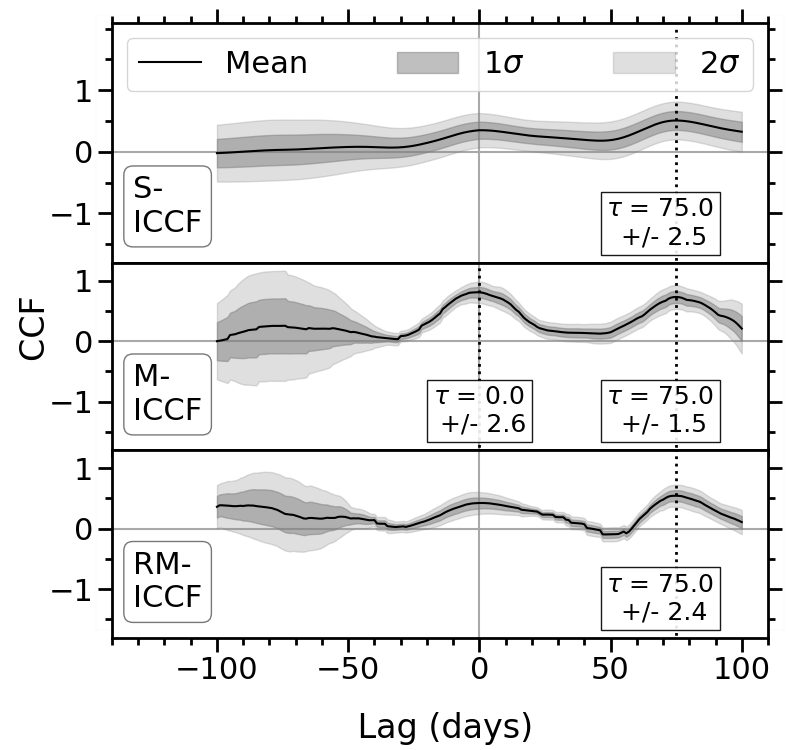} \\
                \includegraphics[width=\textwidth]{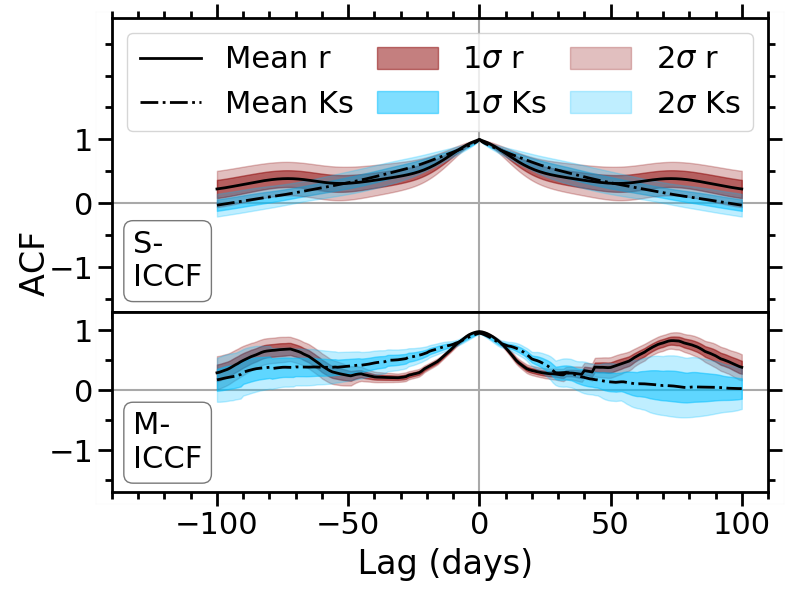}
             \end{minipage}} \\
        \subfloat[][CCFs and ACFs of the 2017 \emph{r} and \emph{Ks} light \\ curves. \label{fig:rKs_CCF_2017}]{
            \begin{minipage}{\textwidth}
                \includegraphics[width=\textwidth]{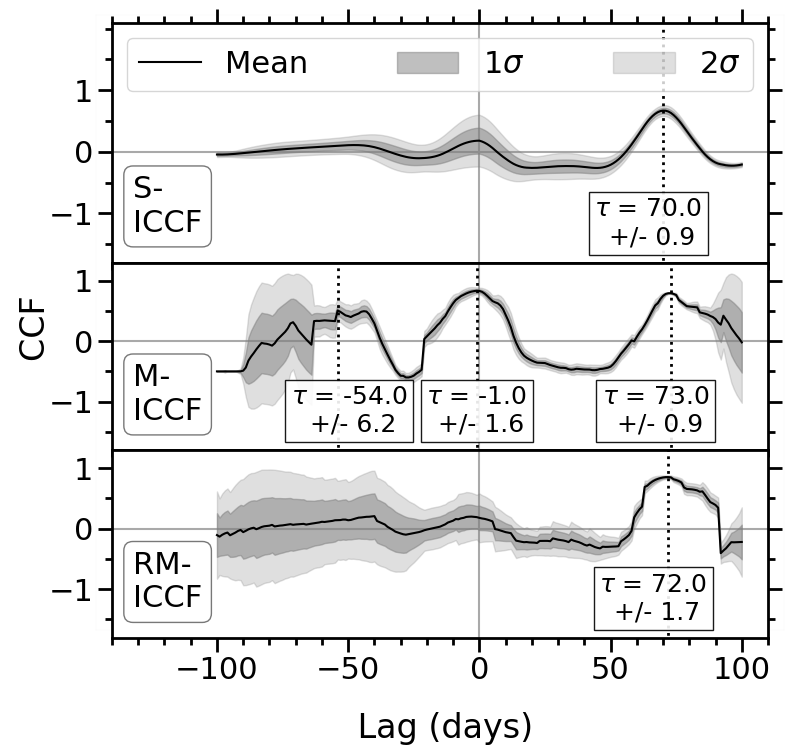} \\
                \includegraphics[width=\textwidth]{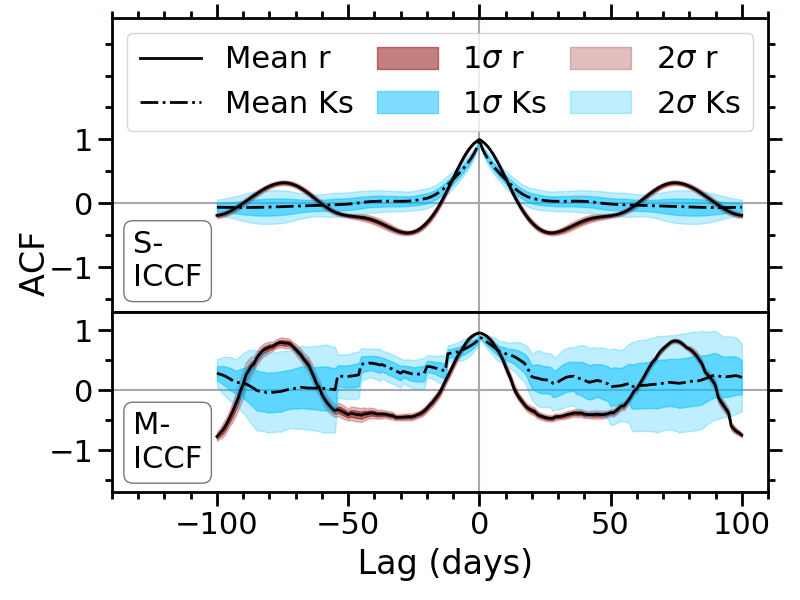}
            \end{minipage}}
        \caption{\textit{Top Panels}: Mean CCFs of \emph{r} and \emph{Ks} light curves in the season starting 2017 and in the entire observation period between 2017 and 2019 \textit{Lower Panels}: The corresponding ACFs. Here, the \mbox{M-ICCF} method refers to the interpolated \emph{r} band and the \mbox{RM-ICCF} method refers to the interpolated \emph{Ks} band. \label{fig:rKs_CCF}}
        \vspace{3mm}
    \end{minipage}
\end{figure*}

The entire light curves of each combination of filters in both the optical and NIR were interpolated 10,000 times and cross correlated using the methods described above, as well as the light curves from individual years. Each season was tested with possible observed lags between $\pm$100 days due to the length of the individual season light curves. The CCFs were also compared with the autocorrelation functions (ACFs) of each light curve to determine whether the peaks in the CCFs were a result of a lag between the light curves, or an effect of quasi-periodicity within the individual light curve. Potential lags that were measured from peaks on the mean CCFs were considered positive detections if the CCF values were greater than 0.5. This was chosen as the limit as most non-zero peaks in the ACFs had values smaller than this, with the exception of ACFs where the optical 2017 or 2018 season light curves were included. The possible lags that were classified as positive detections are labelled on the plots, with the uncertainties calculated as the standard deviations of the peak of the CCF for each interpolation around the peak of the mean CCF.

In this section, the cross-correlations of the \emph{r} and \emph{i}, \emph{r} and \emph{Ks} and \emph{J} and \emph{Ks} band light curves are discussed to represent the emission between the optical filters, the optical with NIR filters and the NIR filters with each other, as these were the light curves with the most observations.

\subsubsection{Cross-Correlations of r and i Bands}

The cross-correlation between the \emph{r} and \emph{i} band light curves and their ACFs are displayed in Figure \ref{fig:ri_CCFs_years}, for the entire observational period and the individual season starting 2017. The CCFs of the other individual observation seasons are displayed in Figure \ref{fig:ri_CCFs_other_years} in Appendix \ref{ap:more_lags}. Possible lags are measured from the CCFs of the entire \emph{r} and \emph{i} band light curves in Figure~\ref{fig:ri_CCFs_all} with values of \mbox{0.0 $\pm$ 0.4}, \mbox{-1.0 $\pm$ 0.4} and \mbox{0.0 $\pm$ 0.4} days with the \mbox{S-ICCF}, \mbox{M-ICCF} and \mbox{RM-ICCF} methods respectively, and at \mbox{-72.0 $\pm$ 2.7} and \mbox{72.0 $\pm$ 1.8} days with the \mbox{M-ICCF} and \mbox{RM-ICCF} methods respectively. To further investigate these lags, the CCFs of the individual observation seasons of the \emph{r} and \emph{i} band light curves were analysed to determine whether the lag between light curves remained constant over every year, and to reduce the impact of the interpolations between observation seasons. 

The strong correlation at $\sim$ 0 days is consistently present in the CCFs of each year, except the season starting in 2016 as discussed in Appendix \ref{ap:more_lags}, with an overall mean value of -0.1$\pm$0.2 days. This implies that the emission in both filters is co-temporal, or any delay between the emission regions is on timescales smaller than the cadences of observations. The $\sim \pm$ 75 day lags are also observed in all of the 2017 CCFs which are displayed in Figure~\ref{fig:ri_CCF_2017}, and in 2018 \mbox{M-ICCF} and \mbox{RM-ICCF} in Figure~\ref{fig:ri_CCF_2018}, however this is assumed to be due to the shape of the light curves, as in the 2017 light curves there are 2 peaks separated by $\sim$ 75 days, and in the 2018 season the light curves follow a decline that tapers off for a short period before declining again. This is also seen in the corresponding ACFs of each light curve in the 2017 and 2018 seasons, therefore implying that it is not a delay between emission regions but a consequence of aliasing.

\subsubsection{Cross-Correlations of J and Ks Bands}

The cross-correlation between the \emph{J} and \emph{Ks} band light curves for the individual season starting in 2017 and the entire light curves are displayed in Figure~\ref{fig:JKs_CCF}, along with their ACFs, and the individual 2018 season is displayed in Figure~\ref{fig:JKs_CCF_2018}. These CCFs show the presence of a strong correlation at $\sim$ 0 days in all seasons with a mean value of -0.8$\pm$1.0 days, as well as possible lags at $\sim$ 55 days in the 2017 season, and at $\sim$ 85 days in the 2018 season. The $\sim$ 0 day correlation is seen consistently in the \mbox{M-ICCF} and \mbox{RM-ICCF} methods over all seasons, however is not observed in the 2017 \mbox{S-ICCF} method, which is assumed to be due to large gaps in the light curves and the interpolations between these observations diluting the overall correlation. The 55.0 $\pm$ 4.5 day lag in the 2017 season is only recorded in the \mbox{RM-ICCF} method. Furthermore, the 2017 CCFs in Figure~\ref{fig:JKs_CCF_2017} are relatively flat, especially in the \mbox{S-ICCF} method which indicates that the lags found are not very distinctive. The $\sim$ 85 day lag in the \mbox{M-ICCF} and \mbox{RM-ICCF} methods of the 2018 season, shown in Figure~\ref{fig:JKs_CCF_2018}, are also present in the ACFs of the \emph{J} and \emph{Ks} light curves, which implies that the lag is not between the different filters but is due to aliasing, for example, the dip at $\sim$ MJD 58300 days and the dips at $\sim$ MJD 58400 days follow similar shapes and therefore correlate with each other. It therefore follows that this peak is more pronounced in the \mbox{M-ICCF} and \mbox{RM-ICCF} methods as the interpolations between observations could dilute the correlation found here. As this lag is only present in this scenario and not in the 2017 season or overall light curve, it can be assumed that it is not a delay between the \emph{J} and \emph{Ks} band light curves.

\subsubsection{Cross-Correlations of r and Ks Bands}

The results of the cross-correlations between the \emph{r} and \emph{Ks} band light curves are displayed in Figure~\ref{fig:rKs_CCF} along with their ACFs for the 2017 season and the entire overlapping observational period. The CCFs and ACFs of the 2018 season are displayed in Figure~\ref{fig:rKs_CCF_2018}. The \emph{r} and \emph{Ks} band CCFs show the presence of strong correlations at $\sim$ 0 days, with a mean of 0.0$\pm$1.2 days, and at $\sim$ $\pm$ 75 days, however these measured lags are not consistent over all methods and all seasons of the \emph{r} and \emph{Ks} band light curves, therefore, further analysis of these lags was performed.

In the \mbox{S-ICCF} methods, the light curves are not as well correlated at 0 days, which could be due to the interpolated epochs equally impacting the CCFs, and therefore reducing the overall correlation. Furthermore, there is a lower correlation at 0 days for \mbox{RM-ICCF} method when the 2017 season is used, but this is suspected to be due to the lack of observations in the NIR when the optical light curve displays the first of 2 peaks. As the NIR light curve is interpolated and epochs extracted matching the observed optical epochs, the interpolations during the unobserved month will reduce the overall correlation.

The $\sim$ $\pm$ 75 day lags only appear when the 2017 \emph{r} band observation season, which contains two peaks separated by $\sim$ 75 days, is included in the CCF. This can be seen in the ACFs, as the ACF of the \emph{r} band in the 2017 season from both \mbox{S-ICCF} and \mbox{M-ICCF} methods also contains peaks at $\sim$ $\pm$ 75 days, as does the ACF of the entire \emph{r} band light curve, although the correlations are not as strong. The \emph{Ks} band ACFs do not include the $\sim$ $\pm$ 75 day lags however, which is likely due to a lack of observations in the NIR band during the month that corresponded to the first peak in the \emph{r} band. The presence of the \mbox{$\sim$ $\pm$ 75 day} lags was further investigated in Appendix~\ref{ap:test75daylag} using the SF of the \emph{r} and \emph{Ks} bands from observations between 2017-2019. By simulating light curves using the method described by \cite{TimmerKonig1995} with a range of power spectra with varied properties including break frequencies, slopes, and white noise amplitudes, it is found that the $\sim$ 75 day timescale does not depend on specific properties of the light curve of \mbox{PKS 0027-426}, but occurred for a random $\sim$ 12 and 14\% of the 10,000 simulated light curves in the \emph{r} and \emph{Ks} bands respectively.

\section{Spectral Variability}
\label{Sect:Colour}

\begin{figure*}
    \begin{minipage}{\textwidth}
        \captionof{table}{The slopes, Spearman rank coefficients, probability of no correlation, and colour trend of the DES filters in the colour vs \emph{r} magnitudes plots in Figure~\ref{fig:colmag}.}
    	\subfloat[table][The slopes, Spearman rank coefficients, probability of no correlation, and colour trend for each combination of DES filters in the colour vs \emph{r} magnitudes plots in Figure~\ref{fig:optcol}. \label{tab:optcol}]{
    	\begin{minipage}{0.46\textwidth}
            \begin{tabular}{C{0.08\textwidth}C{0.23\textwidth}C{0.11\textwidth}C{0.2\textwidth}C{0.18\textwidth}}
                \hline
                    Colour & Slope of Colour vs \emph{r} mag & $\rho$-value & $p$-value & Colour Trend \\
                \hline
                   \emph{g}-\emph{r} & -0.30 $\pm$ 0.02 & -0.60 & 2.45 $\times 10^{-15}$ & RWB \\
                   \emph{g}-\emph{i} & -0.36 $\pm$ 0.02 & -0.52 & 5.50 $\times 10^{-11}$ & RWB \\
                   \emph{g}-\emph{z} & -0.25 $\pm$ 0.02 & -0.33 & 1.67 $\times 10^{-8}$ & RWB \\
                   \emph{r}-\emph{i} & -0.06 $\pm$ 0.01 & -0.26 & 1.73 $\times 10^{-3}$ & RWB \\
                   \emph{r}-\emph{z} & 0.06 $\pm$ 0.01 & 0.31 & 1.25 $\times 10^{-7}$ & BWB\\
                   \emph{i}-\emph{z} & 0.12 $\pm$ 0.01 & 0.73 & 5.06 $\times 10^{-48}$ & BWB \\
                \hline
            \end{tabular}
            \vspace{0.5cm}
        \end{minipage}}
    	\hspace{8mm}
    	\subfloat[table][The slopes, Spearman rank coefficients, probability of no correlation, and colour trend for each season of DES in the \emph{g}-\emph{z} colour vs \emph{r} magnitude plots in Figure~\ref{fig:gz_sep_years}. \label{tab:gz_sep_years}]{
    	\begin{minipage}{0.48\textwidth}
            \begin{tabular}{C{0.08\textwidth}C{0.23\textwidth}C{0.11\textwidth}C{0.2\textwidth}C{0.18\textwidth}}
                \hline
                    Season & Slope of \emph{g-z} vs \emph{r} mag & $\rho$-value & $p$-value & Colour Trend \\
                \hline
                   2013 & 0.74 $\pm$ 0.06 & 0.8 & 5.77 $\times 10^{-13}$ & BWB \\
                   2014 & 0.06 $\pm$ 0.06 & 0.09 & 0.52 & SWB \\
                   2015 & 0.12 $\pm$ 0.09 & 0.17 & 0.18 & SWB \\
                   2016 & -0.14 $\pm$ 0.02 & -0.77 & 3.58 $\times 10^{-12}$ & RWB \\
                   2017 & -0.26 $\pm$ 0.02 & -0.78 & 3.68 $\times 10^{-11}$ & RWB \\
                \hline
            \end{tabular}
    	\end{minipage}}
	\end{minipage}
    \vspace{1mm}
	\begin{minipage}{\textwidth}
        \subfloat[figure][Optical colour variability plots of each colour combination of DES light curves for their entire observational periods.  \label{fig:optcol}]{
        	\includegraphics[width=0.49\textwidth]{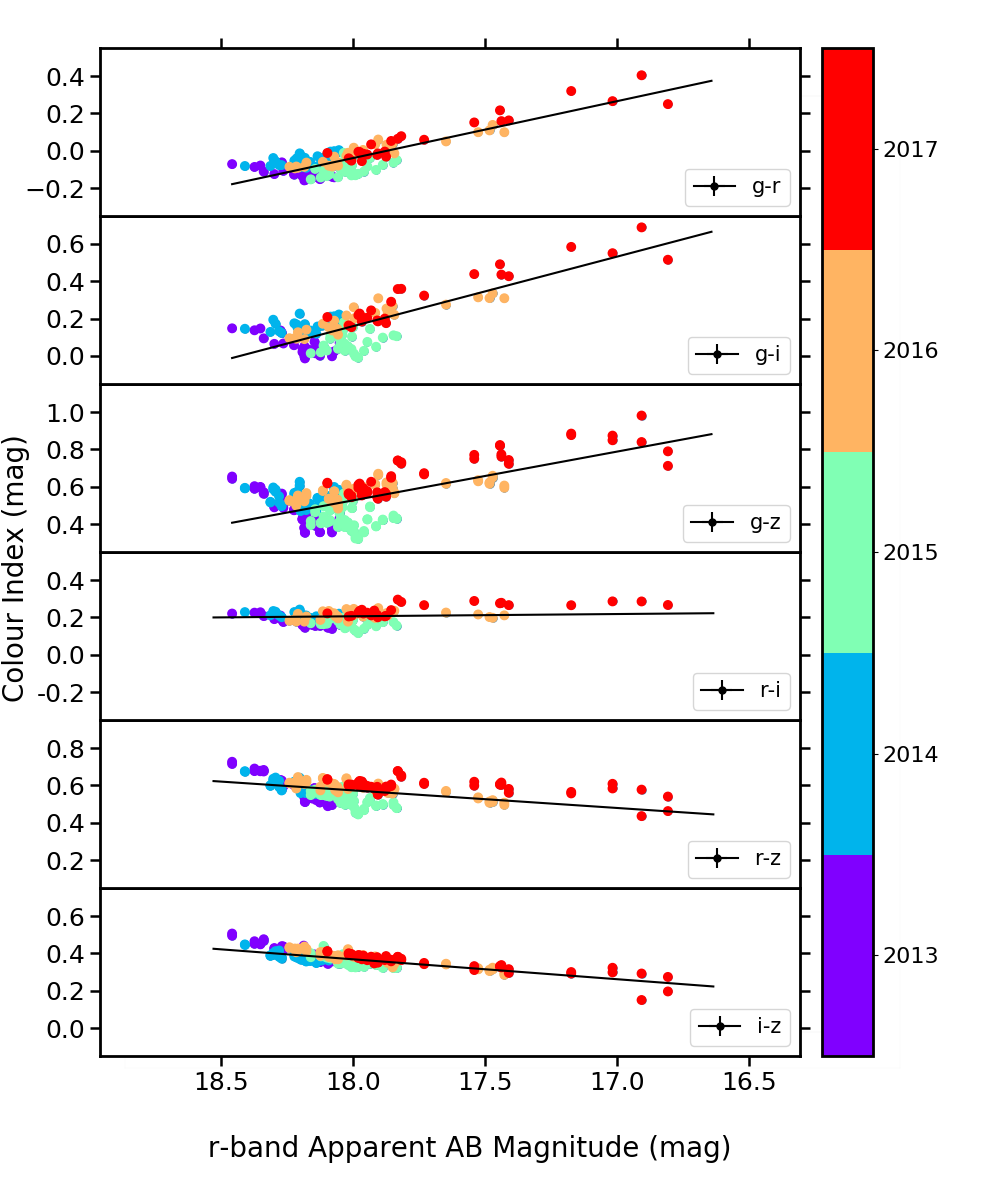}}
        \hfill
        \subfloat[figure][Optical \emph{g}-\emph{z} colour variability for each observation season of DES.
            \newline \label{fig:gz_sep_years}]{\includegraphics[width=0.49\textwidth]{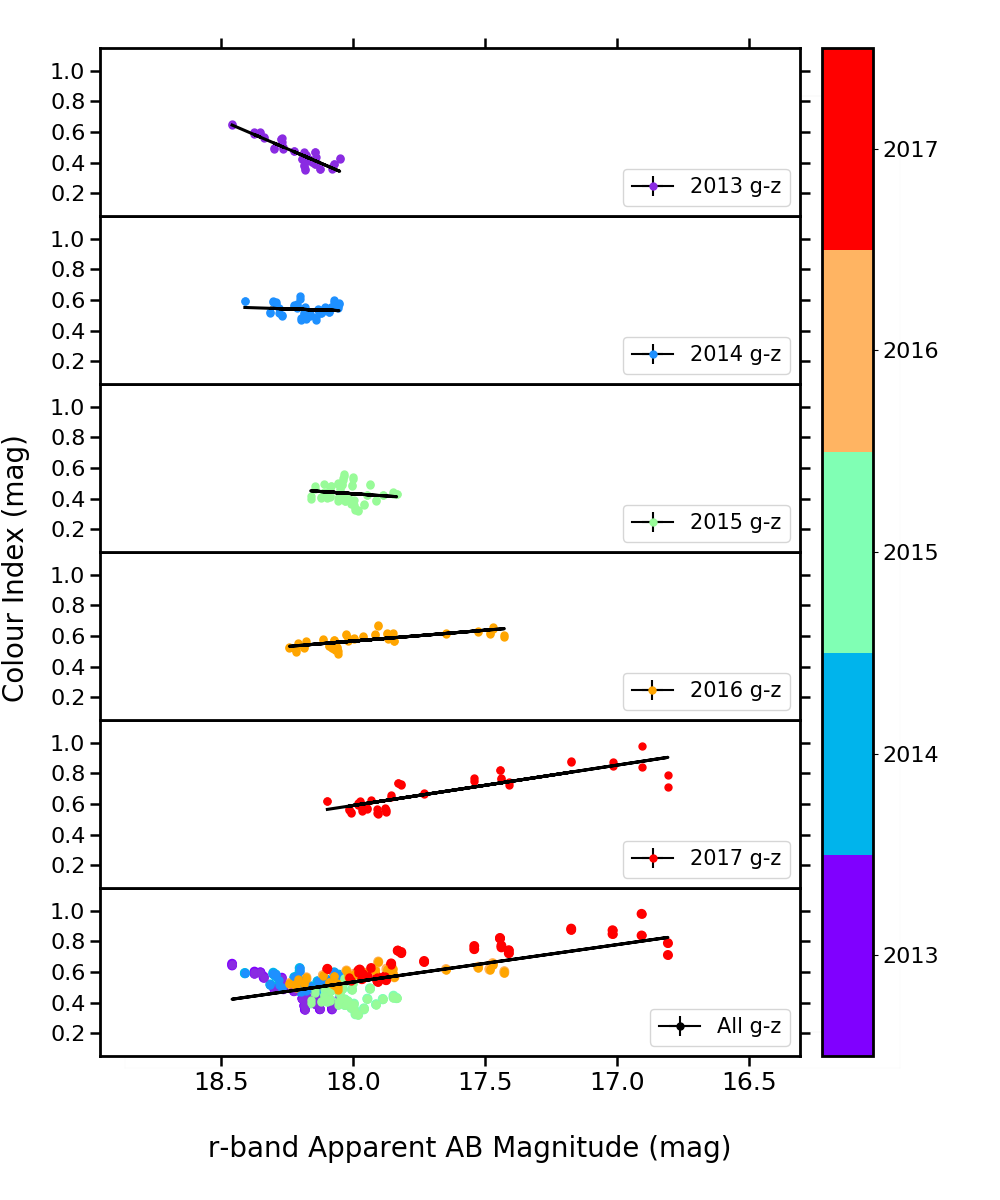}}
        \captionof{figure}{Optical colour variability plots of PKS~0027-426 in DES. The colours of the data points correspond to each observation season. \label{fig:colmag}}
    \end{minipage}
\end{figure*}

\subsection{Optical Colour Variability from DES}

The optical colour behaviour of PKS~0027-426 was studied by measuring the colour indices of each combination of the observed \emph{griz} bands, starting with DES data. Each colour index from DES was calculated using quasi-simultaneous observations from the same instrument that were first corrected for galactic extinction using the measurements from \cite{schlafly2011}, and had an average time difference of $\sim$ 12 minutes between observations in the different bands, and a maximum time difference of $\sim$ 18.5 minutes. 

The optical colour indices (\emph{g-r}, \emph{g-i}, \emph{g-z}, \emph{r-i}, \emph{r-z} and \emph{i-z}) were plotted against the \emph{r} magnitude in Figure~\ref{fig:optcol}, and the colour behaviour was quantified using the slope of the plot, the Spearman rank correlation coefficient ($\rho$-values) and the probability of no correlation ($p$-values). The slope here is calculated as described by \cite{Kelly2007} using a Bayesian method of linear regression, excluding outliers as explained in Appendix~\ref{sec:app_des_colmag}. The Spearman rank correlation coefficient is a non-parametric measure of the strength and direction of the relationship between variables, which returns a value between $\pm$1, where values of 0 corresponds to no correlation, and correlations of $\pm$1 is an exact monotonic relationship. A large $p$ value indicates a high probability of no correlation and a small $p$ value indicates a low probability that the correlation is due to random noise. Positive slopes and Spearman rank correlation coefficients imply the colour increases with decreasing brightness (increasing magnitude), which corresponds to a BWB trend, while negative slopes and Spearman rank correlation coefficients imply RWB trends. These possible colour trends were deemed significant if the linear slopes were consistent within 3$\sigma$. The plots with colour trends that were not significantly BWB or RWB within 3$\sigma$ have corresponding $p$-values that were relatively large ($p > 0.01$) and $\rho$-values that were relatively small ($|\rho| < 0.4$) compared to those with significant colour trends, which therefore imply a SWB trend. The data points in Figure~\ref{fig:optcol} were coloured according to observation season. Table~\ref{tab:optcol} contains the slope of each colour-magnitude plot, the Spearman rank correlation coefficient, the probability that no correlation is present between the colour and \emph{r} magnitude, and the colour trend. 

The overall \emph{g}-\emph{r}, \emph{g}-\emph{i} and \emph{g}-\emph{z} colours are shown to be RWB by the strong anti-correlations present, with slopes of \mbox{-0.30 $\pm$ 0.02}, \mbox{-0.36 $\pm$ 0.02} and \mbox{-0.25 $\pm$ 0.02} respectively, and $\rho$ values of \mbox{-0.60}, \mbox{-0.52} and \mbox{-0.33} respectively. The \emph{r}-\emph{i} colour has a slope of \mbox{-0.06 $\pm$ 0.01} and \mbox{$\rho$ = -0.26} which indicates a slight RWB trend, but also has a relatively high probability of no correlation with a $p$ value of \mbox{1.73 $\times 10^{-3}$}, which implies this RWB trend is not as significant as in the \emph{g}-\emph{r}, \emph{g}-\emph{i} and \emph{g}-\emph{z}. Finally, the \emph{r}-\emph{z} and \emph{i}-\emph{z} colours show positive correlations which implies the source becomes BWB with slopes of \mbox{0.06 $\pm$ 0.01} and \mbox{0.12 $\pm$ 0.01} and $\rho$ values of 0.31 and 0.73 respectively. 

The colour behaviours of each season are also shown to vary; for example, Figure~\ref{fig:gz_sep_years} displays plots of the colour behaviour of \emph{g}-\emph{z} in the different observations seasons of DES. Table~\ref{tab:gz_sep_years} contains the slope, the Spearman rank correlation coefficient, the probability of no correlation for each season of \emph{g}-\emph{z}, and the colour trend. The 2013 season shows a strong BWB trend, with slope and $\rho$ values of \mbox{0.74 $\pm$ 0.06} and 0.8 respectively. The 2014 and 2015 seasons both display a small positive correlation with slopes of \mbox{0.06 $\pm$ 0.06} and \mbox{0.12 $\pm$ 0.09} respectively, and $\rho$-values of 0.09 and 0.17 respectively, however the positive slopes are not significant within 3$\sigma$ uncertainties, and they also have a large probability of no correlation with $p$ values of 0.52 and 0.18 respectively, which indicates the colour behaviour follows a SWB trend. The seasons starting in 2016 and 2017 demonstrate RWB trends, with slopes of \mbox{-0.14 $\pm$ 0.02} and \mbox{-0.26 $\pm$ 0.02}, and $\rho$ values given as \mbox{-0.77} and \mbox{-0.78} respectively. The \emph{g}-\emph{z} colour behaviour over the entire observational period is shown to follow a RWB trend due to the 2016 and 2017 seasons which are the brightest, most variable seasons and therefore dominate the correlation. The colour-magnitude plots and tables containing slopes of each season of the remaining combination of optical DES \emph{griz} filters are given in Appendix~\ref{sec:app_des_colmag}, and a summary of the colour trends of each combination of filters in each observation season is given in Table~\ref{tab:coloursummary}.

\subsection{Optical Colour-Magnitude Plots from Interpolation}

\begin{figure*}
    \begin{minipage}{0.47\textwidth}
        \centering
        \captionof{table}{Colour behaviour of each combination of optical \emph{griz} filters in each season. \label{tab:coloursummary}}
        \begin{tabular}{C{0.1\columnwidth}C{0.12\columnwidth}C{0.12\columnwidth}C{0.12\columnwidth}C{0.12\columnwidth}C{0.12\columnwidth}}
        \hline
           Colour & 2013 & 2014 & 2015 & 2016 & 2017 \\
        \hline
           \emph{g}-\emph{r} & BWB & RWB & SWB & RWB & RWB \\
           \emph{g}-\emph{i} & BWB & SWB & SWB & RWB & RWB \\
           \emph{g}-\emph{z} & BWB & SWB & SWB & RWB & RWB \\
           \emph{r}-\emph{i} & BWB & SWB & SWB & SWB & RWB \\
           \emph{r}-\emph{z} & BWB & BWB & BWB & BWB & BWB \\
           \emph{i}-\emph{z} & BWB & BWB & BWB & BWB & BWB \\
        \hline
        \end{tabular}
    \end{minipage}
    \hspace{0.3cm}
    \begin{minipage}{0.47\textwidth}
        \centering
        \captionof{table}{Comparison between the slopes of the colour-magnitude plots and those made from 10,000 interpolated light curves. The uncertainties of the interpolated slopes are the 1$\sigma$ uncertainties. \label{tab:histsDES5yr}}
        \begin{tabular}{C{0.1\columnwidth}C{0.25\columnwidth}C{0.25\columnwidth}C{0.25\columnwidth}}
        \hline
           Colour Index  & Actual Slope & Mean Slope from Filter 1 & Mean Slope from Filter 2 \\
        \hline
           \emph{g}-\emph{r} & -0.30 $\pm$ 0.02 & -0.39 $\pm$ 0.09 & -0.18 $\pm$ 0.13 \\
           \emph{g}-\emph{i} & -0.36 $\pm$ 0.02 & -0.45 $\pm$ 0.09 & -0.24 $\pm$ 0.14  \\
           \emph{g}-\emph{z} & -0.25 $\pm$ 0.02 & -0.32 $\pm$ 0.09 & -0.21 $\pm$ 0.05 \\
           \emph{r}-\emph{i} & -0.06 $\pm$ 0.01 & -0.18 $\pm$ 0.13 & 0.06 $\pm$ 0.13  \\
           \emph{r}-\emph{z} & 0.06 $\pm$ 0.01 & -0.05 $\pm$ 0.13 & 0.09 $\pm$ 0.05  \\
           \emph{i}-\emph{z} & 0.12 $\pm$ 0.01 & 0.01 $\pm$ 0.14 & 0.15 $\pm$ 0.05  \\
        \hline
        \end{tabular}
    \vspace{0.5cm}
    \end{minipage}
    \begin{minipage}{\textwidth}
        \subfloat[figure][The distributions from 10,000 interpolations of each combination of DES light curves for their entire observational periods.\label{fig:hists5years}]{\includegraphics[width=0.49\textwidth]{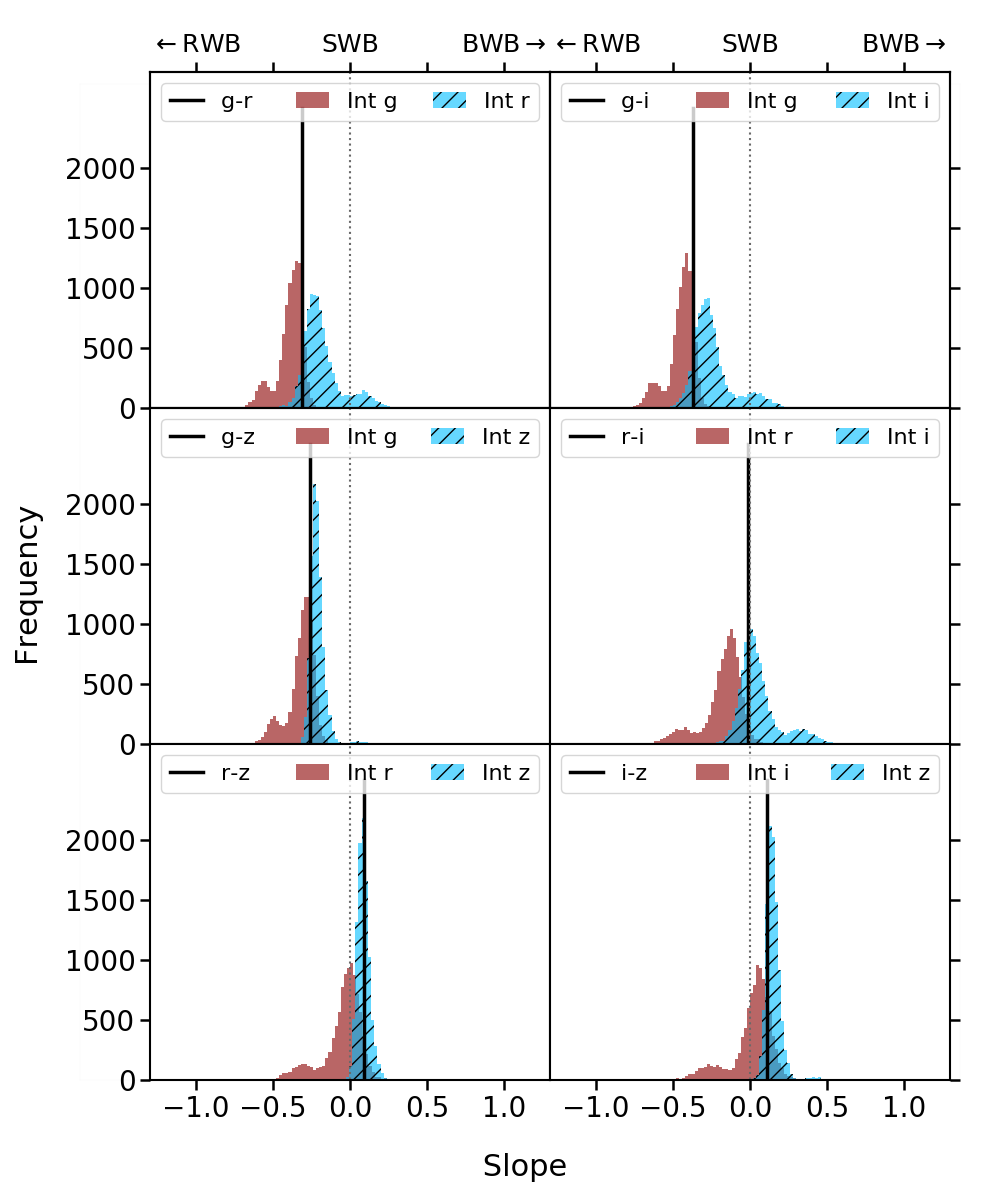}}
        \hfill
        \subfloat[figure][The distributions from 10,000 interpolations of each season of light curve of g-z.\label{fig:hists1yrDES}]{\includegraphics[width=0.49\textwidth]{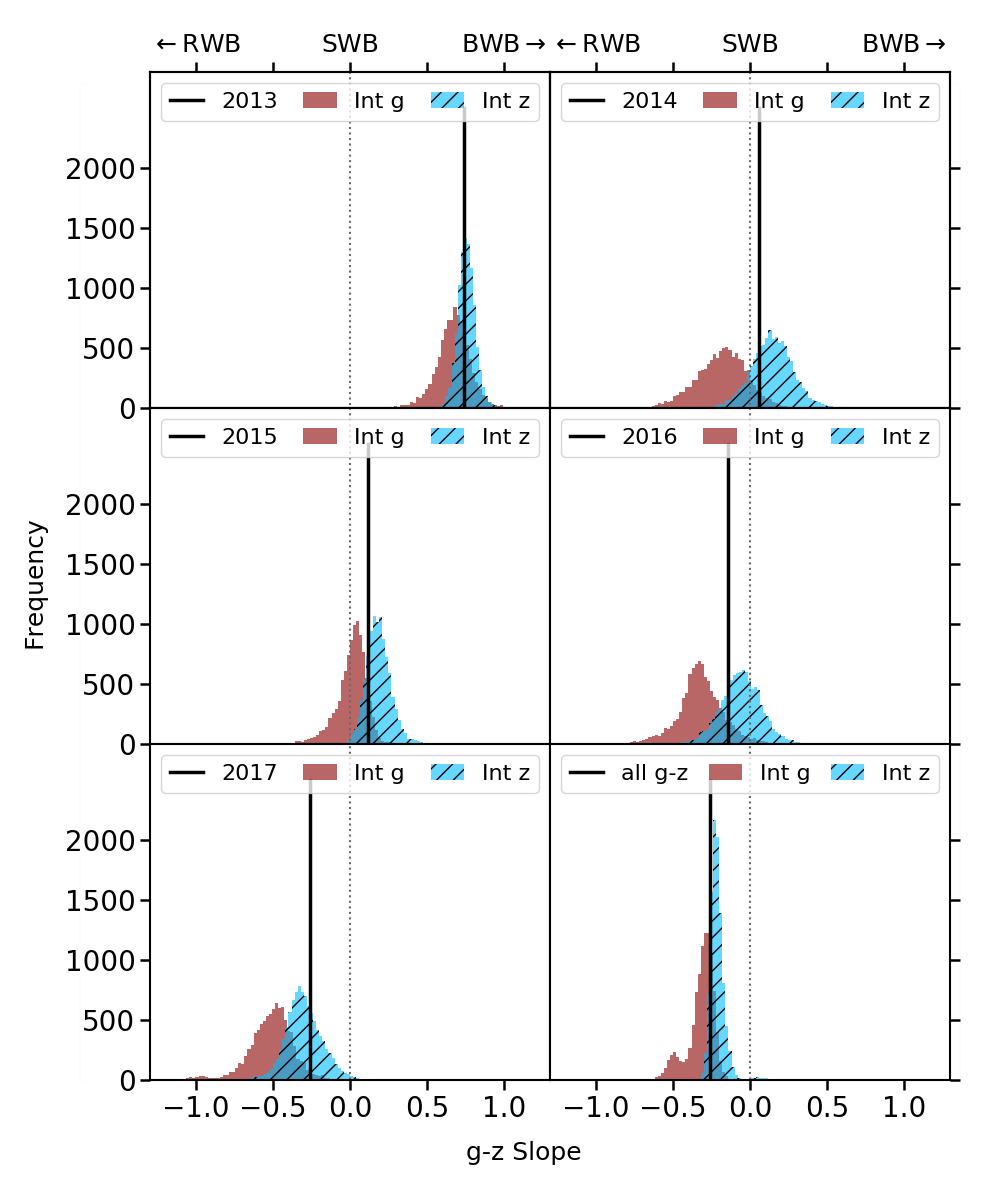}}
        \caption{The distributions of the slopes returned from the colour-magnitude plots created from 10,000 interpolations of each light curve compared to the slopes measured from Figure~\ref{fig:optcol}.}
    \end{minipage}
\end{figure*}

The colour indices of PKS~0027-426 were also studied for each combination of observations in the optical and the NIR, however, the optical and NIR were not observed simultaneously or even quasi-simultaneously as is necessary for studying the colour behaviour of highly variable objects such as blazars. Therefore, one of the light curves was interpolated so that quasi-simultaneous observations could be used. 

To test the reliability of obtaining the colour behaviour from interpolated light curves, the optical colour variability from interpolated DES light curves were studied first, by removing half of the observations (randomly selected) from one light curve, and interpolating the remaining light curve using the structure function method to produce a light curve with a cadence of 1 day, as explained in Section~\ref{Sect:Time}. The dates matching the observations from the second filter were then used to recreate the colour-magnitude plots.

This method was repeated 10,000 times for each light curve. The distribution of slopes given from the colour-magnitude plots from interpolated DES light curves are shown in Figure~\ref{fig:hists5years} and the results of these were then compared to the values obtained from Figure~\ref{fig:optcol} in Table~\ref{tab:histsDES5yr}. The results from each of the interpolated light curves are shown to be consistent with each other and within 1$\sigma$ uncertainties of the results from Figure~\ref{fig:optcol}. It is shown in Figure~\ref{fig:hists5years} that distributions from interpolating filter 1 return a smaller slope of the \mbox{filter 1 - filter 2} colour index vs r band magnitude than the distributions from interpolating filter 2, and often the result from interpolating filter 1 and filter 2 return a smaller and larger slope than the actual measured slope from Figure \ref{fig:optcol} respectively. This could be due to the interpolations underestimating some of the larger variability in the light curves, which therefore means that when filter 1 is interpolated it could be less variable than filter 2 during these regions of large variability and therefore the \mbox{filter 1 - filter 2} colour is smaller, and vice versa for filter 2. Furthermore, some of the histograms in Figure~\ref{fig:hists5years} also display a secondary, smaller peak, which corresponds to a slope less than the slope of the main peak when filter 1 is interpolated and a slope greater than the slope of the main peak when filter 2 is interpolated. Analysis of this peak has shown that it occurs when the brightest points in the light curve are removed before interpolation, specifically the dates between MJD 58014 and 58025. This results in a smaller peak to the left when filter 1 is interpolated as the interpolated light curve is then shown to vary less than filter 2, hence the slope becomes steeper, and the smaller peak to the right when filter 2 is interpolated as in this case the slope becomes shallower as the subtracted filter varies less. 

To further investigate this method, it was then replicated for each colour combination in each individual season of DES, for example, Figure~\ref{fig:hists1yrDES} shows the distribution of the slope of \emph{g}-\emph{z} in each season of DES. It was found that 82\% of the mean slopes from interpolations of individual years for all combination of filters were consistent with the slopes from individual years within 1$\sigma$ and all were consistent within 1.5$\sigma$. Comparisons between the slope obtained using all observations and using the interpolation method are displayed in Appendix~\ref{sec:app_des_interps}, for each season in each colour combination.

\begin{figure}
    \centering
    \includegraphics[width=\columnwidth]{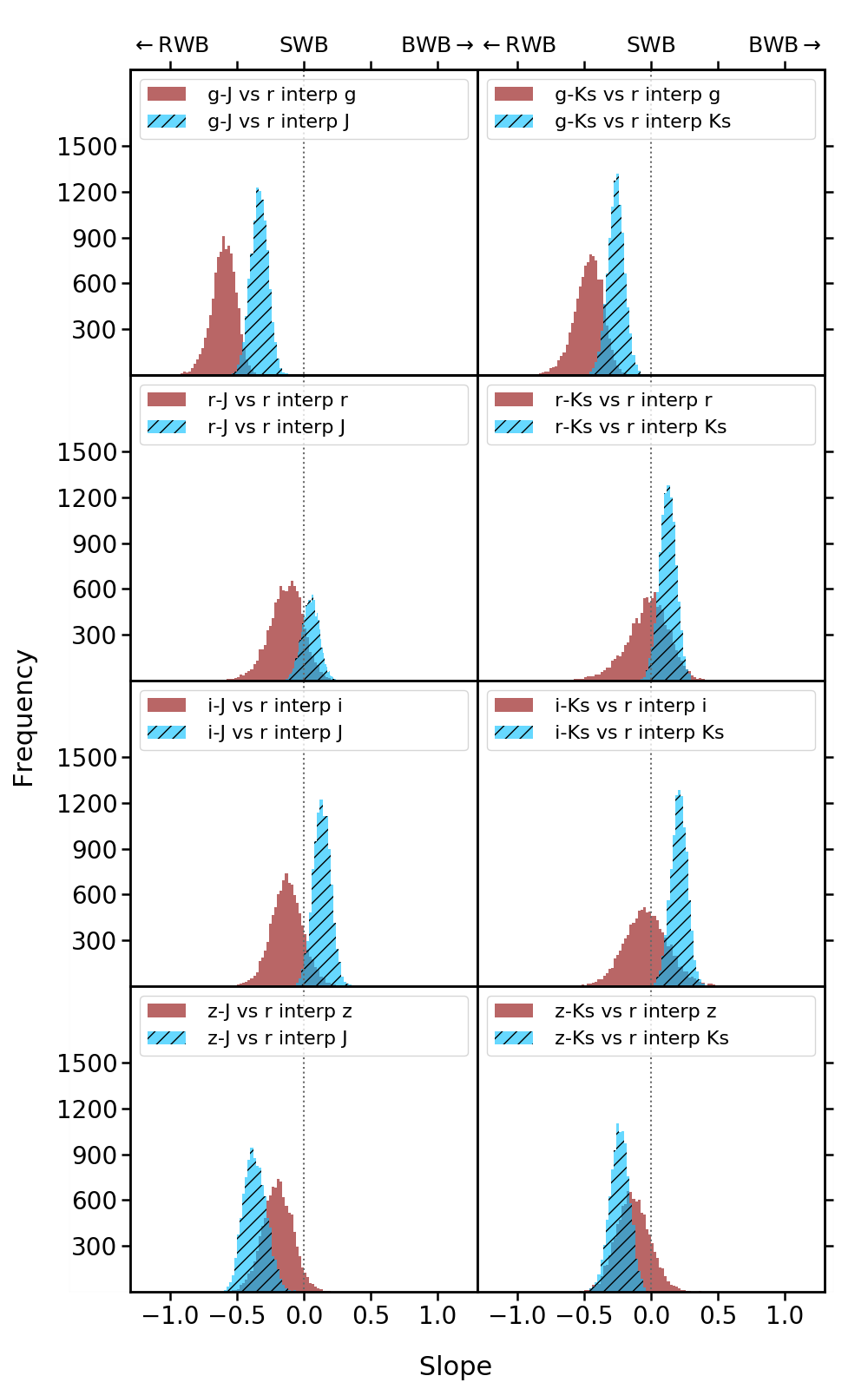}
    \caption{\centering The distributions of the slope of the 2017-18 optical-NIR colour vs \emph{r} magnitude plots returned from 10,000 interpolations of each light curve.}
    \label{fig:nircolvopt}
\end{figure} 

\newpage

\subsection{Optical-NIR Colour-Magnitude Plots from Interpolation}

The interpolation method was shown to be consistent for the DES colour indices, so it was therefore used to measure each combination of optical and NIR in the 2017 and the 2018 seasons. The results of the 2017 season slope of optical-NIR colour against \emph{r} band are displayed in Figure~\ref{fig:nircolvopt}. VEILS did not observe in the 2017 season between MJD 57993 and 58044, during which time a peak was present in the optical, so therefore the light curves were restricted to MJD greater than 58044 to prevent large portions of interpolations impacting the results as explained in Appendix~\ref{sec:app_201718_interps}. Figure~\ref{fig:nircolvopt} was plotted against the \emph{r} band to show comparable colour trends with the optical colours in Figure~\ref{fig:optcol}, however, this required additional interpolation of the \emph{r} band when the dates matching the NIR light curves were extracted. The other interpolated optical and \emph{r} band light curves in this situation therefore did not necessarily follow the same variations during the interpolations which could have an effect on the slope of the interpolated optical - NIR colour vs \emph{r} band magnitude plot. The results in Figure~\ref{fig:nircolvopt} show that 50\% of the interpolated slopes of the colour vs \emph{r} magnitude plot are consistent within 1$\sigma$ of each other, and all are consistent within 1.8$\sigma$. Furthermore, it was found that the plots of \emph{g}-\emph{J} and \emph{g}-\emph{Ks} vs \emph{r} have negative slopes within 5$\sigma$ and 4$\sigma$ uncertainties respectively, and the plots of \emph{z}-\emph{J} and \emph{z}-\emph{Ks} vs \emph{r} have negative slopes within 1.7$\sigma$ and 1$\sigma$ uncertainties respectively, which implies RWB trends. The plots of \emph{r-J}, \emph{r-Ks}, \emph{i}-\emph{J} and \emph{i}-\emph{Ks} do not show a conclusive trend. All slopes of the 2018 season, which are displayed in Appendix~\ref{ap:2018colmag}, show inconclusive colour behaviours.

\section{Discussion}
\label{Sect:Discuss}

In the previous sections, the multi-wavelength variability of PKS~0027-426, a FSRQ at z~=~0.495, was studied using observations in the optical and NIR to attempt to further understand the inner regions of AGN that cannot be spatially resolved.

\subsection{Temporal Variability}

The optical and NIR light curves of PKS~0027-426 were cross correlated amongst themselves and with each other to determine possible lags between the light curves emission. In this paper, the \emph{r} and \emph{i}, \emph{r} and \emph{Ks},  and \emph{J} and \emph{Ks} CCFs were discussed as representations of optical-optical, optical-NIR and NIR-NIR correlations as they had the most observations. 

Over each combination of filters, the most consistent correlation present was at $\sim$~0~days, which implies that the emission is simultaneous or any time delay between the light curves is on timescales less than the cadence of the surveys used (which have mean values of $\sim$ 6 days in the optical and $\sim$ 11 days in the NIR). Many studies of other blazars have also shown strong correlations between the optical and NIR light curves with time lags shorter than 1 day which implies that the source of the emission processes are co-spatial (e.g. \citealt{Bonning2012}, \citealt{DAmmando2013}, \citealt{Gupta2017}, \citealt{Kushwaha2017}). This could be due to the synchrotron radiation in the inner jet originating in similar regions for the optical and NIR. As this possible detected lag is smaller than the cadences of observations, multiple intra-day observations would be necessary to further constrain it.

Longer lags have also been found between the optical and NIR light curves in other blazars, for example, \cite{Safna2020} found significant time delays for three FSRQs on the order of 10-100 days. Similarly, \cite{Li2018} found that the NIR light curves variations lagged the optical by a few weeks in PKS~0537-441. Additional non-zero lags were also measured for PKS~0027-426 inconsistently across the light curves, for example, in the 2017 season, when optical light curves were included in the cross-correlation, an observed lag of $\sim \pm$~75 days (which corresponds to a rest frame lag of $\sim \pm$~50 days) was often recorded. However, further analysis of these lags shows that they are unlikely to be a delay between the emission regions and instead are caused by aliasing in the light curves. 

\begin{figure*}
    \centering
    \subfloat[figure][Mean and RMS Mg~II spectra and synthesised light curve compared to the DES \emph{g} band. \label{fig:MgII_spec_lcs}]{\includegraphics[width=0.49\textwidth]{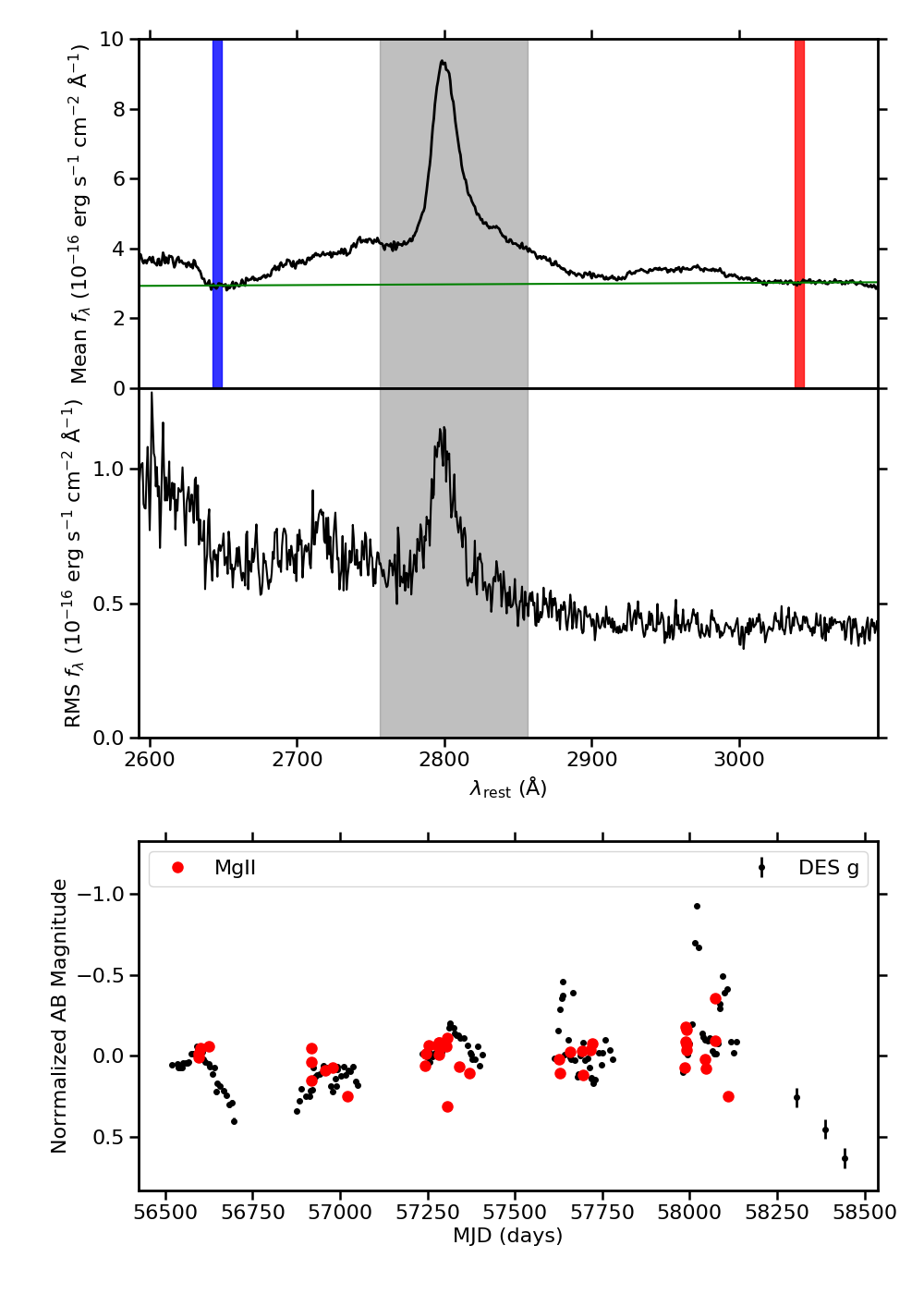}}
    \hfill
    \subfloat[figure][Mean and RMS continuum spectra and synthesised light curve compared to the DES \emph{g} band. \label{fig:Cont_spec_lcs}]{\includegraphics[width=\columnwidth]{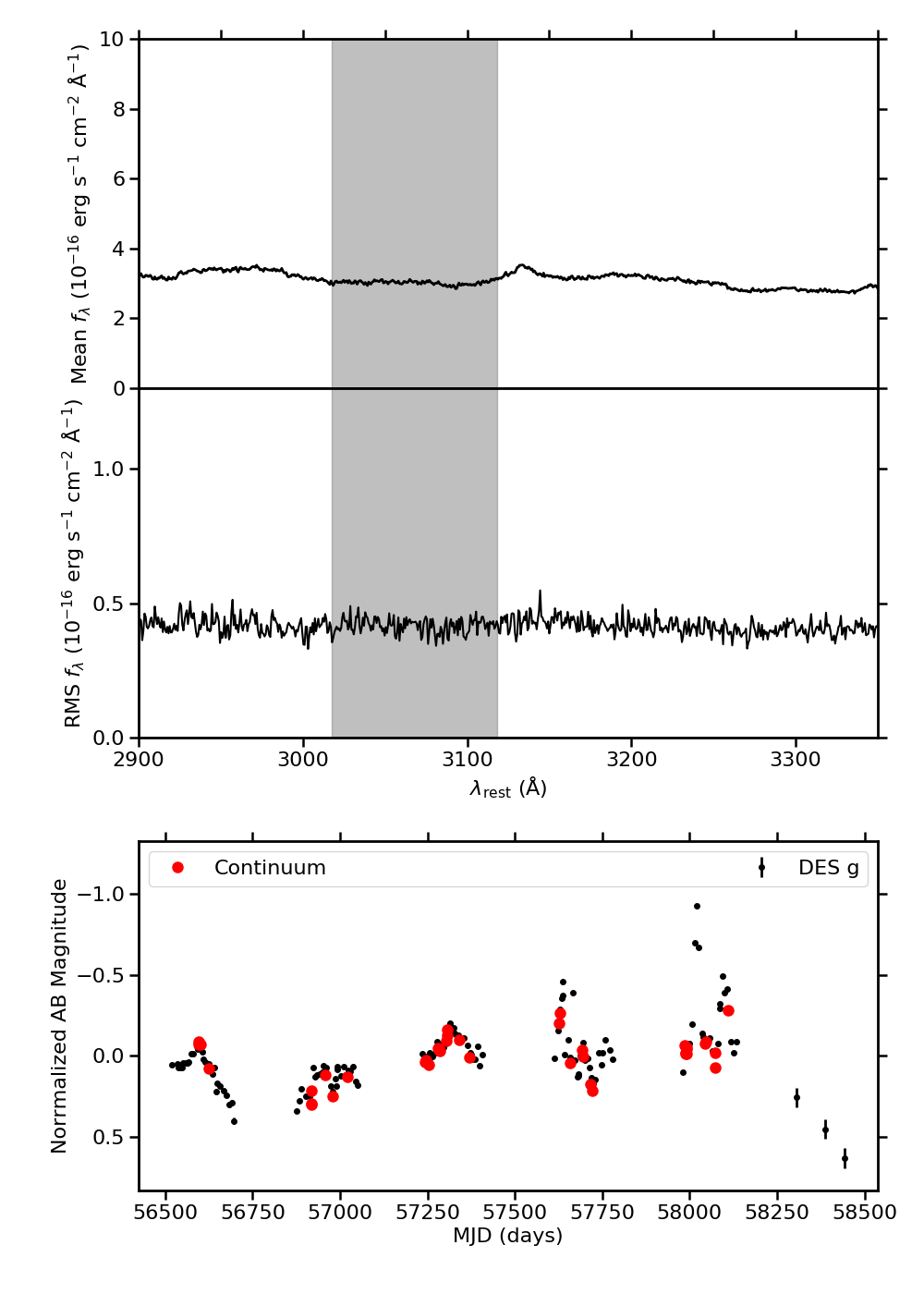}}
    \caption{\centering \textit{Top Panel}: The regions of OzDES spectra used to create the synthesised light curves (grey shaded regions). The Mg~II line is continuum subtracted by fitting a line for the approximate continuum (green line) between points on either side of the emission line region (the blue and red shaded regions). \textit{Middle panel}: RMS Spectra. \textit{Lower Panel}: Synthesised light curves created from spectra compared to the DES photometric light curves.}
\end{figure*}

\subsection{Spectral Variability}
\label{sect:Discuss-Spect}

The spectral variability of PKS~0027-426 was studied for each combination of optical and NIR light curves by calculating the slopes of colour vs magnitude plots, and was found to demonstrate a complex colour behaviour. 

The colour trend for each combination of the DES \emph{griz} filters, which could be studied directly due to the quasi-simultaneous observations, were shown to change both over time and depending on the colours used. For example, in the plot of \emph{g}-\emph{z} vs \emph{r} band magnitude in Figure~\ref{fig:gz_sep_years}, it changes from BWB in the 2013 season to SWB in the 2014 and 2015 seasons and RWB in the 2016 and 2017 seasons. Furthermore, RWB, SWB and BWB trends were observed simultaneously depending on the combination of filters used to calculate the colour, as shown in Figure~\ref{fig:optcol}, for example, the overall colour behaviour of the \emph{g}-\emph{r}, \emph{g}-\emph{i}, \emph{g}-\emph{z} and \emph{r}-\emph{i} demonstrates RWB trends, while the \emph{r}-\emph{z} and \emph{i}-\emph{z} follow a BWB trend. 

The spectral behaviour of the observations from VEILS and VOILETTE could not be directly measured from their light curves as they were not observed even quasi-simultaneously. Instead, one of the light curves was interpolated so that epochs matching the other light curve could be extracted. This method was first tested by comparing the mean colour vs \emph{r} magnitude slopes measured by removing 50$\%$ of the data and interpolating one of the DES light curves with the original slope measured for that colour, and was shown to be consistent for all DES colours across all seasons within 1.5$\sigma$ uncertainties. It was therefore used to obtain measurements of the optical-NIR spectral behaviour using VEILS and VOILETTE. It was found that the \emph{g}-\emph{J}, \emph{g}-\emph{Ks}, \emph{z}-\emph{J} and \emph{z}-\emph{Ks} slopes for the 2017 season were negative withing 5$\sigma$, 4$\sigma$, 1.7$\sigma$ and 1$\sigma$ respectively, which indicates RWB behaviour. The remaining colours showed inconclusive trends within their uncertainties. 

The possible colour trends observed in blazars have previously been explained independently. For example, a RWB colour behaviour could be explained in terms of the contribution of thermal emission from the accretion disk, which is more slowly varying than the variable jet emission (e.g. \citealt{Bonning2012}). Similarly, the BWB trend can be explained in terms of a faster varying blue component with a slower varying red component (e.g. \citealt{Fiorucci2004}). Alternatively, the BWB colour behaviour has been explained by a one component synchrotron model by \cite{Fiorucci2004}, who suggest that the more intense the energy release, the higher the particle's energy. Additionally, the BWB trend has been explained in terms of the shock-in-jet model, which suggests that accelerated electrons at the front of the shock lose energy while propagating away, and the higher frequency electrons lose energy faster due to synchrotron cooling, therefore making the higher frequency bands more variable (e.g. \citealt{Kirk1998}, \citealt{Agarwal2019}).

The change in colour behaviour over different periods of time shown in Figure~\ref{fig:gz_sep_years} has similarly been seen by \cite{Bonning2012}, who found that individual flares in other blazars can behave differently to the overall colour behaviours, which suggests that different jet components become important at different times. Furthermore, \cite{Raiteri2008} find that 3C~454.3 shows a RWB trend until the blazar reaches a saturation magnitude and turns into a BWB trend in bright states. In the \emph{g-z} colour-magnitude plots of PKS~0027-426, the 2016 and 2017 seasons are brighter than the earlier seasons and do demonstrate a different colour trend, but the 2013 season also follows a different colour behaviour while covering a similar magnitude range to the 2014 and 2015 seasons, which means a saturation magnitude is not likely to be the entire explanation in this scenario. Furthermore, while the 2016 and 2017 seasons demonstrate flares, the 2014 and 2015 seasons do not show any dramatic change in magnitude and yet still follow a SWB trend instead of the BWB trend in the 2013 seasons, thus implying individual flares are not solely responsible for the change in colour behaviour over time. 

The varying colour behaviour in different combination of filters has also been observed in 3C~345 by \cite{Wu2011}, who explained this phenomenon in terms of the emission features from the accretion disc or BLR, such as the Mg~II line, which vary less than the non-thermal continuum and dominate the flux at the shorter wavelength (i.e. \emph{g} band). This theory was investigated for PKS~0027-426 using the spectra from OzDES that were observed over the same periods as DES.

\subsubsection{Possible Contamination of Spectral Variability from Emission Lines}

Concurrent spectra of PKS~0027-426 were observed on 37 epochs with OzDES along side the photometric DES observations, so therefore the theory presented by \cite{Wu2011} could be examined. If emission features from the accretion disc varied less and dominated the flux, then it would be expected that synthesised light curves created from the emission lines in the \emph{g} band would be shown to be less variable than both synthesised light curves from different sections of the continuum and the photometric light curves from DES. 

To create synthesised light curves from emission lines within the spectra, they were first continuum subtracted. For example, the Mg~II line is displayed in the upper panels of Figure~\ref{fig:MgII_spec_lcs}, and exists within a region of the spectra known as the small blue bump, in which the Fe II emission lines contribute substantially to the underlying continuum. The continuum was approximately subtracted from the Mg~II line by fitting a line between points on either side of the emission line, depicted by the blue and red shaded regions, and subtracting this approximate continuum, depicted by the green line, from the spectra. The amplitude variability of the synthesised light curves were compared for emission lines in different filters and with synthesised light curves created from the continuum to then test whether the emission lines are less variable than the continuum emission. The fraction of the total flux in each synthesised DES band light curve that comes from the individual emission lines was also calculated to test whether the emission lines dominated in the filters.

\begin{table}
	\begin{minipage}{\columnwidth}
		\centering
		\footnotesize{
        \caption{The amplitude variations of the synthesised light curves created from OzDES spectra. \label{tab:spec_lines_var}}
        \begin{tabular}{C{0.27\textwidth}C{0.2\textwidth}C{0.17\textwidth}C{0.17\textwidth}}
        \hline
            Emission Line or Region of the Spectra & Rest Frame Wavelength Range (\AA) & Amplitude Variation (mag) & Percentage of Total flux (\%) \\
        \hline
           Synthesised \emph{g} band & 2500 - 3780 & 0.54 & 100 \\
           Synthesised \emph{r} band & 3620 - 4930 & 0.79 & 100 \\
           Synthesised \emph{i} band & 4530 - 5830 & 0.93 & 100 \\
        \hline
           Mg~II & 2760 - 2860 & 0.67 & 5 \\
           Blue Continuum & 3020 - 3120 & 0.59 & 9 \\
           H$\beta$ & 4830 - 4930 & 0.51 & 5 \\
        \hline
        \end{tabular}}
	\end{minipage}
\end{table}

The lower panels of Figure~\ref{fig:MgII_spec_lcs} and \ref{fig:Cont_spec_lcs} display the normalised light curves created for the Mg~II line and a section of the continuum in the \emph{g} band respectively. It can be seen that the synthesised light curve from a section of the continuum follows similar variability to the photometric DES \emph{g} band light curve, which is to be expected, however, the synthesised light curve from the continuum subtracted Mg~II line varies differently to the photometric \emph{g} band light curve. Furthermore, Table~\ref{tab:spec_lines_var} contains the amplitude variations for the Mg~II and blue continuum synthesised light curves, as well as from synthesised H$\beta$ light curves. The light curves from the emission lines are shown to vary similarly to the continuum region and to the synthesised DES light curves. Furthermore, the Mg~II line is more variable than the H$\beta$ line, and the synthesised light curve from the blue continuum region, which disagrees with the theory presented by \cite{Wu2011} as they suggest that the Mg~II line should be the least variable and should dominate in the \emph{g} band. Table~\ref{tab:spec_lines_var} also contains the percentage of the flux in each synthesised light curve that contributes to the overall flux of the synthesised DES light curves, and shows that the Mg~II and H$\beta$ emission lines contribute to 5\% of the overall flux.

\subsubsection{Multiple Contributing Components to the Overall Emission}

The differences in colour behaviour that occur simultaneously with different combination of optical \emph{griz} filters could be explained by the multiple components that contribute to the overall optical emission.

\paragraph{Decomposing the Spectra into Red and Blue Components}
\label{sec:model_spec}
~ \\

The change in spectral behaviour between different filter light curves could be explained by the presence of multiple different coloured components that contribute to the overall optical emission. The different wavelength ranges could be dominated by a different coloured component which would mean when one component varies differently to the other, the colour behaviour will not follow the same trend for each combination of filters.

This was investigated using the optical broadband spectra which were made using the optical DES light curves. The overall emission is assumed to follow a power law of $f_{\lambda} \propto \lambda^{\alpha_\lambda}$, where $f_{\lambda}$ is the flux density at wavelength $\lambda$ and $\alpha_\lambda$ is the spectral index. The total flux is here is assumed to be a combination of the flux from a red component and a blue component, which each follow their own power laws and have spectral indices of $\alpha_\text{red}$ and $\alpha_\text{blue}$ respectively. The broadband spectra of the red and blue emission for each season were modelled and summed to match the shape of the average spectra of the 3 dimmest epochs that were observed within their 1$\sigma$ uncertainties. The spectral index of one of the components was then varied so that the new total spectra matched the shape of the average spectra of the brightest 3 epochs that were observed within their 1$\sigma$ uncertainties.

\begin{figure*}
    \begin{minipage}{\textwidth}
        \subfloat[figure][ Modelled Spectra of the 2013 Season. \label{fig:S1_fit} ~ \\]{\includegraphics[width=0.49\textwidth]{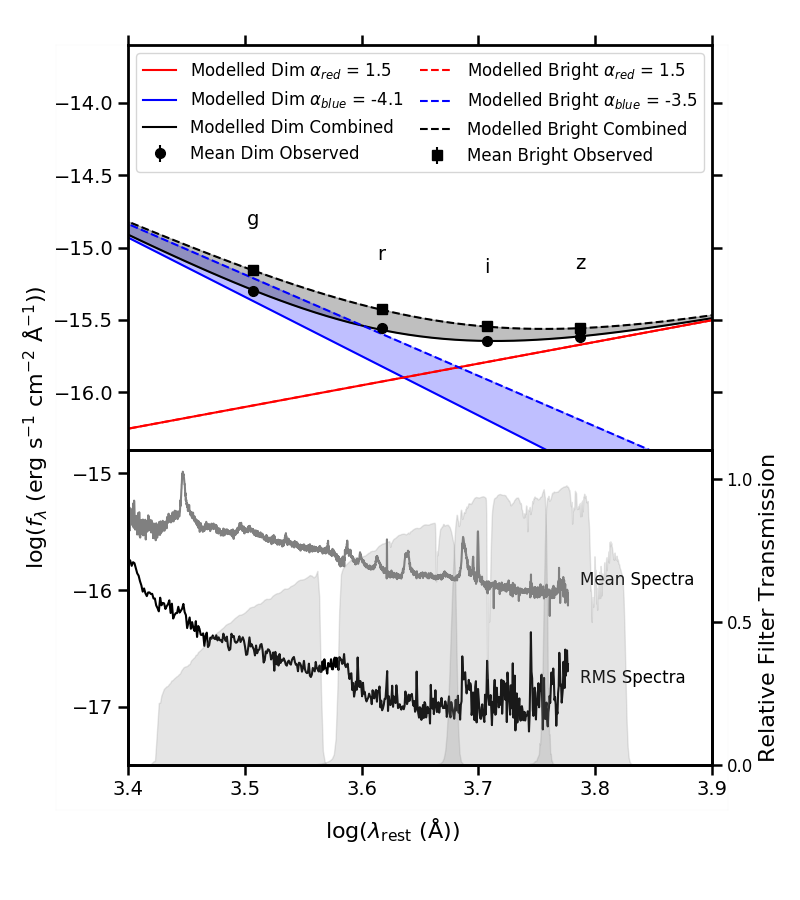}}
        \hfill
        \subfloat[figure][Modelled Spectra of the 2014 Season. \label{fig:S2_fit} ~ \\]{\includegraphics[width=0.49\textwidth]{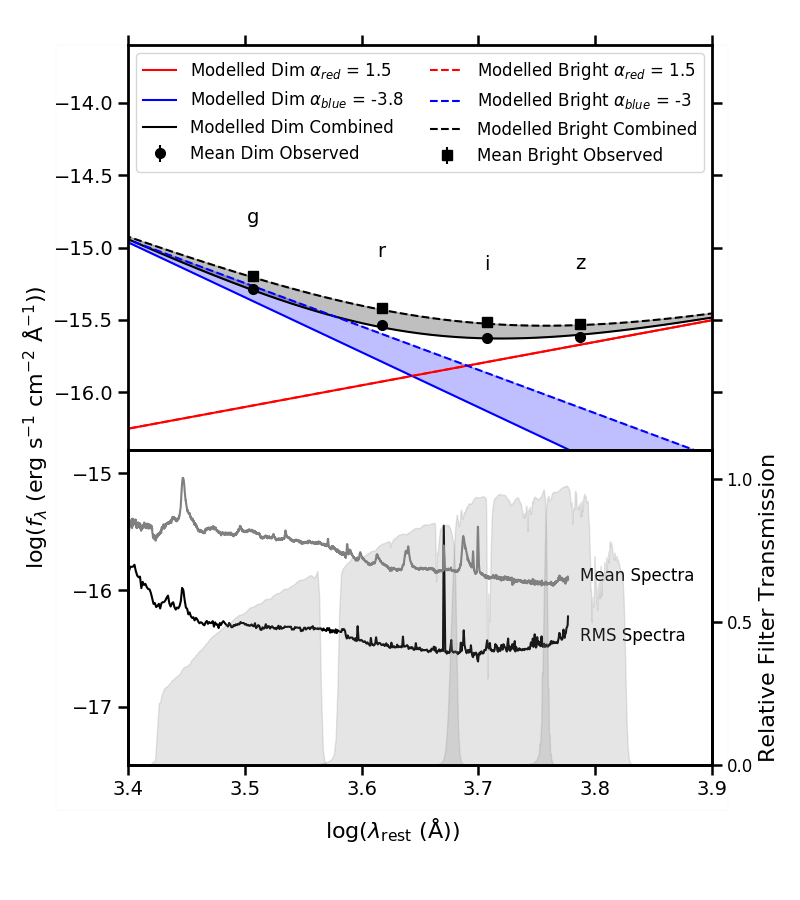}}
        \\
        \subfloat[figure][ Modelled Spectra of the 2016 Season. \label{fig:S4_fit} ~ \\]{\includegraphics[width=0.49\textwidth]{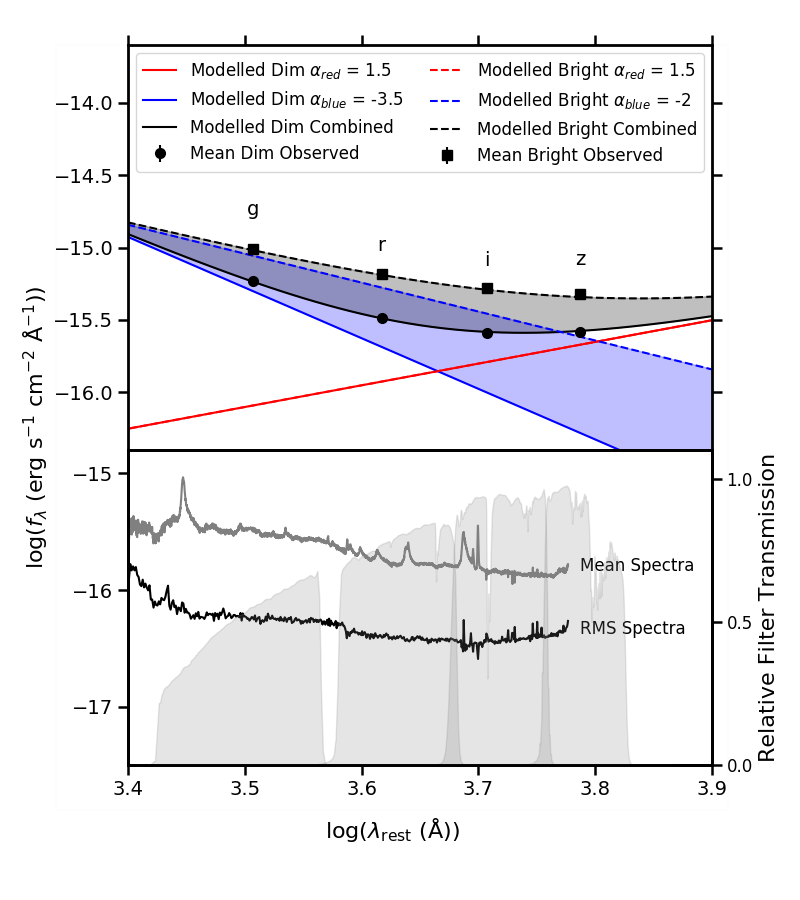}}
        \hfill
        \subfloat[figure][ Modelled Spectra of the 2017 Season. \label{fig:S5_fit} ~ \\]{\includegraphics[width=0.5\textwidth]{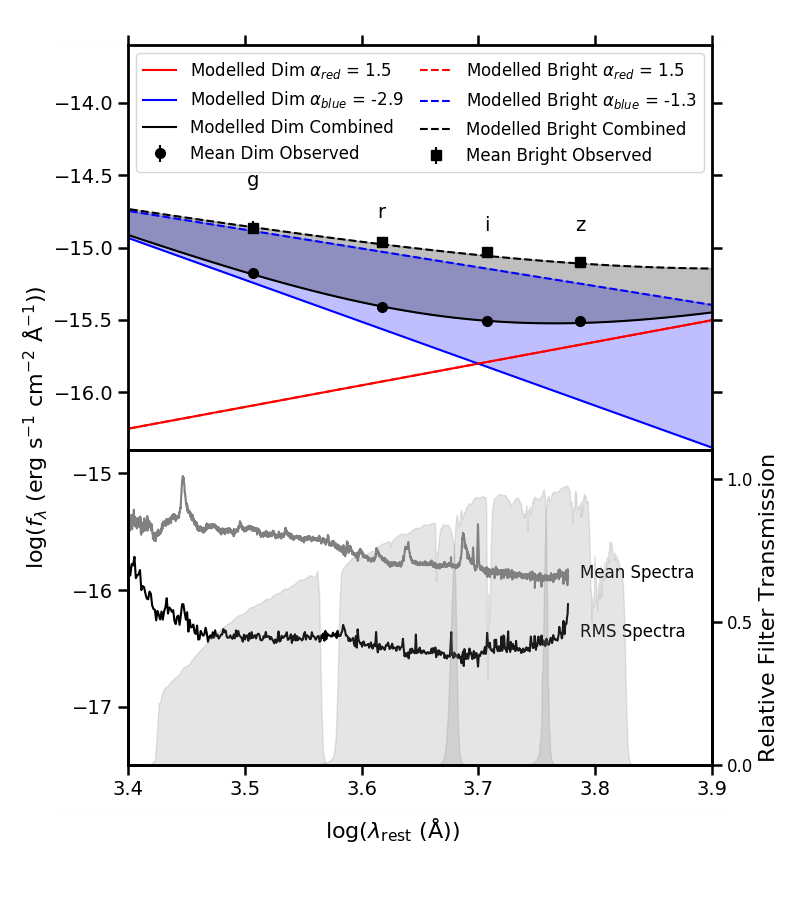}}
        \caption{\textit{Upper panel}: Modelled broadband spectra of the red and blue emission that combine to match the average brightest and dimmest epochs in each observation season compared to the observed broadband spectra. The solid lines correspond to the modelled spectra of the dimmest epochs, and the dashed lines correspond to the modelled spectra of the brightest epochs. 
        \textit{Lower panel}: Mean and smoothed RMS OzDES spectra for each season plotted over the DES filter Transmissions.
        \label{fig:Modelled_AD_Sync}}
    \end{minipage}
\end{figure*}

Figure~\ref{fig:Modelled_AD_Sync} displays the mean brightest and mean dimmest spectra of the \emph{griz} bands in the seasons starting 2013, 2014, 2016 and 2017, along with examples of the modelled red and blue emission that are combined to fit the observed spectral shape. The 2015 season and the model over the entire DES observational period are displayed in Figure~\ref{fig:Modelled_AD_Sync_More}. In these figures, the red emission was kept constant with the equation $\text{log}(f_\lambda) = 1.5 \ \text{log}(\lambda_\text{rest}) -21.35$, where \mbox{$\alpha_\text{red}$ = 1.5}, and the blue emission was varied to match the mean brightest spectra. The value of $\alpha_\text{red}$ was chosen somewhat arbitrarily here to demonstrate how the change in the blue slope can effect the shape of the overall spectra, however the value is not unique to the broadband spectra as shown in Appendix \ref{ap:more_models}, which explores alternative values including plots in which the blue slope is fixed and the red component is varied to match the change in spectra with brightness. The mean and RMS OzDES spectra of each season are plotted below for comparison. Table~\ref{tab:modelled_ad_sync} gives the equations of the lines that are modelled for the blue and red components in each season to match the mean brightest and dimmest spectra. The change in the blue slope is shown to increase towards the later seasons where the object has previously been shown to be more variable in Table~\ref{tab:perampvar}.

\begin{table}
	\begin{minipage}{\columnwidth}
        \centering
        \footnotesize{
        \caption{The slopes (spectral indices, $\alpha$) and intercepts (int) of the red and blue components that are modelled in Figure \ref{fig:Modelled_AD_Sync} to match the mean observed brightest and dimmest broadband spectra in each season. \label{tab:modelled_ad_sync}}        
        \begin{tabular}{C{0.06\textwidth}C{0.06\textwidth}C{0.09\textwidth}C{0.09\textwidth}C{0.1\textwidth}C{0.12\textwidth}C{0.12\textwidth}}
            \hline
                Season & $\alpha_\text{\ red}$ & $\text{int}_\text{\ red}$ & $\alpha_\text{\ blue,dim}$ & $\text{int}_\text{\ blue,dim}$ & $\alpha_\text{\ blue,bright}$ & $\text{int}_\text{\ blue,bright}$ \\
            \hline
               2013 & 1.5 & -21.35 & -4.1 & -0.99 & -3.5 & -2.94 \\
               2014 & 1.5 & -21.35 & -3.8 & -2.04 & -3 & -4.75 \\
               2015 & 1.5 & -21.35 & -3.6 & -2.62 & -2.9 & -4.98 \\
               2016 & 1.5 & -21.35 & -3.5 & -3.03 & -2 & -8.04 \\
               2017 & 1.5 & -21.35 & -2.9 & -5.07 & -1.3 & -10.33 \\
               All & 1.5 & -21.35 & -3.9 & -1.70 & -1.3 & -10.33 \\
            \hline
        \end{tabular}}
	\end{minipage}
\end{table}

\begin{table*}
    \caption{Slope of log($A_\lambda$) and log($S_\lambda$) vs log($\lambda$) from Figure \ref{fig:comp_decomps} between the \emph{g-r}, \emph{r-i} and \emph{i-z} filters for each season of DES and for the entire observational period. \label{tab:comp_decomps}}
	\subfloat[table][Slope of log($A_\lambda$) vs log($\lambda$) from Figure \ref{fig:comp_decomps}.]{\begin{minipage}{0.48\textwidth}
        \begin{tabular}{C{0.2\textwidth}C{0.2\textwidth}C{0.2\textwidth}C{0.2\textwidth}}
        \hline
            Season & \emph{g-r} & \emph{r-i} & \emph{i-z}  \\
        \hline
           2013 & -2.37 $\pm$ 0.01 & -1.19 $\pm$ 0.01 & 0.22 $\pm$ 0.01 \\
           2014 & -2.15 $\pm$ 0.01 & -1.04 $\pm$ 0.01 & 0.09 $\pm$ 0.01 \\
           2015 & -2.28 $\pm$ 0.01 & -1.23 $\pm$ 0.01 & -0.02 $\pm$ 0.01 \\
           2016 & -2.00 $\pm$ 0.02 & -1.03 $\pm$ 0.02 & 0.03 $\pm$ 0.02 \\
           2017 & -1.66 $\pm$ 0.03 & -0.84 $\pm$ 0.03 & -0.17 $\pm$ 0.03\\
           All & -2.06 $\pm$ 0.01 & -1.05 $\pm$ 0.01 & 0.01 $\pm$ 0.01 \\
        \hline
        \end{tabular}
        \end{minipage}}
	\hspace{1mm}
	\subfloat[table][Slope of log($S_\lambda$) vs log($\lambda$) from Figure \ref{fig:comp_decomps}.]{\begin{minipage}{0.48\textwidth}
        \begin{tabular}{C{0.2\textwidth}C{0.2\textwidth}C{0.2\textwidth}C{0.2\textwidth}}
        \hline
            Season & \emph{g-r} & \emph{r-i} & \emph{i-z}  \\
        \hline
           2013 & -2.90 $\pm$ 0.05 & -2.22 $\pm$ 0.05 & -2.77 $\pm$ 0.06 \\
           2014 & -1.61 $\pm$ 0.05 & -1.29 $\pm$ 0.05 & -0.83 $\pm$ 0.05 \\
           2015 & -1.92 $\pm$ 0.05 & -1.37 $\pm$ 0.06 & -1.29 $\pm$ 0.06 \\
           2016 & -0.92 $\pm$ 0.06 & -0.99 $\pm$ 0.06 & -0.98 $\pm$ 0.06 \\
           2017 & -0.34 $\pm$ 0.09 & -0.57 $\pm$ 0.08 & -1.00 $\pm$ 0.08 \\
           All & -0.57 $\pm$ 0.07 & -0.58 $\pm$ 0.07 & -0.83 $\pm$ 0.07 \\
        \hline
        \end{tabular}
        \end{minipage}}
\end{table*}

These simple models in Figure \ref{fig:Modelled_AD_Sync} demonstrate that if one component is varying when the source gets brighter, the overall variability observed between filters can be different due to the dominating emission process in each filter. For example, when the blue emission varies, the overall variation observed in the filters that are more strongly impacted by the red emission is diluted due to the strong constant red emission, whereas in the filters where the blue component dominates, the overall variability will better reflect the blue emission's variability, which could therefore explain why the colour behaviour has been shown to change between different combinations of optical filters. 

The blue and red components used here could correspond to physical processes such as the thermal emission from the accretion disk and the synchrotron emission from the jet respectively, as \cite{Wills1992} suggest that the thermal emission generally dominates in the optical-UV region, however, in FSRQs, when bright, the spectrum could be dominated by the synchrotron component towards the longer optical wavelengths and the IR. The spectral indices of the accretion disk and synchrotron emission from the jet have previously been predicted to be $\alpha_{\lambda,\text{AD}} \approx$ -7/3 \citep{Kishimoto2008} and $\alpha_{\lambda, \text{Sync}} \approx$ -0.5 \citep{Wills1992}, which are not consistent with the steep slopes used in this analysis, however Appendix \ref{ap:more_models} demonstrates that the values used for the red and blue slopes here are not unique. This analysis assumes the presence of only one variable component which may be oversimplifying it, so therefore, an alternative approach is applied in the following section.

\paragraph{Decomposing the Spectra into the Variable and Non-Variable Components}
~
\newline

\begin{figure}
        \centering
        \includegraphics[width=0.5\textwidth]{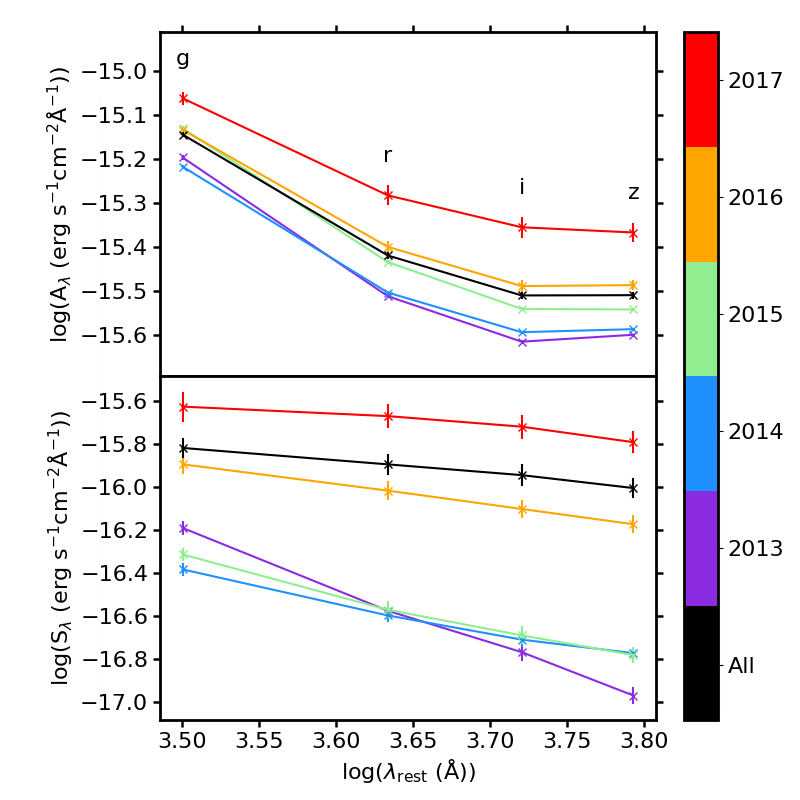}
        \caption{\small Decomposition of the spectra into the variable (S$_\lambda$) and non-variable (A$_\lambda$) components for each season of DES and for the entire observational period. \label{fig:comp_decomps} ~ \\}   
\end{figure}

The previous section assumes the presence of two different coloured components that contribute to the overall optical emission, however, it is simplified and only implies that one component varies while the other remains constant. To explore this further, the light curves in each wavelength range, $f_\lambda(t)$, are instead decomposed into the variable and non-variable components using the separable model given in Equation \ref{eq:decomp}. 

\begin{equation}
\label{eq:decomp}
    f_\lambda(t) = A_\lambda + S_\lambda X(t)
\end{equation}
Where $A_\lambda$ is the spectra of the mean light (i.e. the non-variable component), $S_\lambda$ is the spectra of the variable component and X(t) is the light curve that has been normalized such that $\langle X(t) \rangle = 0$ and $\langle X^2(t) \rangle = 1$. 

Figure \ref{fig:comp_decomps} displays the $A_\lambda$ and $S_\lambda$ spectra covering the DES \emph{griz} filters in each observation season in DES and for the overall observation period, and Table \ref{tab:comp_decomps} gives the slopes between each filter for the $A_\lambda$ and $S_\lambda$ spectra, where the slopes and uncertainties correspond to the mean slope and the standard deviation obtained from bootstrapping the light curves. Appendix \ref{ap:OzDES_decomp} contains similar analysis for the OzDES spectra in each season. 

The constant component, $A_\lambda$, is shown to follow roughly the same shape for each observations season, where the slope of log($A_\lambda$) vs log ($\lambda$) is steepest between the \emph{g} and \emph{r} bands and flattest between the \emph{i} and \emph{z} bands. This implies that there are multiple different coloured components that contribute to the non-variable emission, as there is a strong blue component effecting the \emph{g} and \emph{r} region of the spectra and a redder component that is flattening the spectra between the \emph{r} and \emph{i} and the \emph{i} and \emph{z} bands. 

The slope of the variable component, $S_\lambda$, is also shown to change between seasons, with the 2013 season having the steepest slope between all filters and the 2017 season the flattest. The shape of the spectra here also indicates the presence of more than one spectral component, including a steep blue component but also a red component that contributes to the change of slope of the variable spectra with increasing wavelength in the later seasons.

This method of decomposing into the spectra into variable and non-variable components therefore supports the assumption in the previous section that both a blue and red spectral component contribute to the overall optical emission, but it also demonstrates that both coloured components are likely to contribute to the overall variable and non-variable emission. The contribution of multiple different coloured components could therefore explain the change in colour behaviour that is observed simultaneously with different combinations of optical DES filters.

\section{Summary}
\label{Sect:Concl}

We studied the multi-wavelength temporal and spectral variability of the FSRQ PKS~0027-426, using optical observations from DES (2013-2018) and VOILETTE (2018-2019) in the \emph{griz} bands, and NIR observations from VEILS (2017-2019) in the \emph{JKs} bands. The results are summarised below:

\begin{enumerate}
    \item The temporal variability was studied using cross-correlation analysis of the optical and NIR light curves, and the most consistent correlation over all combination of light curves was found at $\sim$ 0 days, which implies that the emission is simultaneous or any delay between light curves occurs on timescales smaller that the cadence of observations.
    \item The spectral variability was studied for each combination of optical DES \emph{griz} light curves using the slopes of the colour vs r magnitude plots. The overall colour trends are shown to vary when different combinations of filters are used, from RWB trends in the \emph{g-r}, \emph{g-i}, \emph{g-z} and \emph{r-i} to BWB in the \emph{r-z} and \emph{i-z}. 
    \item The spectral variability was also shown to vary over each observation season, for example, in the \emph{g-z}, the colour behaviour follows a BWB trend in the 2013 season, a SWB trend in the 2014 and 2015 seasons and a RWB trend in the 2016 and 2017 seasons.
    \item Using OzDES spectra from 2013-2018, we investigated the possible explanation for the changing colour behaviour with different combinations of filters provided by \cite{Wu2011} for 3C~345, in which emission features from the accretion disk or BLR dominate the flux at shorter wavelengths and vary less than the non-thermal continuum, and found that our results disagreed as the emission lines were not less variable than the continuum. 
    \item The variations in colour behaviour across different combinations of filters was instead explained as a result of each filter containing a different ratio of the multiple different coloured components that combine to give the overall optical emission. These red and blue components are thought to vary differently, which could therefore cause the emission between filters to vary.  
    \item The optical and NIR spectral variability was also studied, however, as the observations were not simultaneous, one of the light curves was interpolated to extract matching epochs. This method was shown to be reliable for the DES data after half of the observations were removed, with consistent results given within 1.5$\sigma$ uncertainties.
\end{enumerate}

\section*{Acknowledgements}

EG and SFH acknowledge support from the Horizon 2020 ERC Starting Grant \textit{DUST-IN-THE-WIND} (ERC-2015-StG-677117). EG  acknowledges support from STFC for funding this PhD. 

This research is in part based on observations from Paranal Observatory for the European Southern Observatory (ESO) observing programmes 198.A-2005 and 1101.A-0327. 

This research has made use of the NASA/IPAC Extragalactic Database (NED), which is operated by the Jet Propulsion Laboratory, California Institute of Technology, under contract with the National Aeronautics and Space Administration.

This paper makes use of observations taken using the Anglo-Australian Telescope under programme A/2013B/12. We  acknowledge  the traditional  owners  of  the  land  on  which  the  AAT  stands, the Gamilaraay people, and pay our respects to elders past and present.

Funding for the DES Projects has been provided by the U.S. Department of Energy, the U.S. National Science Foundation, the Ministry of Science and Education of Spain, the Science and Technology Facilities Council of the United Kingdom, the Higher Education Funding Council for England, the National Center for Supercomputing Applications at the University of Illinois at Urbana-Champaign, the Kavli Institute of Cosmological Physics at the University of Chicago, the Center for Cosmology and Astro-Particle Physics at the Ohio State University, the Mitchell Institute for Fundamental Physics and Astronomy at Texas A\&M University, Financiadora de Estudos e Projetos, Fundação Carlos Chagas Filho de Amparo à Pesquisa do Estado do Rio de Janeiro, Conselho Nacional de Desenvolvimento Científico e Tecnológico and the Ministério da Ciência, Tecnologia e Inovação, the Deutsche Forschungsgemeinschaft and the Collaborating Institutions in the Dark Energy Survey.

The Collaborating Institutions are Argonne National Laboratory, the University of California at Santa Cruz, the University of Cambridge, Centro de Investigaciones Energéticas, Medioambientales y Tecnológicas-Madrid, the University of Chicago, University College London, the DES-Brazil Consortium, the University of Edinburgh, the Eidgenössische Technische Hochschule (ETH) Zürich, Fermi National Accelerator Laboratory, the University of Illinois at Urbana-Champaign, the Institut de Ciències de l’Espai (IEEC/CSIC), the Institut de Física d’Altes Energies, Lawrence Berkeley National Laboratory, the Ludwig-Maximilians Universität München and the associated Excellence Cluster Universe, the University of Michigan, the National Optical Astronomy Observatory, the University of Nottingham, The Ohio State University, the University of Pennsylvania, the University of Portsmouth, SLAC National Accelerator Laboratory, Stanford University, the University of Sussex, Texas A\&M University, and the OzDES Membership Consortium.

This study is based in part on observations at Cerro Tololo Inter-American Observatory, National Optical Astronomy Observatory, which is operated by the Association of Universities for Research in Astronomy (AURA) under a cooperative agreement with the National Science Foundation.

The DES data management system is supported by the National Science Foundation under grant nos AST-1138766 and AST-1536171. The DES participants from Spanish institutions are partially supported by MINECO under grants AYA2015-71825, ESP2015-66861, FPA2015-68048, SEV-2016-0588, SEV-2016-0597, and MDM-2015-0509, some of which include ERDF funds from the European Union. IFAE is partially funded by the CERCA programme of the Generalitat de Catalunya. Research leading to these results has received funding from the European Research Council under the European Union’s Seventh Framework Programme (FP7/2007-2013) including ERC grant agreements 240672, 291329, and 306478. We acknowledge support from the Brazilian Instituto Nacional de Ciência e Tecnologia (INCT) e-Universe (CNPq grant no. 465376/2014-2).

This manuscript has been authored by Fermi Research Alliance, LLC under Contract No. DE-AC02-07CH11359 with the U.S. Department of Energy, Office of Science, Office of High Energy Physics. The United States Government retains and the publisher, by accepting the article for publication, acknowledges that the United States Government retains a non-exclusive, paid-up, irrevocable, world-wide license to publish or reproduce the published form of this manuscript, or allow others to do so, for United States Government purposes.

\section*{Data Availability}

The data underlying this article are available in the article and in its online supplementary material.




\bibliographystyle{mnras}
\bibliography{bib} 

\section*{Affiliations}

$^{1}$ School of Physics and Astronomy, University of Southampton,  Southampton, SO17 1BJ, UK \\
$^{2}$ Department of Biological and Physical Sciences, South Carolina State University, Orangeburg, SC 29117,USA.\\
$^{3}$ SUPA School of Physics and Astronomy, University of St. Andrews, North Haugh, KY16 9SS, Scotland, UK \\
$^{4}$ Department of Physics, Faculty of Science, Kyoto Sangyo University, Kamigamo-motoyama, Kita-ku, Kyoto 603-8555, Japan \\
$^{5}$ Laborat\'orio Interinstitucional de e-Astronomia - LIneA, Rua Gal. Jos\'e Cristino 77, Rio de Janeiro, RJ - 20921-400, Brazil \\
$^{6}$ Fermi National Accelerator Laboratory, P. O. Box 500, Batavia, IL 60510, USA \\
$^{7}$ Instituto de F\'{i}sica Te\'orica, Universidade Estadual Paulista, S\~ao Paulo, Brazil \\
$^{8}$ Centro de Investigaciones Energ\'eticas, Medioambientales y Tecnol\'ogicas (CIEMAT), Madrid, Spain \\
$^{9}$ Institute of Astronomy, University of Cambridge, Madingley Road, Cambridge CB3 0HA, UK  \\
$^{10}$ Kavli Institute for Cosmology, University of Cambridge, Madingley Road, Cambridge CB3 0HA, UK \\
$^{11}$ CNRS, UMR 7095, Institut d'Astrophysique de Paris, F-75014, Paris, France \\
$^{12}$ Sorbonne Universit\'es, UPMC Univ Paris 06, UMR 7095, Institut d'Astrophysique de Paris, F-75014, Paris, France \\
$^{13}$ Department of Physics \& Astronomy, University College London, Gower Street, London, WC1E 6BT, UK \\
$^{14}$ Kavli Institute for Particle Astrophysics \& Cosmology, P. O. Box 2450, Stanford University, Stanford, CA 94305, USA \\
$^{15}$ SLAC National Accelerator Laboratory, Menlo Park, CA 94025, USA \\
$^{16}$ Instituto de Astrofisica de Canarias, E-38205 La Laguna, Tenerife, Spain \\
$^{17}$ Universidad de La Laguna, Dpto. Astrofísica, E-38206 La Laguna, Tenerife, Spain \\
$^{18}$ INAF, Astrophysical Observatory of Turin, I-10025 Pino Torinese, Italy \\
$^{19}$ Center for Astrophysical Surveys, National Center for Supercomputing Applications, 1205 West Clark St., Urbana, IL 61801, USA \\
$^{20}$ Department of Astronomy, University of Illinois at Urbana-Champaign, 1002 W. Green Street, Urbana, IL 61801, USA \\
$^{21}$ Institut de F\'{\i}sica d'Altes Energies (IFAE), The Barcelona Institute of Science and Technology, Campus UAB, 08193 Bellaterra (Barcelona) Spain \\
$^{22}$ Astronomy Unit, Department of Physics, University of Trieste, via Tiepolo 11, I-34131 Trieste, Italy \\
$^{23}$ INAF-Osservatorio Astronomico di Trieste, via G. B. Tiepolo 11, I-34143 Trieste, Italy \\
$^{24}$ Institute for Fundamental Physics of the Universe, Via Beirut 2, 34014 Trieste, Italy \\
$^{25}$ Observat\'orio Nacional, Rua Gal. Jos\'e Cristino 77, Rio de Janeiro, RJ - 20921-400, Brazil \\
$^{26}$ School of Mathematics and Physics, University of Queensland,  Brisbane, QLD 4072, Australia \\
$^{27}$ Santa Cruz Institute for Particle Physics, Santa Cruz, CA 95064, USA \\
$^{28}$ Institute of Theoretical Astrophysics, University of Oslo. P.O. Box 1029 Blindern, NO-0315 Oslo, Norway \\
$^{29}$ Kavli Institute for Cosmological Physics, University of Chicago, Chicago, IL 60637, USA \\
$^{30}$ Department of Physics and Astronomy, University of Leicester, Leicester LE1 7RH, UK \\
$^{31}$ Faculty of Physics, Ludwig-Maximilians-Universit\"at, Scheinerstr. 1, 81679 Munich, Germany \\
$^{32}$ Center for Cosmology and Astro-Particle Physics, The Ohio State University, Columbus, OH 43210, USA \\
$^{33}$ Department of Physics, The Ohio State University, Columbus, OH 43210, USA \\
$^{34}$ Center for Astrophysics $\vert$ Harvard \& Smithsonian, 60 Garden Street, Cambridge, MA 02138, USA \\
$^{35}$ DLR-Institute of Data Science, 07745 Jena, Germany \\
$^{36}$ Max-Planck-Institut für Radioastronomie, 53121 Bonn, Germany \\
$^{37}$ Australian Astronomical Optics, Macquarie University, North Ryde, NSW 2113, Australia \\
$^{38}$ Lowell Observatory, 1400 Mars Hill Rd, Flagstaff, AZ 86001, USA \\
$^{39}$ Sydney Institute for Astronomy, School of Physics, A28, The University of Sydney, NSW 2006, Australia \\
$^{40}$ Centre for Gravitational Astrophysics, College of Science, The Australian National University, ACT 2601, Australia \\
$^{41}$ The Research School of Astronomy and Astrophysics, Australian National University, ACT 2601, Australia \\
$^{42}$ Departamento de F\'isica Matem\'atica, Instituto de F\'isica, Universidade de S\~ao Paulo, CP 66318, S\~ao Paulo, SP, 05314-970, Brazil \\
$^{43}$ Instituci\'o Catalana de Recerca i Estudis Avan\c{c}ats, E-08010 Barcelona, Spain \\
$^{44}$ Physics Department, 2320 Chamberlin Hall, University of Wisconsin-Madison, 1150 University Avenue Madison, WI  53706-1390 \\
$^{45}$ Department of Physics, University of Michigan, Ann Arbor, MI 48109, USA \\
$^{46}$ Department of Astrophysical Sciences, Princeton University, Peyton Hall, Princeton, NJ 08544, USA \\
$^{47}$ Institut d'Estudis Espacials de Catalunya (IEEC), 08034 Barcelona, Spain \\
$^{48}$ Institute of Space Sciences (ICE, CSIC),  Campus UAB, Carrer de Can Magrans, s/n,  08193 Barcelona, Spain \\
$^{49}$ International Centre for Radio Astronomy Research, Curtin University, Bentley, WA 6102, Australia \\
$^{50}$ Computer Science and Mathematics Division, Oak Ridge National Laboratory, Oak Ridge, TN 37831 \\
$^{51}$ Department of Physics, Stanford University, 382 Via Pueblo Mall, Stanford, CA 94305, USA




\appendix

\section{Light Curves}

\begin{table*}
    \centering
    \caption{List of non-varying objects used to calibrate the light curves of \mbox{PKS 0027-426}. \label{tab:r_ref_stars}}
    \begin{tabular}{C{0.04\textwidth}C{0.05\textwidth}C{0.05\textwidth}C{0.05\textwidth}C{0.05\textwidth}C{0.05\textwidth}C{0.05\textwidth}C{0.05\textwidth}C{0.05\textwidth}C{0.05\textwidth}C{0.05\textwidth}C{0.05\textwidth}C{0.05\textwidth}C{0.05\textwidth}}
       \hline
       RA (deg) & Dec (deg) & Mean \emph{g} mag & 1$\sigma$ (\emph{g} mag) & Mean \emph{r} mag & 1$\sigma$ (\emph{r} mag) & Mean \emph{i} mag & 1$\sigma$ (\emph{i} mag) & Mean \emph{z} mag & 1$\sigma$ (\emph{z} mag) & Mean \emph{J} mag & 1$\sigma$ (\emph{J} mag) & Mean \emph{Ks} mag & 1$\sigma$ (\emph{Ks} mag) \\
       \hline
       7.50 & -42.41 & - & - & - & - & - & - & - & - & 18.91 & 0.02 & - & - \\ 
       7.51 & -42.43 & - & - & - & - & - & - & - & - & 19.06 & 0.02 & - & - \\ 
       7.51 & -42.44 & - & - & - & - & - & - & - & - & 19.61 & 0.03 & - & - \\ 
       7.51 & -42.37 & - & - & - & - & - & - & - & - & 19.58 & 0.02 & - & - \\ 
       7.53 & -42.41 & - & - & - & - & - & - & - & - & - & - & 17.10 & 0.03 \\ 
       7.53 & -42.44 & - & - & - & - & - & - & - & - & 18.86 & 0.02 & 19.08 & 0.04 \\ 
       7.54 & -42.38 & - & - & - & - & - & - & - & - & 20.20 & 0.03 & - & - \\ 
       7.54 & -42.35 & - & - & - & - & - & - & - & - & 18.58 & 0.02 & - & - \\ 
       7.54 & -42.42 & - & - & - & - & - & - & - & - & 18.33 & 0.02 & 18.52 & 0.04 \\ 
       7.55 & -42.43 & - & - & - & - & - & - & - & - & - & - & 16.74 & 0.03 \\ 
       7.55 & -42.38 & - & - & - & - & - & - & - & - & 19.52 & 0.02 & 19.85 & 0.13 \\ 
       7.56 & -42.50 & - & - & 18.48 & 0.09 & 18.19 & 0.09 & - & - & - & - & - & - \\ 
       7.56 & -42.40 & 20.75 & 0.03 & 20.32 & 0.07 & - & - & - & - & - & - & - & - \\ 
       7.56 & -42.51 & - & - & 18.38 & 0.07 & 17.84 & 0.08 & - & - & - & - & - & - \\ 
       7.57 & -42.58 & - & - & 20.59 & 0.09 & 19.78 & 0.11 & - & - & - & - & - & - \\ 
       7.57 & -42.49 & - & - & - & - & 20.28 & 0.10 & 20.17 & 0.04 & - & - & - & - \\ 
       7.57 & -42.41 & 20.67 & 0.05 & 20.06 & 0.07 & 19.80 & 0.08 & - & - & - & - & - & - \\ 
       7.57 & -42.44 & - & - & - & - & - & - & 20.51 & 0.04 & - & - & - & - \\ 
       7.57 & -42.53 & 19.90 & 0.06 & 19.77 & 0.05 & 19.66 & 0.05 & 19.50 & 0.04 & - & - & - & - \\ 
       7.57 & -42.39 & - & - & 18.00 & 0.06 & 17.70 & 0.05 & - & - & - & - & - & - \\ 
       7.57 & -42.55 & 17.91 & 0.03 & 18.88 & 0.06 & 17.56 & 0.06 & 17.02 & 0.02 & - & - & - & - \\ 
       7.57 & -42.34 & - & - & 17.31 & 0.05 & 17.04 & 0.05 & 18.64 & 0.05 & - & - & - & - \\ 
       7.58 & -42.52 & 18.48 & 0.01 & 20.77 & 0.13 & 20.22 & 0.10 & - & - & - & - & - & - \\ 
       7.58 & -42.61 & 19.88 & 0.09 & 18.02 & 0.05 & - & - & - & - & - & - & - & - \\ 
       7.58 & -42.56 & - & - & 18.22 & 0.10 & 17.79 & 0.03 & - & - & - & - & - & - \\ 
       7.58 & -42.34 & - & - & 19.78 & 0.05 & 17.10 & 0.13 & 18.93 & 0.04 & 18.41 & 0.02 & - & - \\ 
       7.58 & -42.60 & - & - & - & - & 19.13 & 0.05 & - & - & - & - & - & - \\ 
       7.59 & -42.53 & - & - & 20.70 & 0.12 & 19.47 & 0.04 & - & - & - & - & - & - \\ 
       7.59 & -42.46 & - & - & 20.79 & 0.04 & 20.31 & 0.18 & 20.48 & 0.11 & - & - & - & - \\ 
       7.59 & -42.55 & - & - & - & - & - & - & 19.32 & 0.02 & - & - & - & - \\ 
       7.59 & -42.59 & - & - & - & - & - & - & 17.63 & 0.01 & - & - & - & - \\ 
       7.59 & -42.34 & - & - & 19.51 & 0.07 & 19.76 & 0.04 & 20.37 & 0.07 & - & - & - & - \\ 
       7.59 & -42.54 & 20.61 & 0.03 & 20.59 & 0.06 & 18.24 & 0.07 & 20.19 & 0.06 & - & - & - & - \\ 
       7.60 & -42.34 & - & - & - & - & - & - & 19.95 & 0.03 & - & - & - & - \\ 
       7.60 & -42.44 & 20.06 & 0.01 & 20.88 & 0.06 & 20.27 & 0.06 & 18.51 & 0.02 & - & - & - & - \\ 
       7.60 & -42.59 & - & - & 19.28 & 0.06 & 18.73 & 0.06 & 19.25 & 0.03 & - & - & - & - \\ 
       7.60 & -42.60 & - & - & - & - & - & - & 17.21 & 0.00 & - & - & - & - \\ 
       7.61 & -42.50 & 20.77 & 0.01 & 18.51 & 0.02 & - & - & - & - & - & - & - & - \\ 
       7.61 & -42.45 & - & - & 20.09 & 0.07 & 19.05 & 0.09 & 18.92 & 0.09 & - & - & - & - \\ 
       7.61 & -42.62 & 20.54 & 0.01 & 20.74 & 0.27 & 19.52 & 0.14 & 18.67 & 0.01 & - & - & - & - \\ 
       7.61 & -42.34 & 20.21 & 0.02 & 19.46 & 0.03 & 18.91 & 0.02 & 19.01 & 0.01 & - & - & - & - \\ 
       7.61 & -42.43 & - & - & 19.99 & 0.03 & 19.32 & 0.02 & - & - & - & - & - & - \\ 
       7.61 & -42.35 & 17.73 & 0.01 & - & - & - & - & - & - & - & - & - & - \\ 
       7.61 & -42.32 & - & - & 19.41 & 0.06 & - & - & 19.99 & 0.08 & - & - & - & - \\ 
       7.61 & -42.42 & - & - & - & - & - & - & - & - & - & - & - & - \\ 
       7.62 & -42.51 & - & - & 20.98 & 0.05 & - & - & 19.82 & 0.03 & - & - & - & - \\ 
       7.62 & -42.45 & 18.56 & 0.02 & 17.27 & 0.05 & 20.44 & 0.12 & - & - & - & - & - & - \\ 
       7.62 & -42.59 & - & - & - & - & 19.47 & 0.02 & - & - & - & - & - & - \\ 
       7.62 & -42.32 & 19.37 & 0.04 & 20.77 & 0.04 & 18.34 & 0.02 & 18.03 & 0.02 & - & - & - & - \\ 
       7.62 & -42.55 & 19.57 & 0.02 & 18.68 & 0.04 & 18.01 & 0.02 & - & - & - & - & - & - \\ 
       7.63 & -42.38 & 20.42 & 0.04 & 18.19 & 0.02 & 18.68 & 0.10 & 18.42 & 0.02 & - & - & - & - \\ 
       7.63 & -42.33 & - & - & 18.95 & 0.10 & - & - & - & - & - & - & - & - \\ 
       7.63 & -42.47 & - & - & - & - & 18.13 & 0.04 & 16.74 & 0.03 & - & - & - & - \\ 
       7.63 & -42.58 & 17.49 & 0.02 & 18.57 & 0.05 & 17.39 & 0.03 & 20.08 & 0.06 & - & - & - & - \\ 
       7.63 & -42.51 & 18.52 & 0.02 & 18.79 & 0.04 & - & - & 20.40 & 0.10 & - & - & - & - \\ 
       7.63 & -42.28 & 18.38 & 0.05 & - & - & - & - & - & - & - & - & - & - \\ 
       7.64 & -42.34 & - & - & - & - & 16.42 & 0.07 & 17.65 & 0.02 & - & - & - & - \\ 
       7.64 & -42.31 & - & - & 16.72 & 0.08 & 17.69 & 0.01 & 16.60 & 0.04 & - & - & - & - \\ 
       7.65 & -42.57 & - & - & 17.94 & 0.02 & - & - & 17.99 & 0.01 & - & - & - & - \\ 
       7.65 & -42.35 & - & - & - & - & - & - & - & - & - & - & - & - \\ 
       7.65 & -42.36 & - & - & - & - & 19.50 & 0.06 & - & - & - & - & - & - \\ 
       \hline
    \end{tabular}
\end{table*}

\begin{table*}
    \centering
    \ContinuedFloat
    \caption{Continued}
    \begin{tabular}{C{0.04\textwidth}C{0.05\textwidth}C{0.05\textwidth}C{0.05\textwidth}C{0.05\textwidth}C{0.05\textwidth}C{0.05\textwidth}C{0.05\textwidth}C{0.05\textwidth}C{0.05\textwidth}C{0.05\textwidth}C{0.05\textwidth}C{0.05\textwidth}C{0.05\textwidth}}
       \hline
       RA (deg) & Dec (deg) & Mean \emph{g} mag & 1$\sigma$ (\emph{g} mag) & Mean \emph{r} mag & 1$\sigma$ (\emph{r} mag) & Mean \emph{i} mag & 1$\sigma$ (\emph{i} mag) & Mean \emph{z} mag & 1$\sigma$ (\emph{z} mag) & Mean \emph{J} mag & 1$\sigma$ (\emph{J} mag) & Mean \emph{Ks} mag & 1$\sigma$ (\emph{Ks} mag) \\
       \hline
       7.66 & -42.59 & 19.09 & 0.03 & - & - & 20.19 & 0.09 & 16.92 & 0.02 & - & - & - & - \\ 
       7.66 & -42.41 & - & - & - & - & 16.96 & 0.04 & - & - & - & - & - & - \\ 
       7.66 & -42.36 & 19.10 & 0.01 & 17.23 & 0.05 & 18.47 & 0.03 & - & - & - & - & - & - \\ 
       7.66 & -42.31 & - & - & 18.86 & 0.04 & 17.21 & 0.08 & 19.36 & 0.01 & - & - & - & - \\ 
       7.66 & -42.40 & 19.86 & 0.03 & 17.77 & 0.07 & 20.10 & 0.03 & - & - & - & - & - & - \\ 
       7.66 & -42.44 & - & - & - & - & 16.82 & 0.03 & - & - & - & - & - & - \\ 
       7.66 & -42.51 & 19.53 & 0.02 & 17.61 & 0.04 & 16.69 & 0.06 & 19.09 & 0.00 & - & - & - & - \\ 
       7.67 & -42.43 & 20.61 & 0.05 & - & - & 19.10 & 0.01 & - & - & - & - & - & - \\ 
       7.67 & -42.51 & - & - & 19.36 & 0.02 & 19.59 & 0.03 & 16.40 & 0.02 & - & - & - & - \\ 
       7.68 & -42.46 & 21.06 & 0.06 & 20.75 & 0.02 & 16.87 & 0.02 & 16.36 & 0.01 & - & - & - & - \\ 
       7.68 & -42.28 & - & - & 17.90 & 0.04 & 16.37 & 0.01 & - & - & - & - & - & - \\ 
       7.69 & -42.55 & 21.17 & 0.06 & 16.59 & 0.03 & 18.12 & 0.02 & 19.05 & 0.01 & - & - & - & - \\ 
       7.69 & -42.50 & - & - & 19.09 & 0.02 & 19.33 & 0.03 & 18.31 & 0.01 & - & - & - & - \\ 
       7.69 & -42.32 & 18.17 & 0.03 & 19.96 & 0.03 & 18.71 & 0.02 & 19.42 & 0.04 & - & - & - & - \\ 
       7.70 & -42.50 & 18.85 & 0.01 & 19.58 & 0.02 & - & - & - & - & - & - & - & - \\ 
       7.70 & -42.41 & 17.75 & 0.00 & - & - & - & - & 19.18 & 0.02 & - & - & - & - \\ 
       7.70 & -42.37 & - & - & 20.87 & 0.06 & 19.35 & 0.02 & 17.93 & 0.02 & - & - & - & - \\ 
       7.70 & -42.43 & - & - & 19.77 & 0.02 & 19.40 & 0.05 & 16.83 & 0.01 & - & - & - & - \\ 
       7.70 & -42.44 & 20.83 & 0.04 & - & - & 17.87 & 0.06 & - & - & - & - & - & - \\ 
       7.70 & -42.40 & - & - & 17.55 & 0.03 & 17.04 & 0.02 & 19.16 & 0.03 & - & - & - & - \\ 
       7.71 & -42.49 & 18.93 & 0.00 & 17.22 & 0.02 & 16.96 & 0.02 & 19.26 & 0.05 & - & - & - & - \\ 
       7.71 & -42.39 & - & - & - & - & 19.99 & 0.05 & 20.46 & 0.07 & - & - & - & - \\ 
       7.71 & -42.30 & - & - & 20.49 & 0.04 & - & - & - & - & - & - & - & - \\ 
       7.71 & -42.32 & - & - & 20.58 & 0.04 & - & - & - & - & - & - & - & - \\ 
       7.72 & -42.31 & - & - & 20.75 & 0.05 & 20.16 & 0.05 & 18.38 & 0.03 & - & - & - & - \\ 
       7.72 & -42.33 & 17.31 & 0.02 & 17.50 & 0.03 & 16.76 & 0.03 & - & - & - & - & - & - \\ 
       7.74 & -42.50 & - & - & 18.65 & 0.04 & 18.41 & 0.03 & - & - & - & - & - & - \\ 
       7.74 & -42.42 & - & - & - & - & 19.77 & 0.15 & 18.18 & 0.01 & - & - & - & - \\ 
       \hline
    \end{tabular}
\end{table*}

\subsection{Calibrating the Light Curves of PKS 0027-426 with Nearby, Non-Varying Objects}
\label{ap:ref_stars}

Nearby non-varying sources were used to calibrate the observations of \mbox{PKS 0027-426} to create the light curves shown in Figure~\ref{fig:lc}. Table~\ref{tab:r_ref_stars} contains the list of objects within the same detector that were used to correct the light curves of \mbox{PKS 0027-426}, including their position in RA and Dec (J2000), mean magnitude over the entire observational period in each filter, and standard deviation of the magnitudes in each night from the mean. The object had to be detected in every epoch observed for \mbox{PKS 0027-426} for it to be included, which is why some objects were only used as reference stars in some of the filters. As the NIR observations contained inconsistencies across the detector, the non-varying objects used to correct the NIR light curves were further restricted to within $\sim$ 200 pixels of \mbox{PKS 0027-426}.

\subsection{Light Curve Variability with VEILS Flux-Flux plots}
\label{ap:flux-flux}

Figure \ref{fig:flux_flux_nir} demonstrates a comparison between the flux in all DES \emph{r}, \emph{i} and \emph{z} bands and VEILS \emph{J} and \emph{Ks} bands with the DES \emph{g} band flux in the season starting 2017. Each flux is shown to increase with increasing \emph{g} band flux. The NIR light curves were not observed on the same epochs as the optical light curves, therefore to create this plot, the NIR light curves were interpolated. To prevent the interpolations from impacting the results too much, the light curves were limited to the epochs greater than MJD 58044 as there is an $\sim$ month long gap between observations in the NIR light curve during which a flare is present in the optical. 

Figure \ref{fig:gz_flux_flux} displays the comparison between the DES \emph{g} and \emph{z} bands for each individual observation season. Although the relation over the entire observational season is not exactly linear, the individual seasons do look approximately linear. The slope is shown to get steeper over time, which supports the analysis of the spectral variability in Section~\ref{Sect:Colour} as in the later seasons of the \emph{g-z} plots, the redder filter (\emph{z}) becomes more variable as it gets brighter. 

\subsection{Mean Cadences of Each Observation Season}
\label{ap:cadences}

\begin{figure*}
    \begin{minipage}{\textwidth}
        \centering
        \subfloat[figure][Flux variations in each DES and VEILS filter compared to the DES \emph{g} band for the 2017 season. The NIR light curves were interpolated to extract simultaneous data to the DES \emph{g} band observations.\label{fig:flux_flux_nir}]{\includegraphics[width=0.49\textwidth]{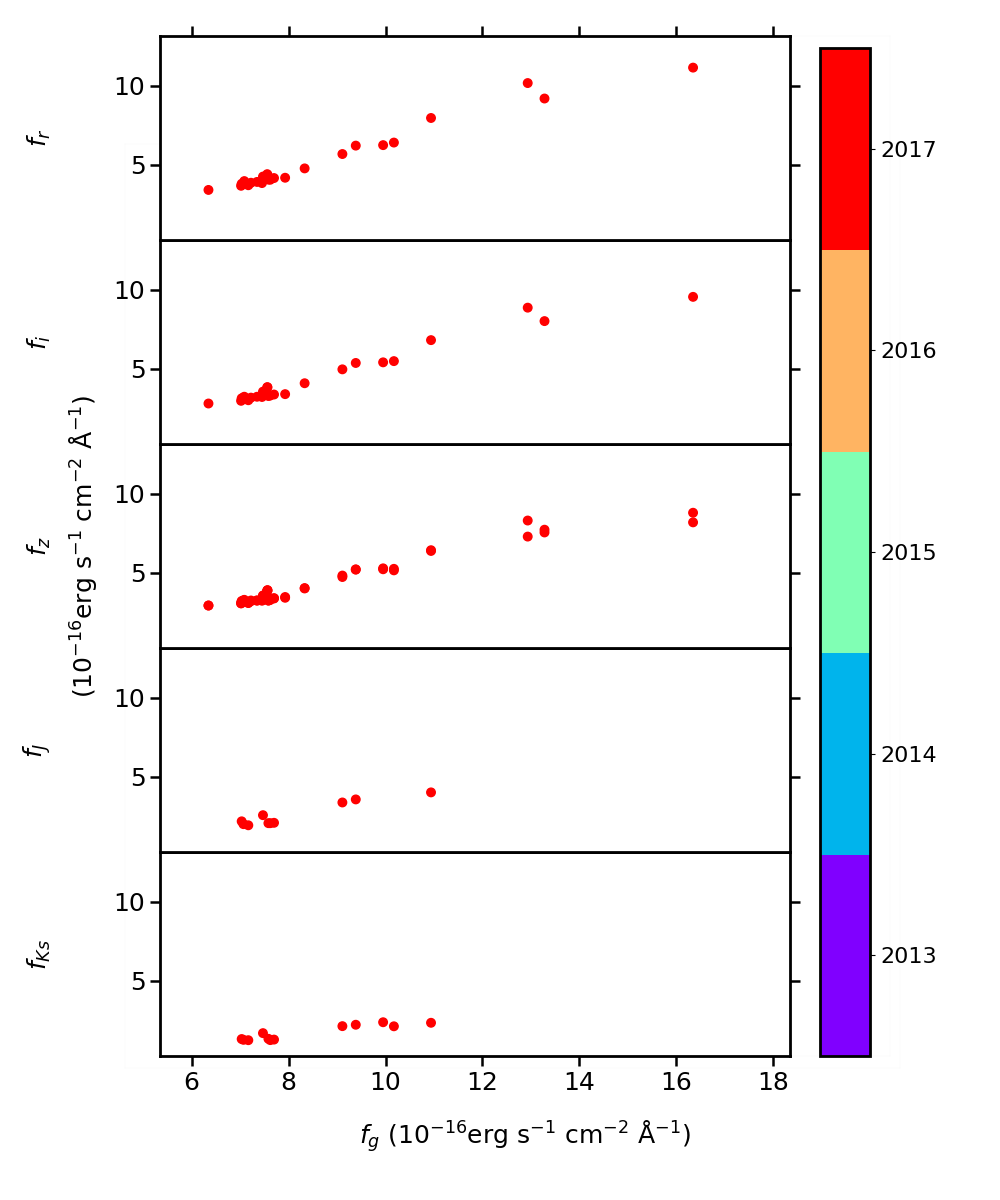}}
        \hfill
        \subfloat[figure][Flux variations in the DES \emph{z} band compared to the DES \emph{g} band for each season season. \label{fig:gz_flux_flux}]{\includegraphics[width=0.49\textwidth]{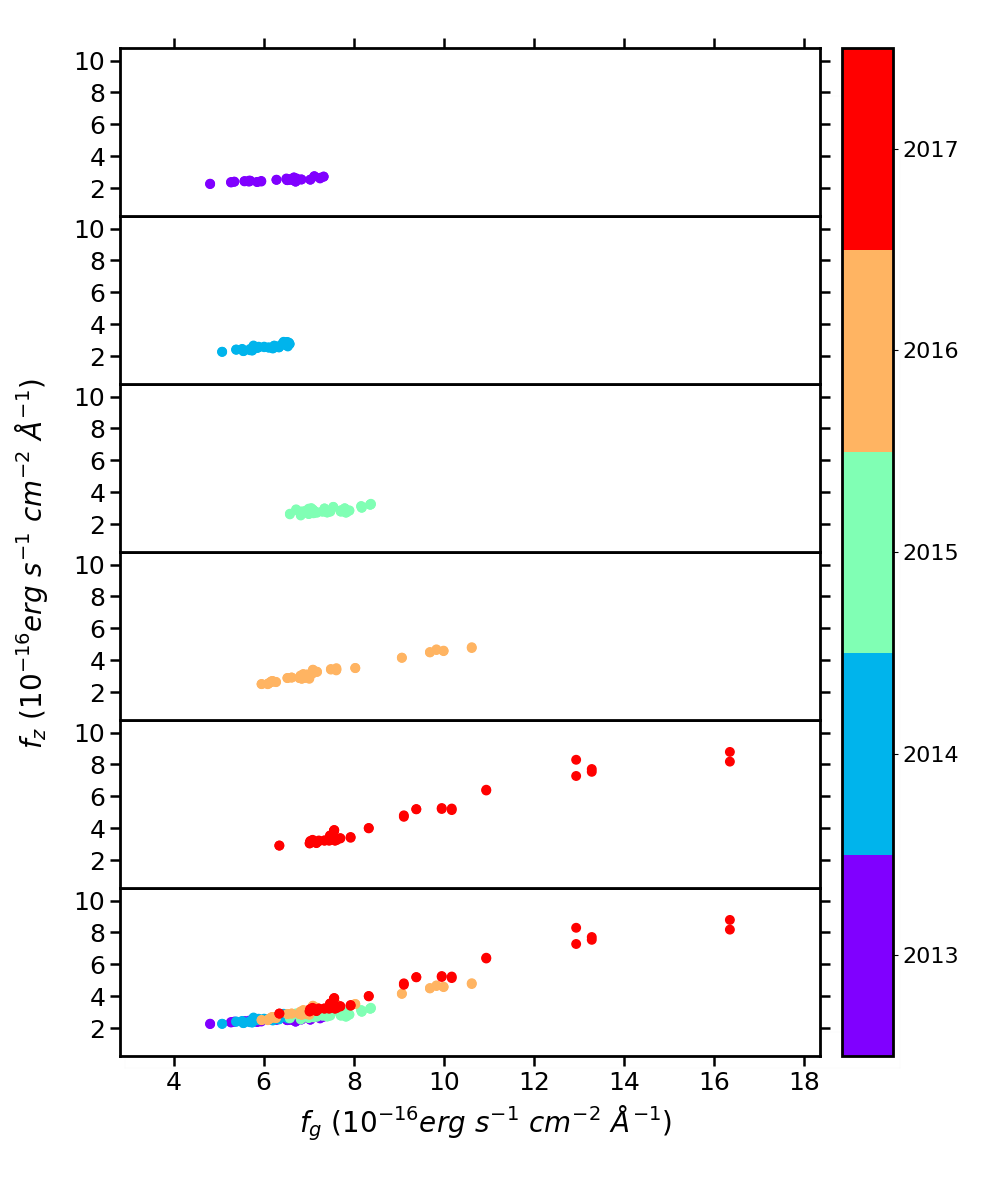}}
        \caption{Flux variations in the different filters compared to the DES \emph{g} band.}
    \vspace{0.5cm}
    \end{minipage}
    \\
    \begin{minipage}{\textwidth}
        \centering
        \footnotesize{
        \captionof{table}{The mean cadences of observations of PKS~0027-426 each season for each filter, where the uncertainty is the standard deviation. \label{tab:mean_cad}}        
        \begin{tabular}{C{0.1\textwidth}C{0.12\textwidth}C{0.12\textwidth}C{0.12\textwidth}C{0.12\textwidth}C{0.12\textwidth}C{0.12\textwidth}}
            \hline
                Year Starting & \emph{g} (nights) & \emph{r} (nights) & \emph{i} (nights) & \emph{z} (nights) & \emph{J} (nights) & \emph{Ks} (nights) \\
            \hline
               2013 & 6.5 $\pm$ 3.0 & 6.5 $\pm$ 3.0 & 6.5 $\pm$ 3.0 & 5.5 $\pm$ 3.0 & - & - \\
               2014 & 6.2 $\pm$ 2.9 & 6.2 $\pm$ 2.9 & 6.4 $\pm$ 3.0 & 6.2 $\pm$ 2.6 & - & - \\
               2015 & 5.7 $\pm$ 2.7 & 5.5 $\pm$ 2.9 & 5.9 $\pm$ 3.2 & 5.5 $\pm$ 2.9 & - & - \\
               2016 & 5.7 $\pm$ 3.0 & 5.7 $\pm$ 2.7 & 5.3 $\pm$ 2.9 & 5.3 $\pm$ 2.9 & - & - \\
               2017 & 6.1 $\pm$ 2.2 & 6.1 $\pm$ 2.2 & 6.1 $\pm$ 2.2 & 6.1 $\pm$ 2.2 & 12.2 $\pm$ 10.5 & 11.7 $\pm$ 6.6 \\
               2018 & 68.9 $\pm$ 14.9 & 12.9 $\pm$ 14.4 & 12.8 $\pm$ 11.3 & 68.8 $\pm$ 14.9 & 10.3 $\pm$ 4.3 & 10.9 $\pm$ 5.0 \\
            \hline
        \end{tabular}}
	\end{minipage}
\end{figure*}

The mean cadence of each observation season in each filter is presented in Table \ref{tab:mean_cad}. PKS~0027-426 was only observed 3 times in the \emph{g} and \emph{z} bands in the 2018 season, hence it has a much larger mean cadence.

\section{Temporal Variability}

\subsection{More CCFs of the Individual Observation Seasons}
\label{ap:more_lags}

\begin{figure*}
    \centering
    \subfloat[figure][CCFs and ACFs of the 2013 \emph{r} and \emph{i} light curves. \label{fig:ri_CCF_2013}]{
        \begin{minipage}{0.32\textwidth}
            \includegraphics[width=\textwidth]{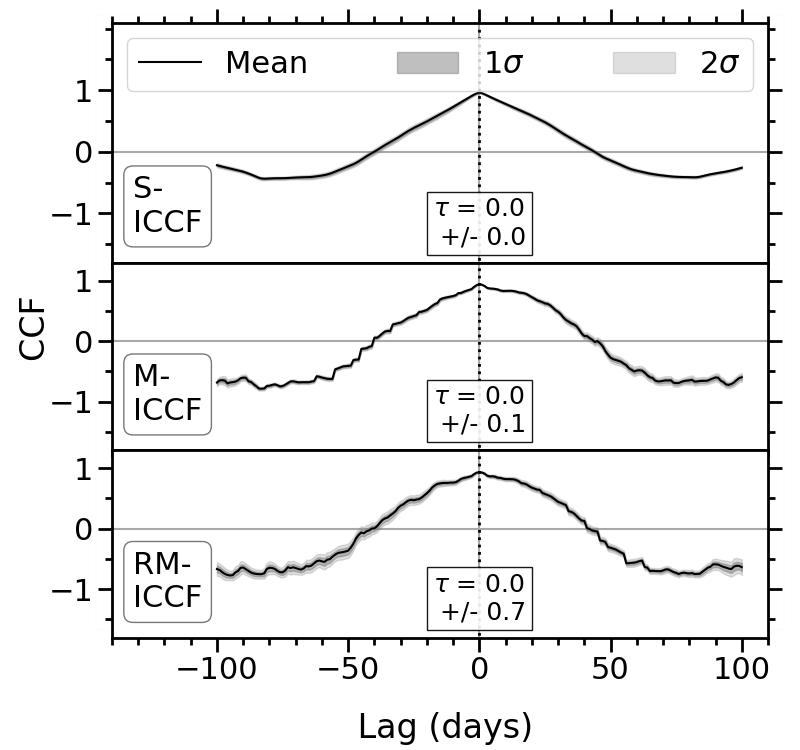}\\
            \includegraphics[width=\textwidth]{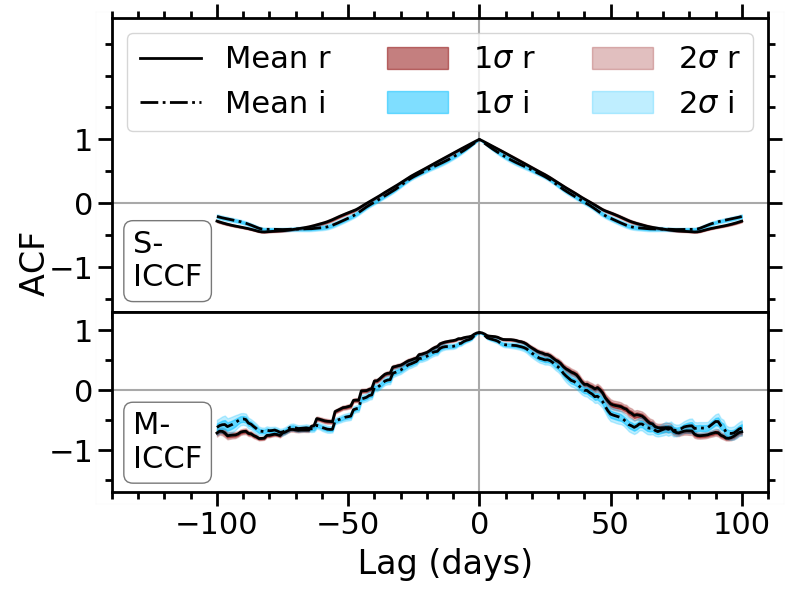}
        \end{minipage}}
    \subfloat[figure][CCFs and ACFs of the 2014 \emph{r} and \emph{i} light curves. \label{fig:ri_CCF_2014}]{
        \begin{minipage}{0.32\textwidth}
            \includegraphics[width=\textwidth]{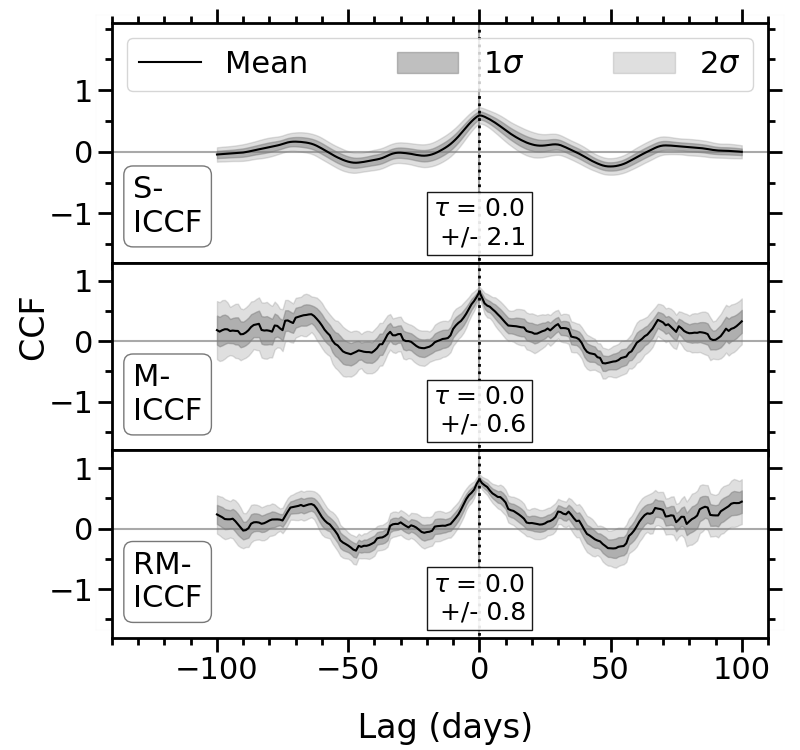} \\
            \includegraphics[width=\textwidth]{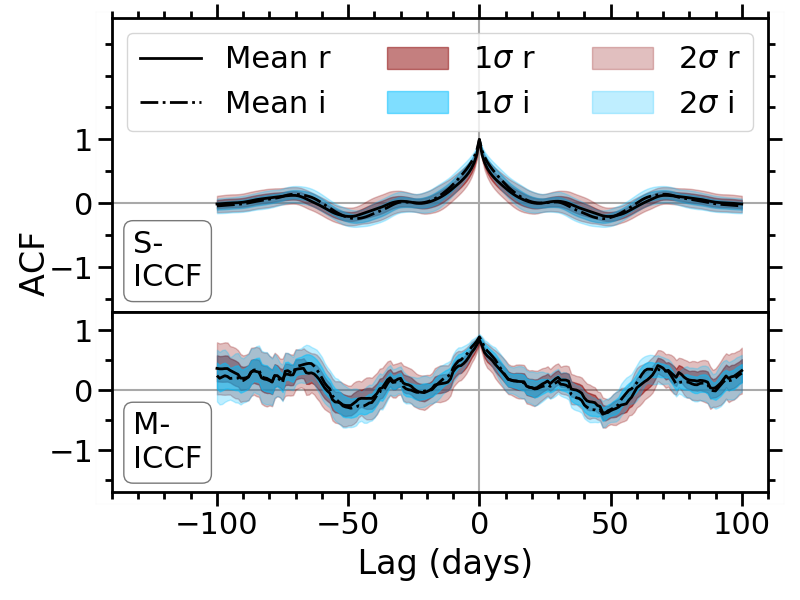}
        \end{minipage}}
    \subfloat[figure][CCFs and ACFs of the 2015 \emph{r} and \emph{i} light curves. \label{fig:ri_CCF_2015}]{
        \begin{minipage}{0.32\textwidth}
            \includegraphics[width=\textwidth]{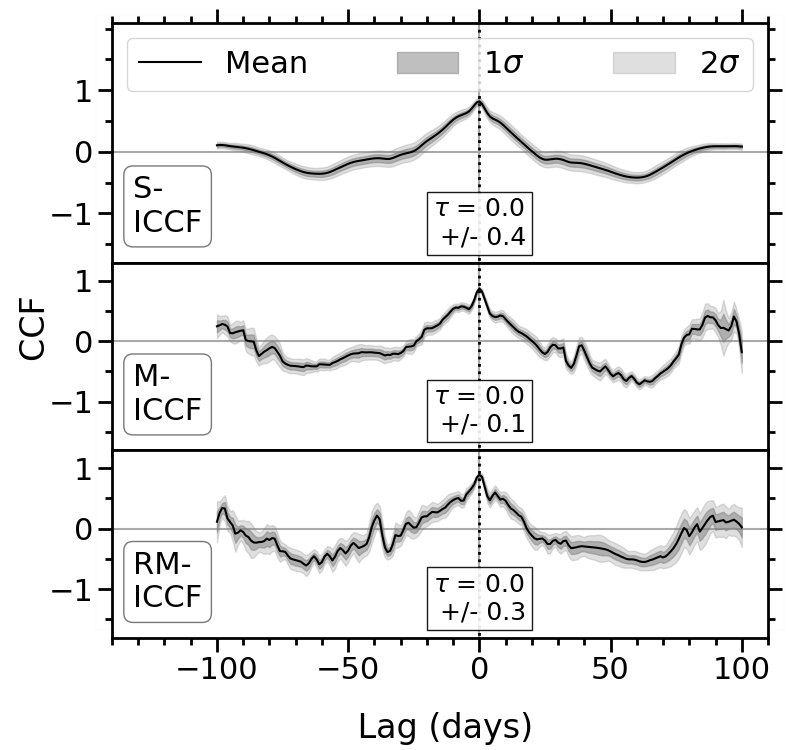} \\
            \includegraphics[width=\textwidth]{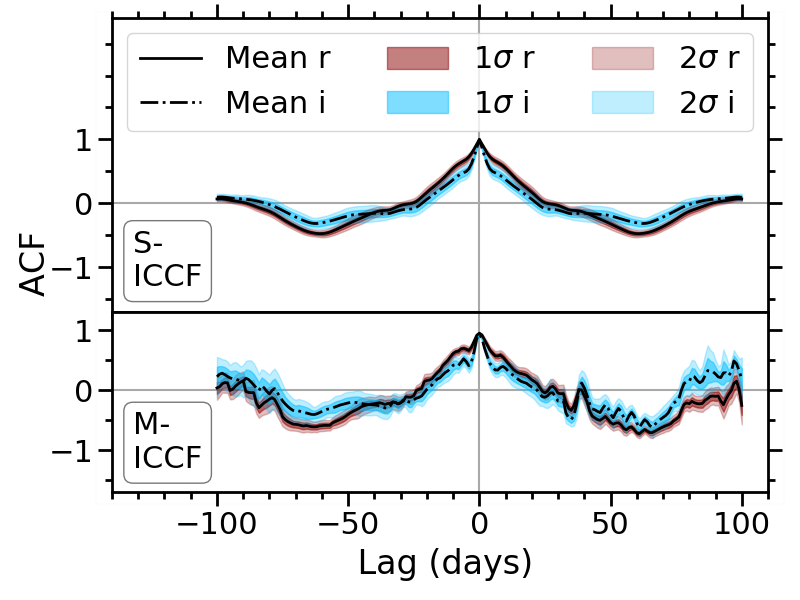}
        \end{minipage}}
    \\
    \subfloat[figure][CCFs and ACFS of the 2016 \emph{r} and \emph{i} light curves. \label{fig:ri_CCF_2016}]{
        \begin{minipage}{0.32\textwidth}
            \includegraphics[width=\textwidth]{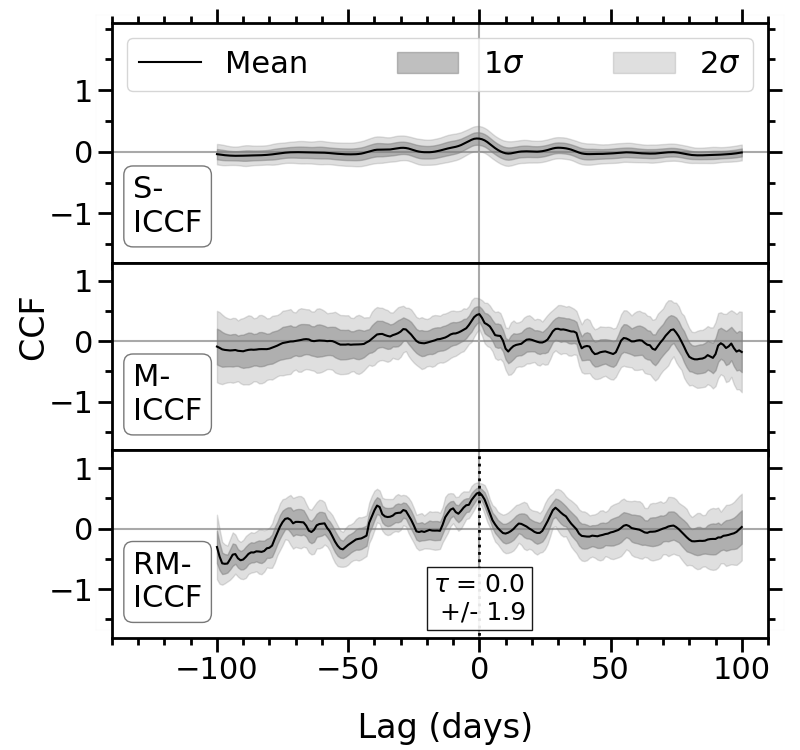} \\
            \includegraphics[width=\textwidth]{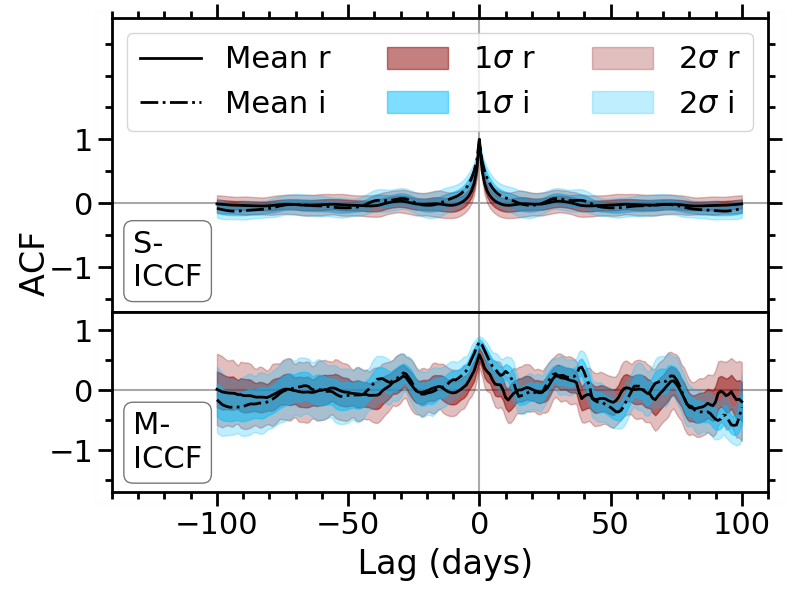}
        \end{minipage}}
    \subfloat[figure][CCFs and ACFs of the 2018 \emph{r} and \emph{i} light curves. \label{fig:ri_CCF_2018}]{
        \begin{minipage}{0.32\textwidth}
            \includegraphics[width=\textwidth]{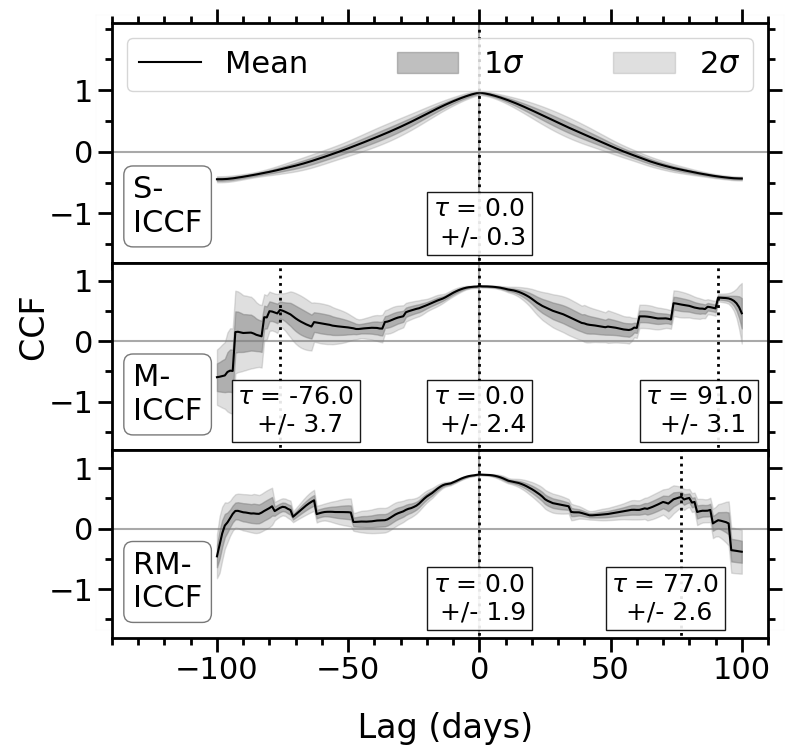} \\
            \includegraphics[width=\textwidth]{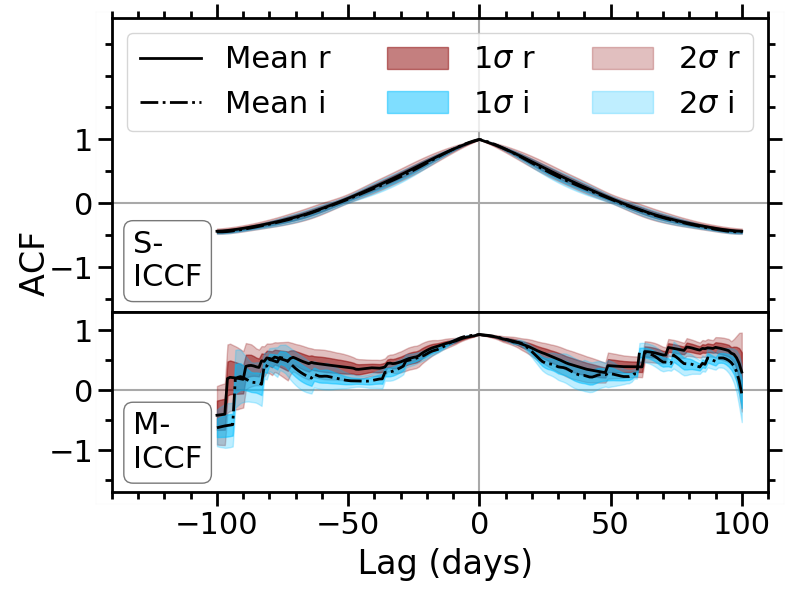}
        \end{minipage}}
    \caption{\textit{Top Panels}: Mean CCFs of the \emph{r} and \emph{i} light curves of the individual years of PKS~0027-426 between 2013-2016 and 2018, made using each method of CCF. \textit{Lower Panels}: The corresponding ACFs \label{fig:ri_CCFs_other_years}. In these CCFs, the \mbox{M-ICCF} method refers to the interpolated \emph{r} band, and the \mbox{RM-ICCF} method refers to the interpolated \emph{i} band. Potential lags corresponding to peak in the mean CCF are labelled, with the uncertainties calculated as the standard deviation of the peak in each CCF around the peak of the mean CCF.}
\end{figure*}

\begin{figure*}
    \begin{minipage}{0.32\textwidth}
        \centering
        \begin{minipage}{\textwidth}
            \includegraphics[width=\textwidth]{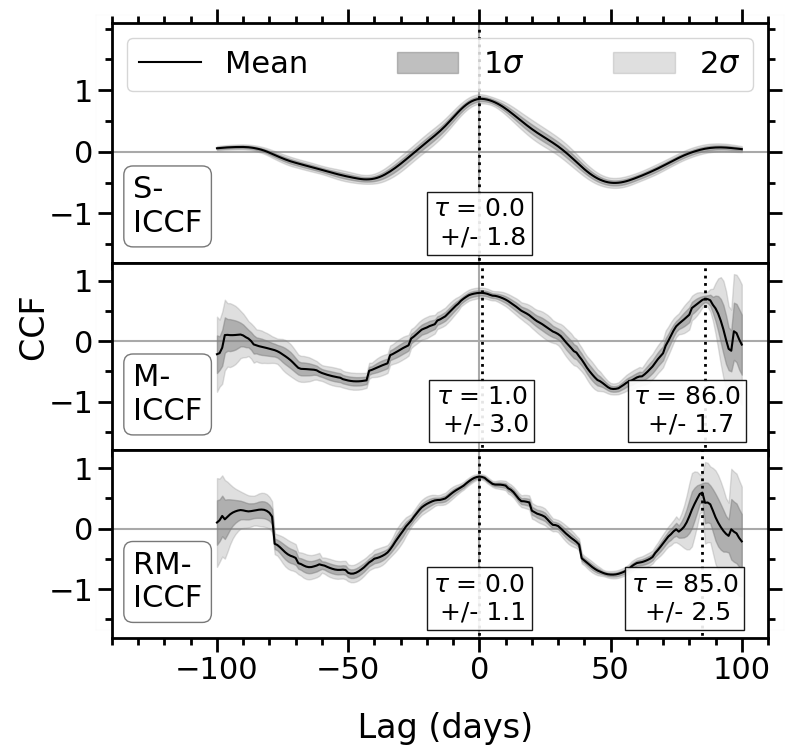}
        \end{minipage}
        \vfill
        \begin{minipage}{\textwidth}
            \includegraphics[width=\textwidth]{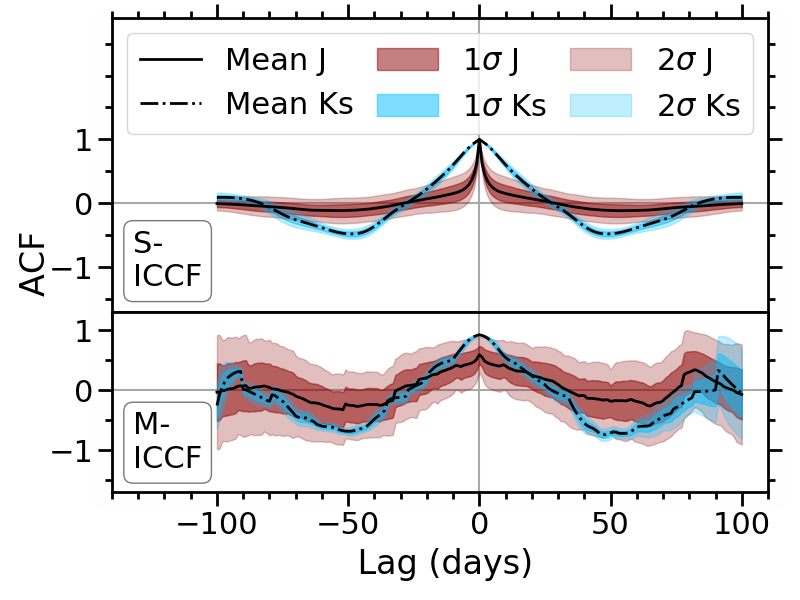}
            \captionof{figure}{\small \textit{Top Panels}: Mean CCFs of the \emph{J} and \emph{Ks} light curves of the 2018 season, made using each method of CCF. \textit{Lower Panels}: The corresponding ACFs. In these CCFs, the \mbox{M-ICCF} method refers to the interpolated \emph{J} band, and the \mbox{RM-ICCF} method refers to the interpolated \emph{Ks} band. \label{fig:JKs_CCF_2018}}
            \vspace{3mm}
        \end{minipage}
        \hspace{0.2cm}
    \end{minipage}
    \hspace{3mm}
    \begin{minipage}{0.32\textwidth}
        \centering
        \begin{minipage}{\textwidth}
            \includegraphics[width=\textwidth]{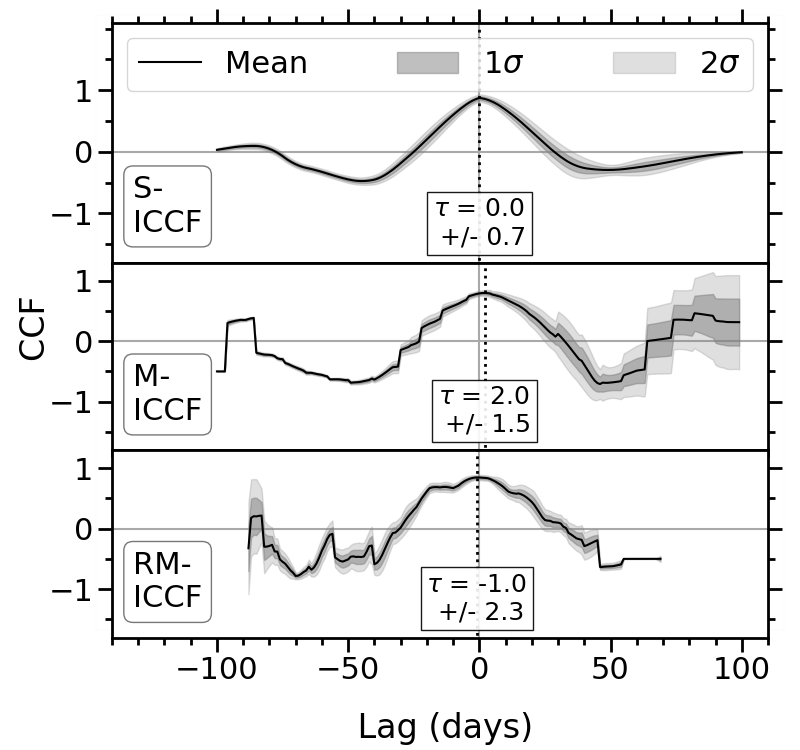}
         \end{minipage}
         \vfill
         \begin{minipage}{\textwidth}
            \includegraphics[width=\textwidth]{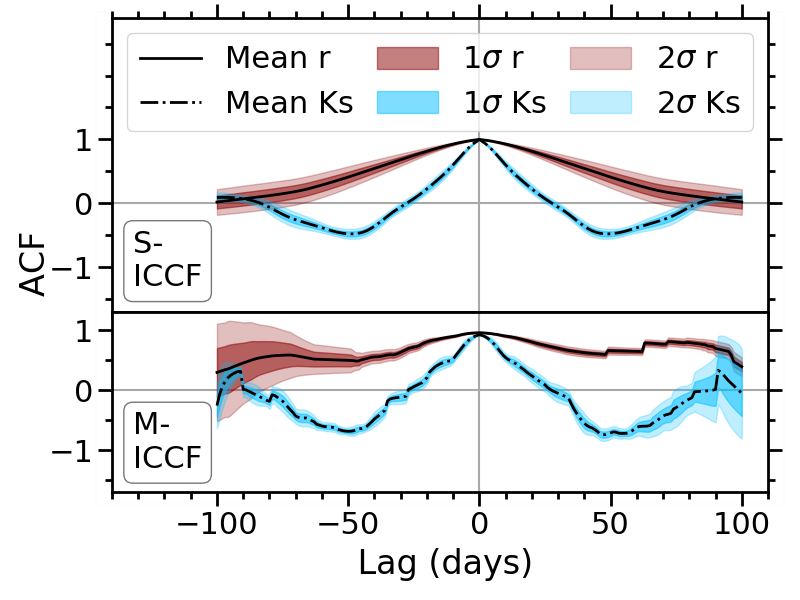}
            \captionof{figure}{\small \textit{Top Panels}: Mean CCFs of the \emph{r} and \emph{Ks} light curves of the 2018 season, made using each method of CCF. \textit{Lower Panels}: The corresponding ACFs. In these CCFs, the \mbox{M-ICCF} method refers to the interpolated \emph{r} band, and the \mbox{RM-ICCF} method refers to the interpolated \emph{Ks} band. \label{fig:rKs_CCF_2018}}
            \vspace{3mm}
         \end{minipage}
    \end{minipage}
\end{figure*}

The \emph{r} and \emph{i} band CCFs of the individual seasons starting 2013-2016 and 2018 are displayed in Figure~\ref{fig:ri_CCFs_other_years} for each CCF method, along with their corresponding ACFs. Each CCF detects a correlation at $\sim$ 0 days, except the 2016 season which instead contains a small peak at $\sim$ 0 days with a value of less than 0.5 in the \mbox{S-ICCF} and \mbox{M-ICCF} methods, and was therefore not counted. The CCFs from 2016 are relatively level, especially in the \mbox{S-ICCF} method, implying that there is no distinctive lag observed in this season, which could be due to the shape of the 2016 light curves, which contain multiple peaks and troughs that would all correlate with each other. The 2018 season CCFs of the \emph{J} and \emph{Ks} band light curves and the \emph{r} and \emph{Ks} band light curves are displayed in Figures~\ref{fig:JKs_CCF_2018} and \ref{fig:rKs_CCF_2018} respectively.

\subsection{Investigating the Possible 75 Day Lag Between Light Curves}
\label{ap:test75daylag}

\begin{figure*}
    \begin{minipage}{0.49\textwidth}
        \centering
        \includegraphics[width=\columnwidth]{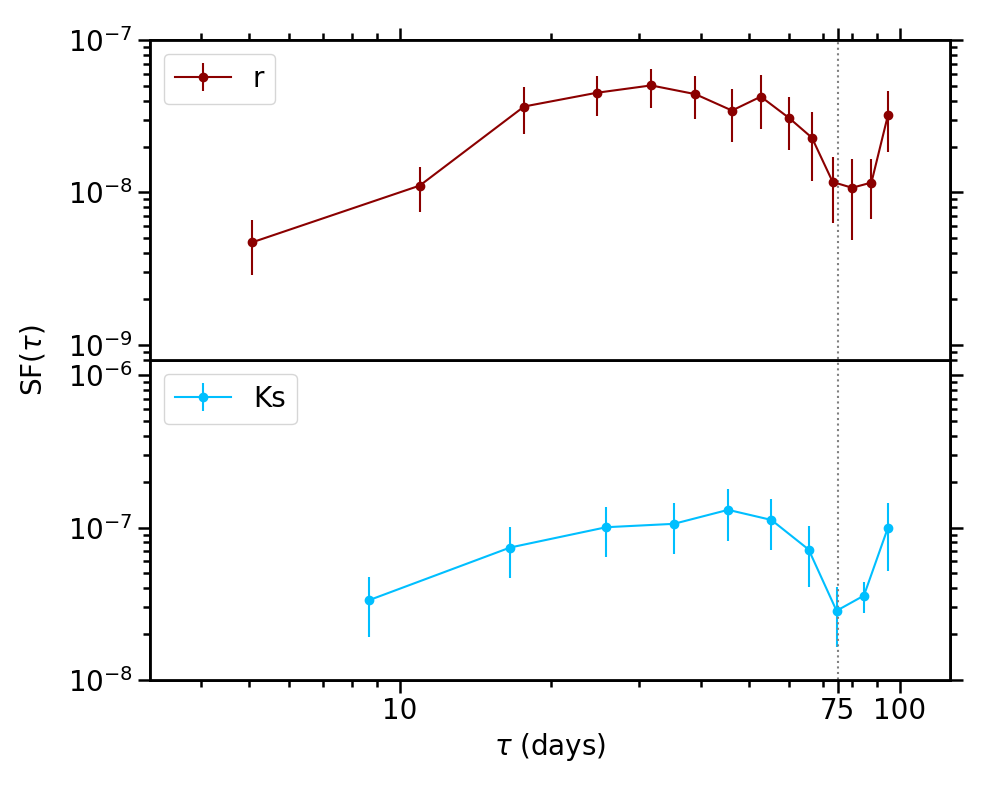}
        \caption{\centering Structure functions of PKS~0027-426 from the \emph{r} and \emph{Ks} band combined 2017 and 2018 light curves.}
        \label{fig:PKS_SF_r_k}
    \end{minipage}
    \hfill
    \begin{minipage}{0.49\textwidth}
        \centering
        \includegraphics[width=\columnwidth]{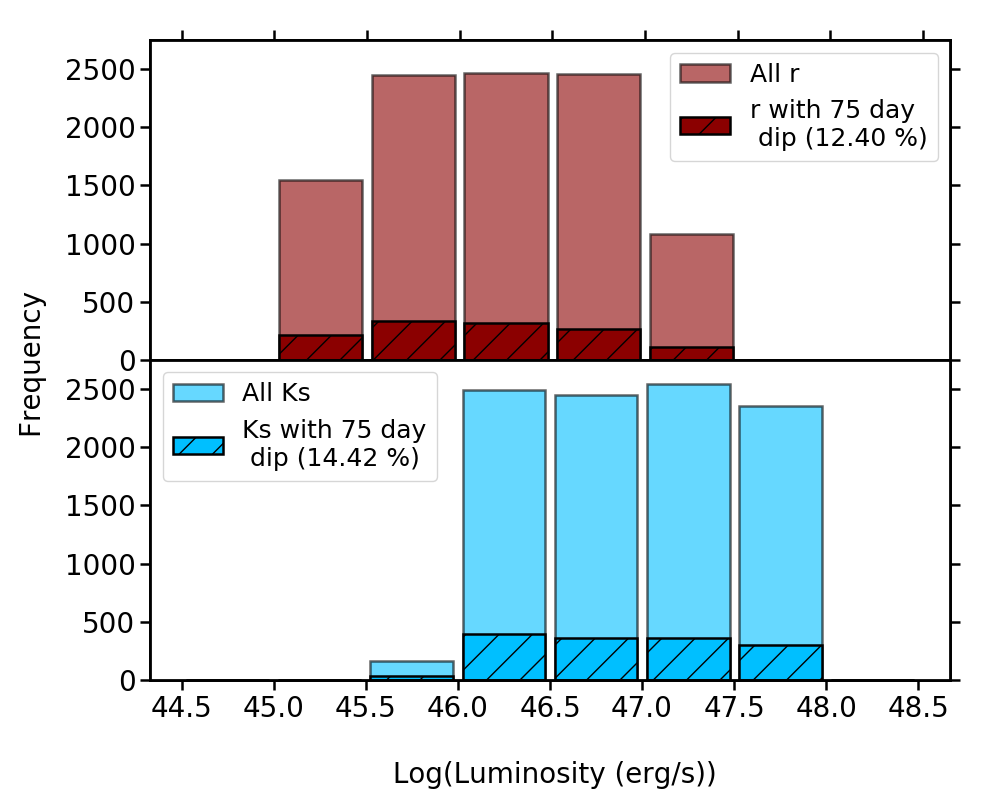}
        \caption{\centering Comparisons between the distributions of the luminosity corresponding to the structure functions from the simulated light curves that demonstrate a dip at $\sim$~75 days with all simulated light curves in \emph{r} and \emph{Ks} bands.}
        \label{fig:sf_lum_dist_r_k}
    \end{minipage}
\end{figure*}

In multiple CCFs a possible lag is detected at $\sim \pm$ 75 days which appears to be due to aliasing. This lag was further investigated by analysing the \emph{r} and \emph{Ks} band SFs from observations between 2017-2019, which are displayed in Figure~\ref{fig:PKS_SF_r_k}, in which an obvious dip is seen at $\sim$ 75 days in both filters. To test whether or not this dip in the SFs was intrinsic to PKS~0027-426, light curves were simulated using the method described by \cite{TimmerKonig1995}, in which a power spectrum is created from the data and is used to produce light curves with similar variability and noise as the data. 10,000 light curves were created using this method for a range of power spectra with varied break frequencies, slopes, and white noise amplitudes, created by varying properties of \mbox{PKS~0027-426} including the luminosity and Eddington luminosity (\citealt{Kelly2011}). The percentage of the SFs from these light curves that also displayed a dip at $\sim$ 75 days were found to be $\sim$~12\% and $\sim$~14\% in \emph{r} and \emph{Ks} bands respectively. Furthermore, the distributions of the SFs with varied inputs that returned the dip at $\sim$ 75 days were plotted and compared to the distribution of SFs from all simulated light curves. The shape was shown to be similar for each value of luminosity in Figure~\ref{fig:sf_lum_dist_r_k}, which therefore implied that the $\sim$ 75 day dip did not depend on specific properties of the light curve of \mbox{PKS 0027-426}, but occurred for a random $\sim$ 12\%.

\section{Spectral Variability}

\subsection{DES Colour - Magnitude Plots for Individual Observation Seasons}
\label{sec:app_des_colmag}

The optical colour-magnitude plots for each season of each combination of DES \emph{griz} filters, \emph{g}-\emph{r}, \emph{g}-\emph{i}, \emph{r}-\emph{i}, \emph{r}-\emph{z} and \emph{i}-\emph{z} are plotted in Figure~\ref{fig:all_sep_years}, and the corresponding tables containing the slopes, Spearman rank correlation coefficients and probability of no correlation are given in Table \ref{tab:sep_years}, to demonstrate how the spectral behaviour changes over time. These colour-magnitude plots are made excluding the outliers which are displayed in Figure~\ref{fig:outliers} for the 2013-2016 seasons of DES, and the corresponding differences in slope between when the outliers are included and excluded is given in Table~\ref{tab:outliers}.

\subsection{Colour - Magnitude Plots from Interpolating DES Light Curves}
\label{sec:app_des_interps}

\begin{figure*}
    \begin{minipage}{\textwidth}
        \centering
        \subfloat[figure][$\Delta$Slope for g-r colour vs r magnitude plots. \label{fig:gr_deltaslope}]{\includegraphics[width=0.49\textwidth]{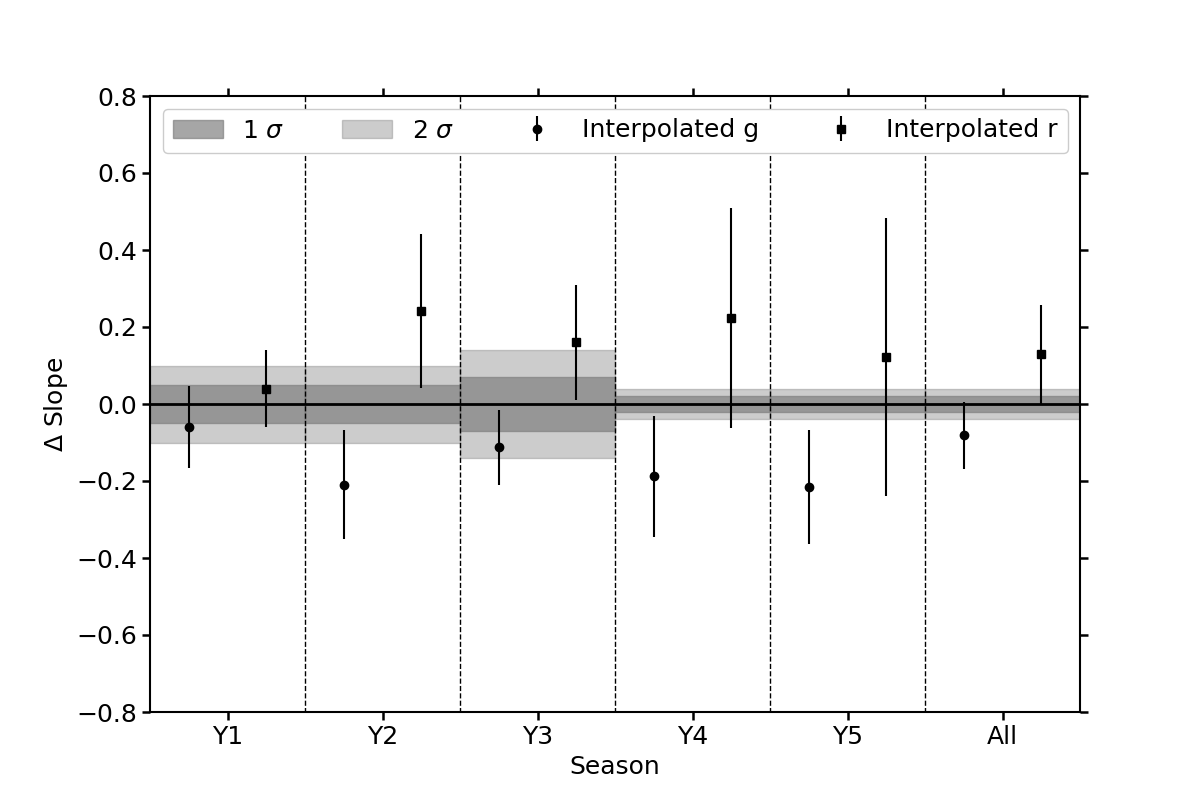}}
        \hfill
        \subfloat[figure][$\Delta$Slope for g-i colour vs r magnitude plots. \label{fig:gi_deltaslope}]{\includegraphics[width=0.49\textwidth]{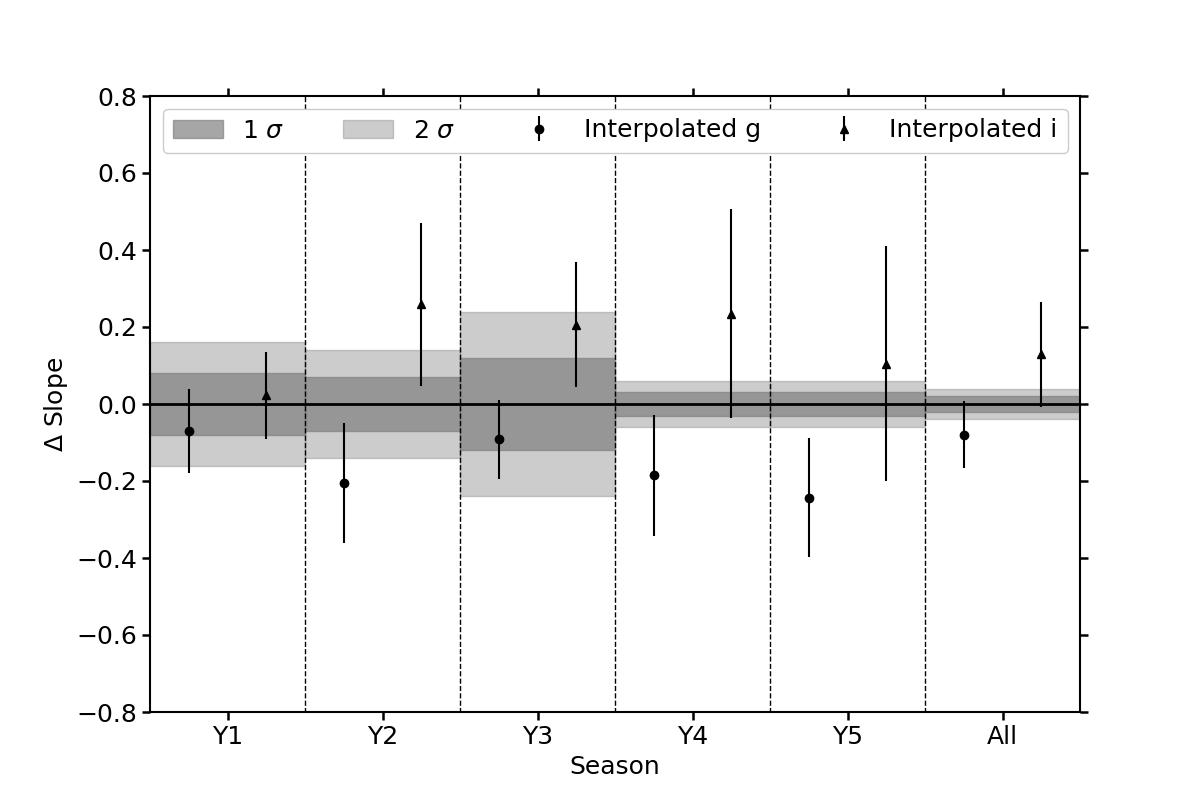}}
        \\
        \subfloat[figure][$\Delta$Slope for g-z colour vs r magnitude plots. \label{fig:gz_deltaslope}]{\includegraphics[width=0.49\textwidth]{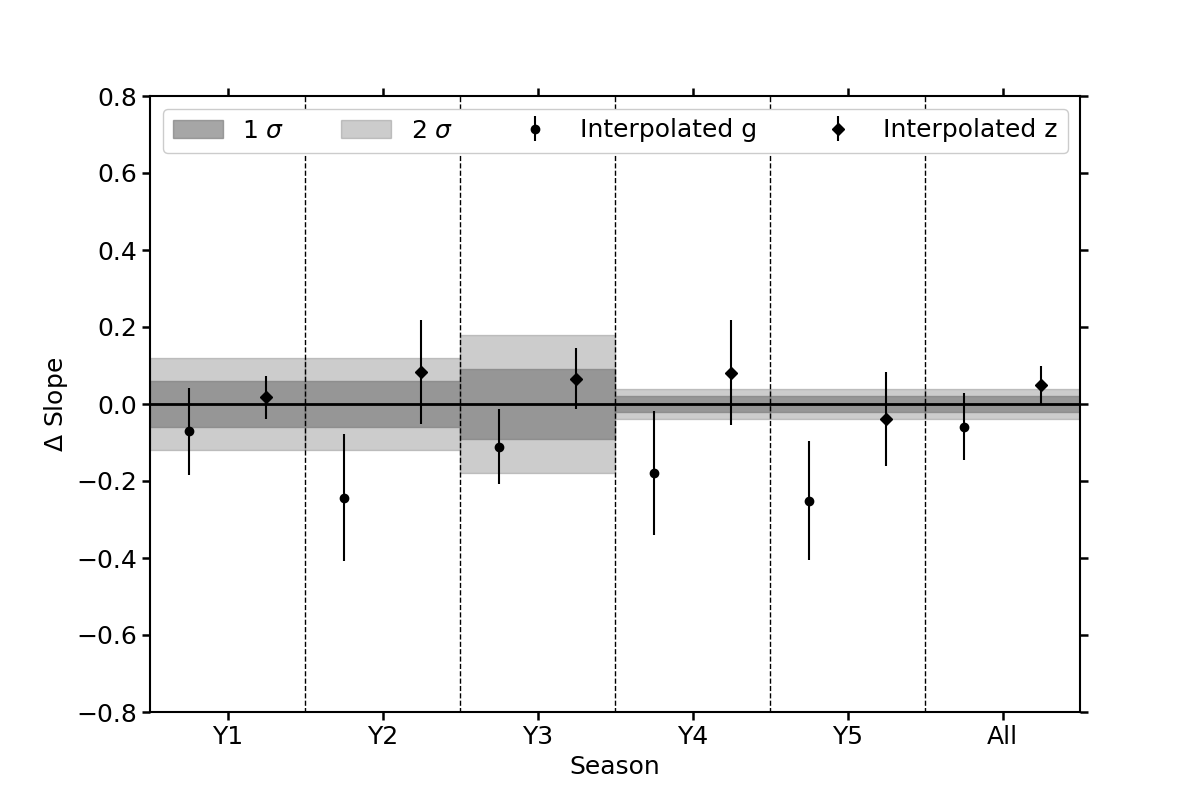}}
        \hfill
        \subfloat[figure][$\Delta$Slope for r-i colour vs r magnitude plots. \label{fig:ri_deltaslope}]{\includegraphics[width=0.49\textwidth]{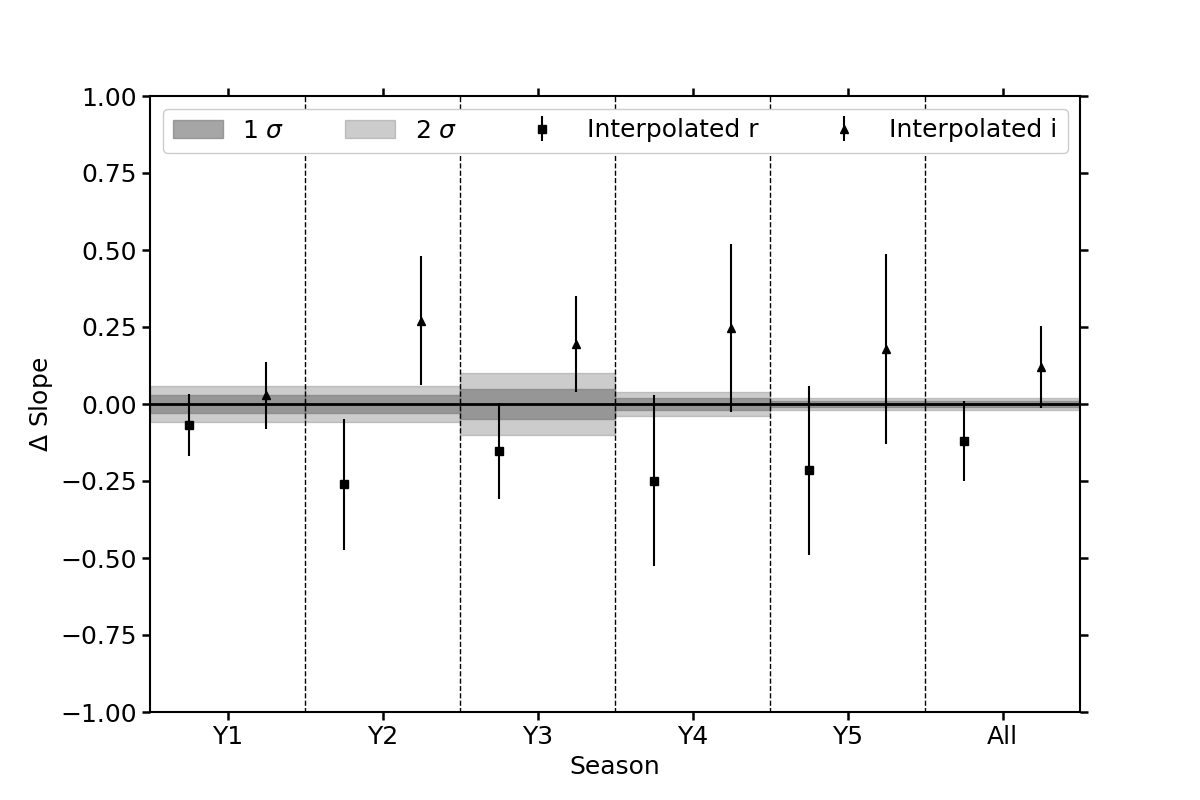}}
        \\
        \subfloat[figure][$\Delta$Slope for r-z colour vs r magnitude plots. \label{fig:rz_deltaslope}]{\includegraphics[width=0.49\textwidth]{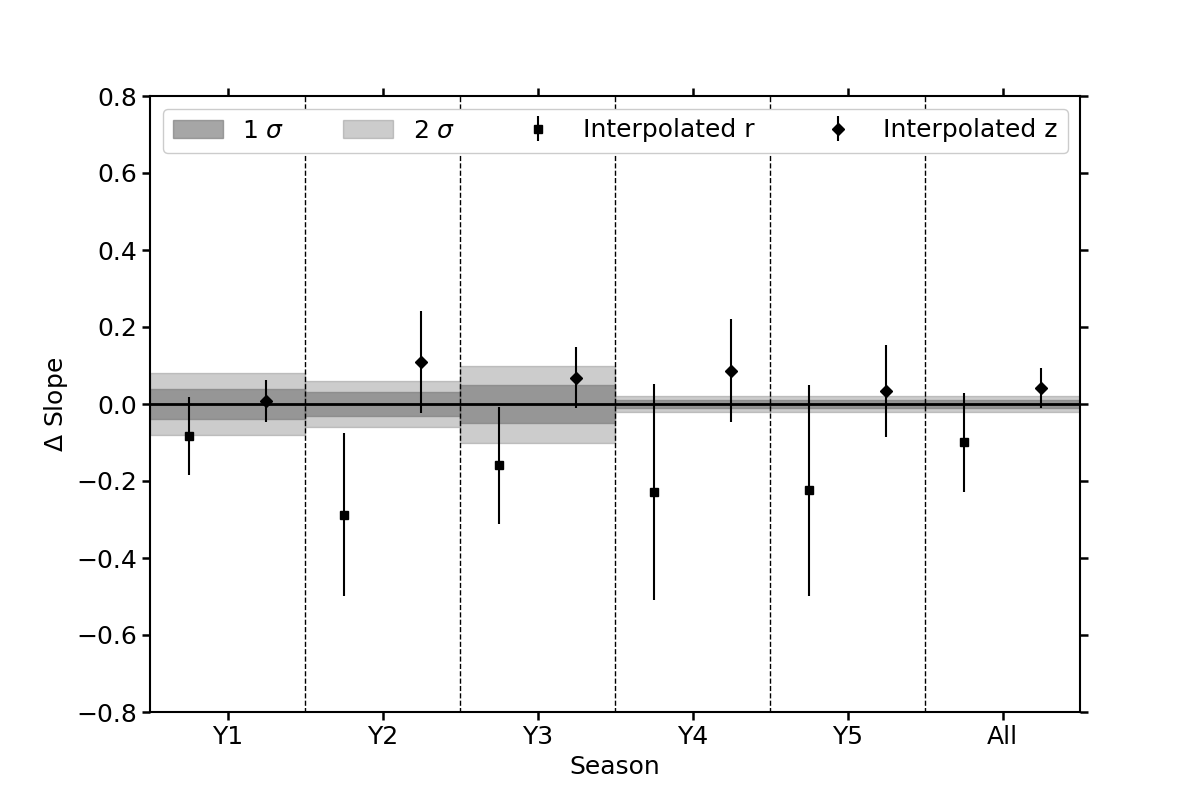}}
        \hfill
        \subfloat[figure][$\Delta$Slope for i-z colour vs r magnitude plots. \label{fig:iz_deltaslope}]{\includegraphics[width=0.49\textwidth]{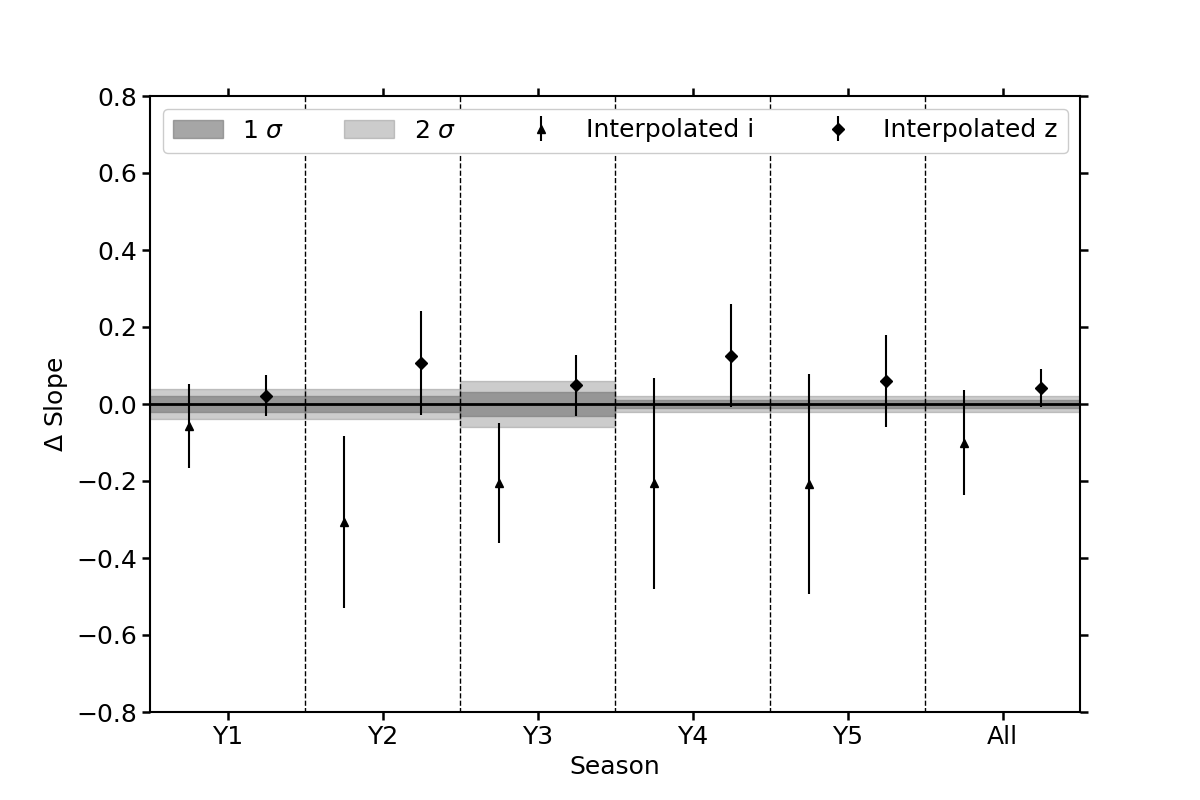}}
        \caption{\small $\Delta$Slope for the DES colour vs r magnitude plots, comparing the slope obtained from the data with the slope obtain from interpolating one of the light curves after removing 50$\%$ of the observations. The shaded regions demonstrate the 1$\sigma$ and 2$\sigma$ uncertainties of the slopes obtained from the data. \label{fig:deltaslope}}
        \vspace{2cm}
    \end{minipage}
\end{figure*}

Figure~\ref{fig:deltaslope} further demonstrates the consistency between calculating the slope of the colour vs r magnitude plots from DES using quasi-simultaneous observations, $S_{\text{Act}}$, and calculating the slope of each interpolated light curve which have had 50$\%$ of the data points removed, $S_{\text{Interp}}$, using $\Delta$S:

\begin{equation}
    \centering
    \Delta S = S_{\text{Act}} - S_{\text{Interp}}
\end{equation}

\begin{figure*}
    \subfloat[figure][Optical \emph{g}-\emph{r} colour variability for each observation season of DES. \newline \label{fig:gr_sep_years}]{\includegraphics[width=0.49\textwidth]{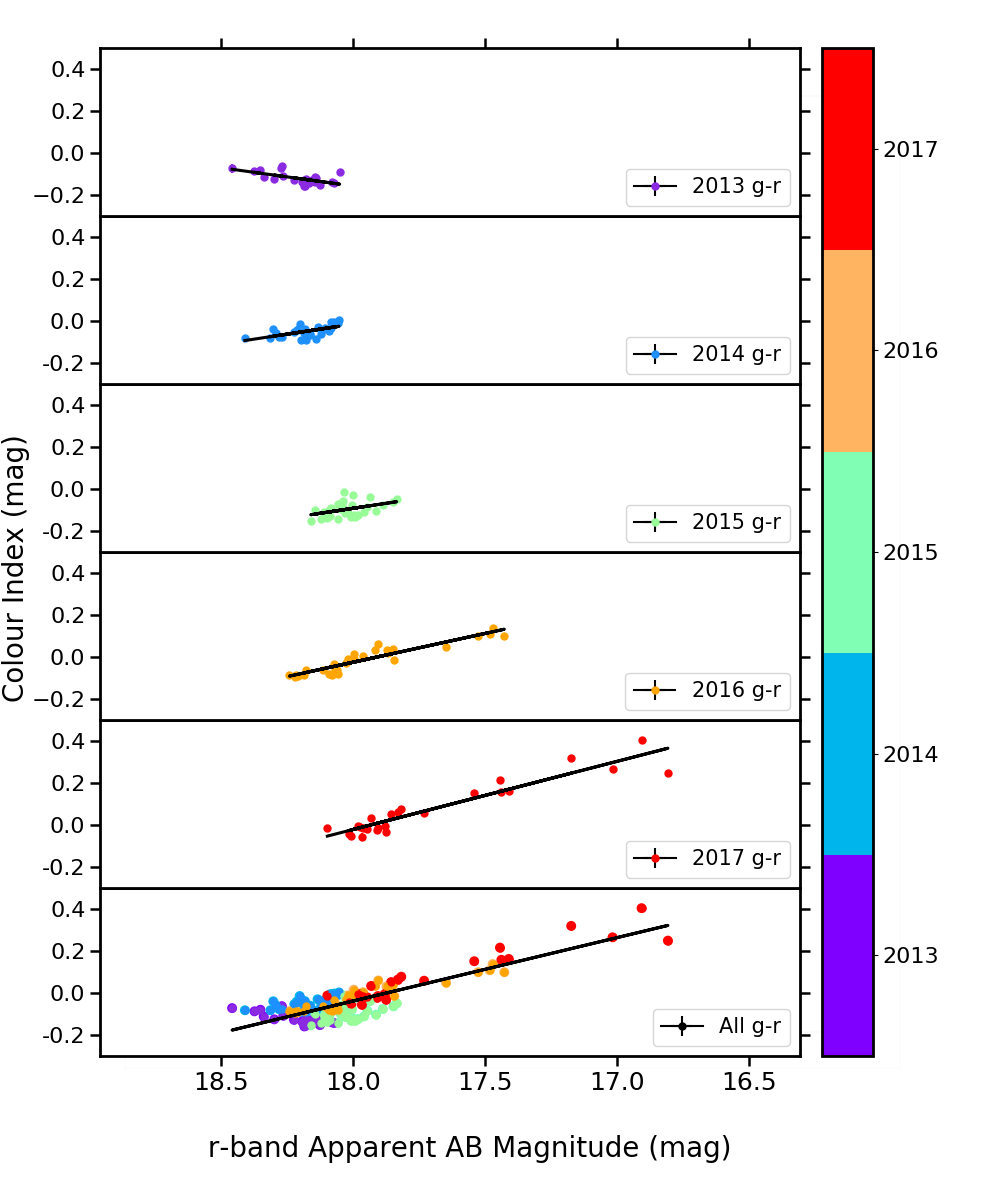}}
    \hfill
    \subfloat[figure][Optical \emph{g}-\emph{i} colour variability for each observation season of DES.\newline \label{fig:gi_sep_years}]{\includegraphics[width=0.49\textwidth]{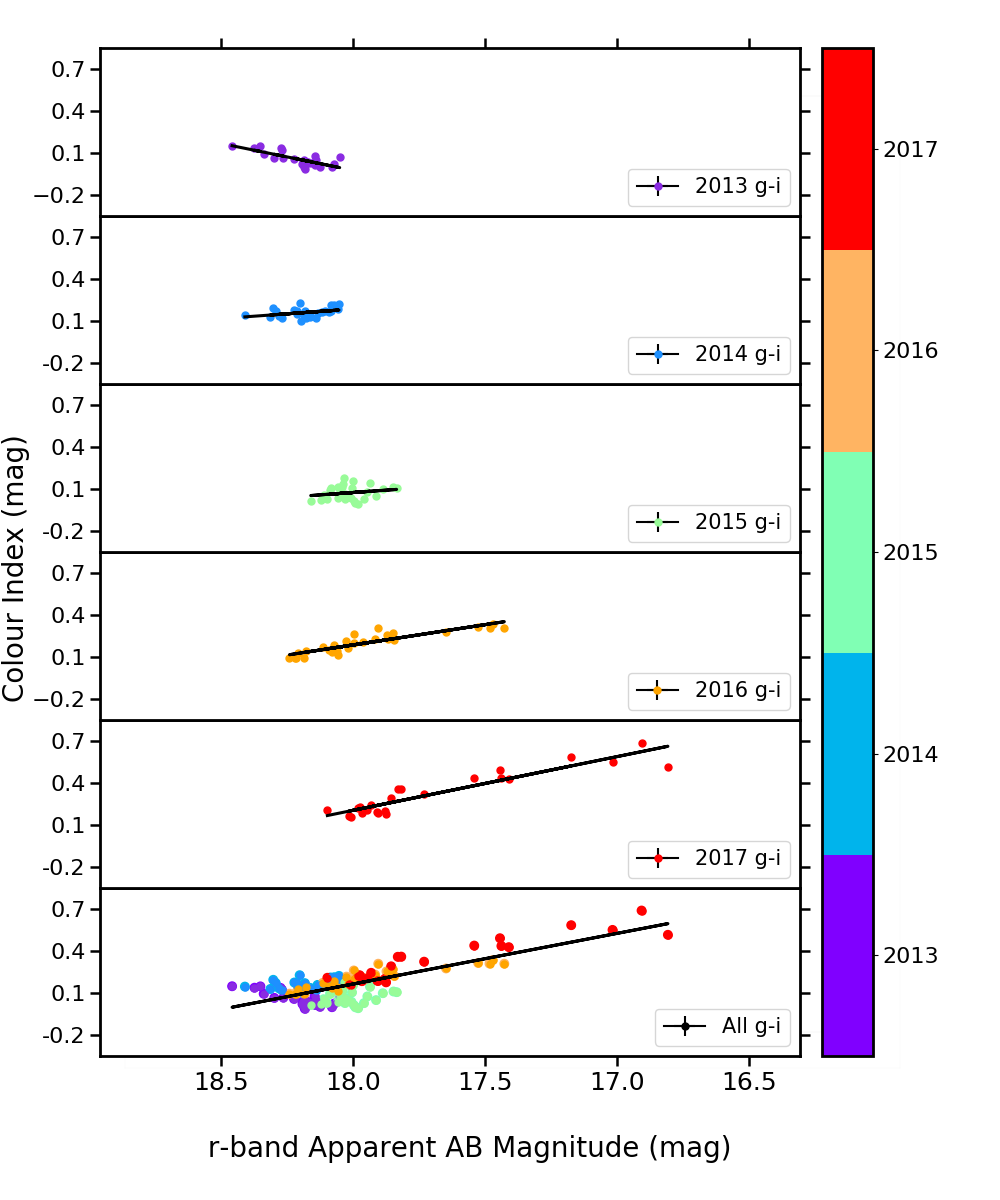}}
    \\
    \subfloat[figure][Optical \emph{r}-\emph{i} colour variability for each observation season of DES. \newline \label{fig:ri_sep_years}]{\includegraphics[width=0.49\textwidth]{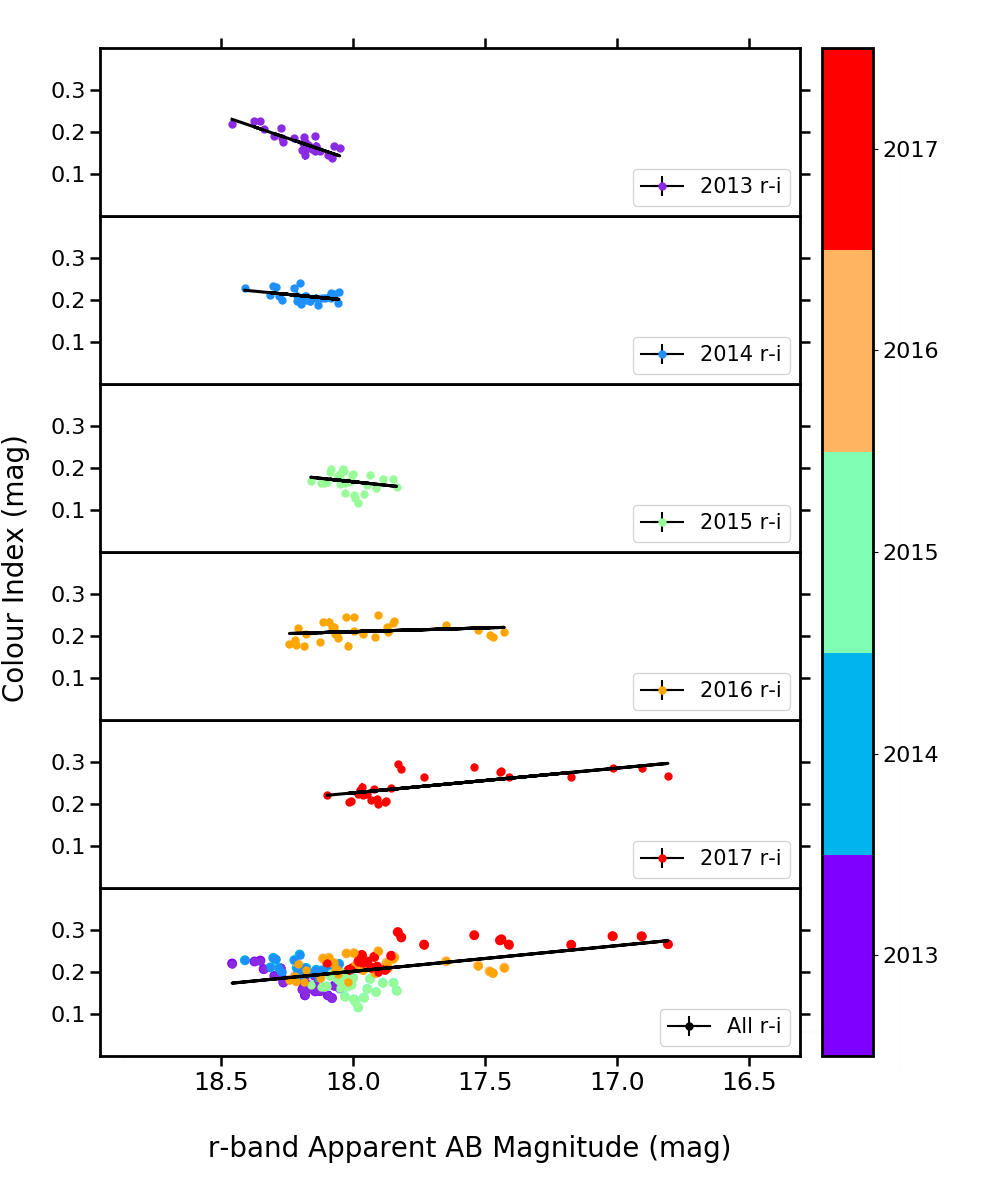}}
    \hfill
    \subfloat[figure][Optical \emph{r}-\emph{z} colour variability for each observation season of DES.\newline \label{fig:rz_sep_years}]{\includegraphics[width=0.49\textwidth]{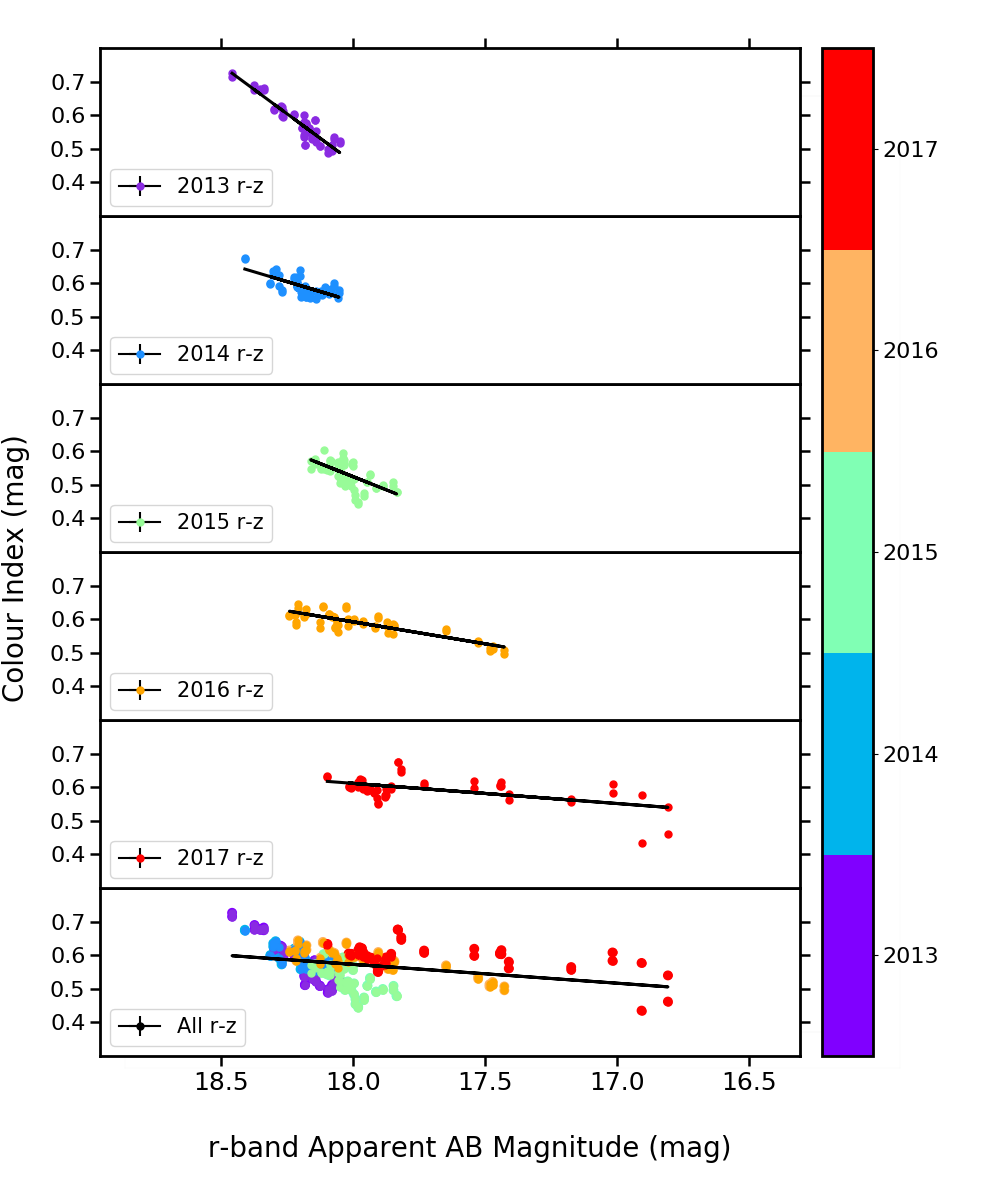}}
    
    \caption{Optical colour variability for each combination of filters in each observation season of DES, where the colour of the data points corresponds to the observation season. \newline \label{fig:all_sep_years}}
\end{figure*}

\begin{figure*}
    \begin{minipage}{\textwidth}
        \ContinuedFloat
        \subfloat[figure][Optical \emph{i}-\emph{z} colour variability for each observation season of DES.\newline \label{fig:iz_sep_years}]{\includegraphics[width=0.49\textwidth]{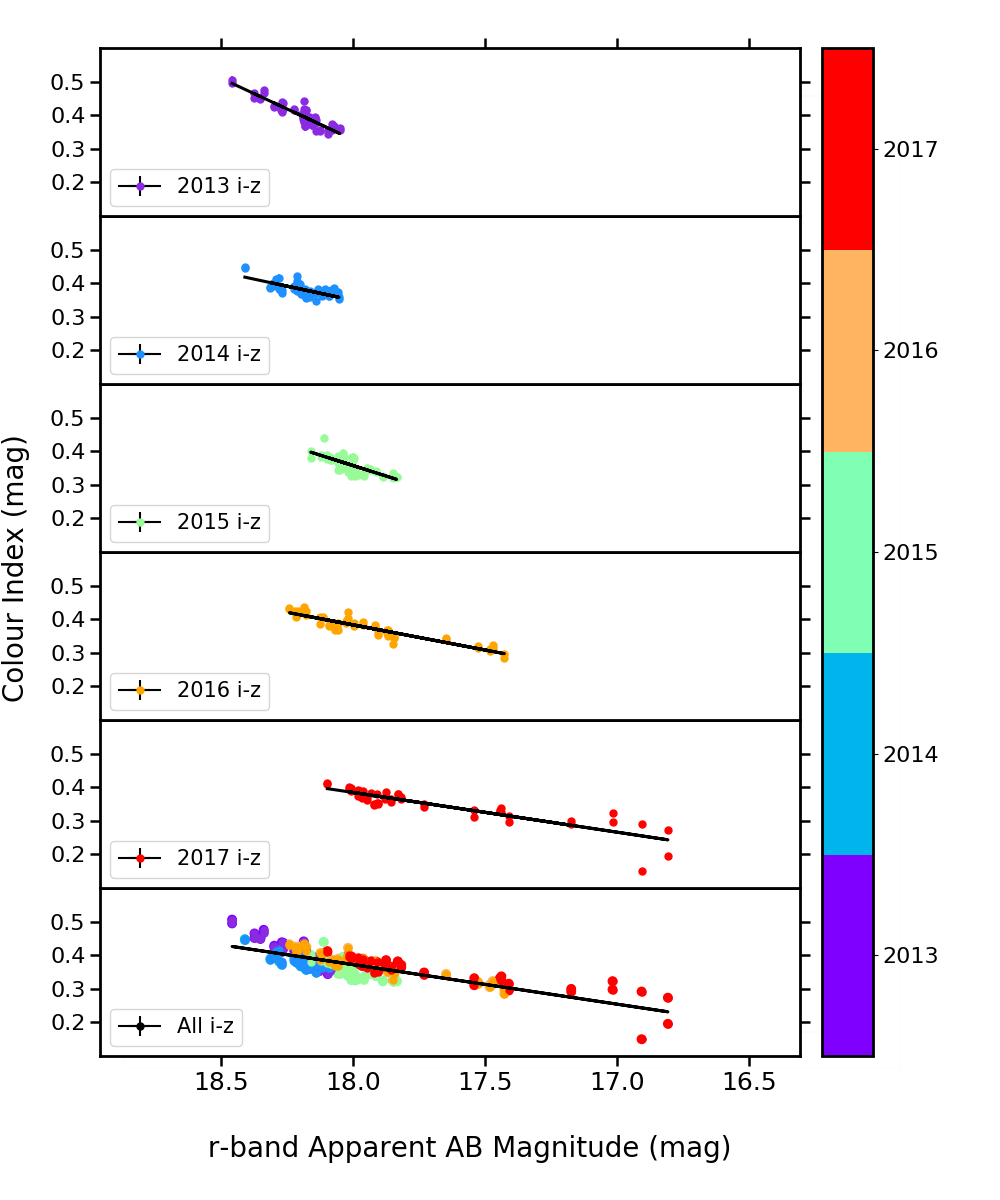}}
        \hspace{8.6cm}
        \caption{Continued.}
    \end{minipage}
    \begin{minipage}{\textwidth}
    \captionof{table}{The slopes, Spearman rank coefficients, probability of no correlation and colour trend for each season of DES in each combination of filters plotted in Figure \ref{fig:all_sep_years}. \label{tab:sep_years}}
    \subfloat[table][From the \emph{g}-\emph{r} colour vs \emph{r} magnitude plots in Figure~\ref{fig:gr_sep_years}. \label{tab:gr_sep_years}]{
    \begin{minipage}{0.49\textwidth}
            \begin{tabular}{C{0.12\textwidth}C{0.2\textwidth}C{0.12\textwidth}C{0.2\textwidth}C{0.12\textwidth}}
            \hline
                Season & Slope of \emph{g-r} vs \emph{r} mag & $\rho$-value & $p$-value & Trend \\
            \hline
               2013 & 0.16 $\pm$ 0.05 & 0.54 & 3.86 $\times 10^{-3}$ & BWB \\
               2014 & -0.19 $\pm$ 0.05 & -0.56 & 2.02 $\times 10^{-3}$ & RWB \\
               2015 & -0.19 $\pm$ 0.07 & -0.4 & 0.03 & SWB \\
               2016 & -0.27 $\pm$ 0.02 & -0.93 & 1.26 $\times 10^{-13}$ & RWB \\
               2017 & -0.32 $\pm$ 0.02 & -0.87 & 1.09 $\times 10^{-8}$ & RWB \\
            \hline
            \end{tabular}
            \vspace{2mm}
    \end{minipage}}
    \hspace{0.8cm}
    \subfloat[table][From the \emph{g}-\emph{i} colour vs \emph{r} magnitude plots in Figure~\ref{fig:gi_sep_years}. \label{tab:gi_sep_years}]{
    \begin{minipage}{0.49\textwidth}
        \begin{tabular}{C{0.12\textwidth}C{0.2\textwidth}C{0.12\textwidth}C{0.2\textwidth}C{0.12\textwidth}}
        \hline
            Season & Slope of \emph{g-i} vs \emph{r} mag & $\rho$-value & $p$-value & Trend \\
        \hline
           2013 & 0.38 $\pm$ 0.07 & 0.63 & 4.73 $\times 10^{-4}$ & BWB \\
           2014 & -0.14 $\pm$ 0.07 & -0.33 & 0.09 & SWB \\
           2015 & -0.14 $\pm$ 0.12 & -0.11 & 0.57 & SWB \\
           2016 & -0.29 $\pm$ 0.03 & -0.92 & 1.30 $\times 10^{-12}$ & RWB \\
           2017 & -0.38 $\pm$ 0.03 & -0.84 & 1.67 $\times 10^{-7}$ & RWB \\
        \hline
        \end{tabular}
        \vspace{2mm}
	\end{minipage}}
	\vfill
	\subfloat[table][From the \emph{r}-\emph{i} colour vs \emph{r} magnitude plots in Figure~\ref{fig:ri_sep_years}. \label{tab:ri_sep_years}]{
	\begin{minipage}{0.49\textwidth}
        \begin{tabular}{C{0.12\textwidth}C{0.2\textwidth}C{0.12\textwidth}C{0.2\textwidth}C{0.12\textwidth}}
        \hline
            Season & Slope of \emph{r-i} vs \emph{r} mag & $\rho$-value & $p$-value & Trend \\
        \hline
           2013 & 0.21 $\pm$ 0.03 & 0.7 & 3.11 $\times 10^{-5}$ & BWB\\
           2014 & 0.06 $\pm$ 0.03 & 0.34 & 0.09 & SWB \\
           2015 & 0.07 $\pm$ 0.05 & 0.35 & 0.06 & SWB \\
           2016 & 0.01 $\pm$ 0.02 & -0.33 & 0.08 & SWB \\
           2017 & -0.06 $\pm$ 0.01 & -0.71 & 5.45 $\times 10^{-5}$ & RWB \\
        \hline
        \end{tabular}
        \vspace{2mm}
	\end{minipage}}
	\hspace{0.8cm}
	\subfloat[table][rom the \emph{r}-\emph{z} colour vs \emph{r} magnitude plots in Figure~\ref{fig:rz_sep_years}. \label{tab:rz_sep_years}]{
	\begin{minipage}{0.49\textwidth}
        \begin{tabular}{C{0.12\textwidth}C{0.2\textwidth}C{0.12\textwidth}C{0.2\textwidth}C{0.12\textwidth}}
        \hline
            Season & Slope of \emph{r-z} vs \emph{r} mag & $\rho$-value & $p$-value & Trend \\
        \hline
           2013 & 0.58 $\pm$ 0.03 & 0.88 & 1.56 $\times 10^{-19}$ & BWB \\
           2014 & 0.23 $\pm$ 0.03 & 0.58 & 3.81 $\times 10^{-6}$ & BWB \\
           2015 & 0.31 $\pm$ 0.05 & 0.67 & 1.03 $\times 10^{-9}$ & BWB \\
           2016 & 0.13 $\pm$ 0.01 & 0.75 & 1.02 $\times 10^{-11}$ & BWB \\
           2017 & 0.06 $\pm$ 0.01 & 0.36 & 9.58 $\times 10^{-3}$ & BWB \\
        \hline
        \end{tabular}
        \vspace{2mm}
	\end{minipage}}
	\vfill
	\subfloat[table][From \emph{i}-\emph{z} colour vs \emph{r} magnitude plots in Figure~\ref{fig:iz_sep_years}. \label{tab:iz_sep_years}]{
	\begin{minipage}{0.49\textwidth}
        \begin{tabular}{C{0.12\textwidth}C{0.2\textwidth}C{0.12\textwidth}C{0.2\textwidth}C{0.12\textwidth}}
        \hline
            Season & Slope of \emph{i-z} vs \emph{r} mag & $\rho$-value & $p$-value & Trend \\
        \hline
           2013 & 0.37 $\pm$ 0.02 & 0.9 & 1.03 $\times 10^{-20}$ & BWB \\
           2014 & 0.17 $\pm$ 0.02 & 0.65 & 1.73 $\times 10^{-7}$ & BWB \\
           2015 & 0.25 $\pm$ 0.03 & 0.77 & 3.68 $\times 10^{-13}$ & BWB \\
           2016 & 0.15 $\pm$ 0.01 & 0.9 & 3.07 $\times 10^{-22}$ & BWB \\
           2017 & 0.12 $\pm$ 0.01 & 0.89 & 9.82 $\times 10^{-19}$ & BWB \\
        \hline
        \end{tabular}
        \vspace{2mm}
	\end{minipage}}
	\hspace{8.7cm}
\end{minipage}
\end{figure*}

\begin{figure*}
    \begin{minipage}{\textwidth}
        \captionof{table}{The change in slope after the outliers are excluded. \label{tab:outliers}}
        \subfloat[table][The 2013 season after the outliers on MJD 56590 are excluded. \label{tab:y1_outlier}]{
        \begin{minipage}{0.46\textwidth}
                \begin{tabular}{C{0.12\textwidth}C{0.37\textwidth}C{0.37\textwidth}}
                \hline
                Colour & Slope with All Epochs & Slope Without Outlier \\
                \hline
                    \emph{g-r} & 0.15 $\pm$ 0.05 & 0.16 $\pm$ 0.05 \\
                    \emph{g-i} & 0.39 $\pm$ 0.07 & 0.38 $\pm$ 0.07 \\
                    \emph{g-z} & 0.85 $\pm$ 0.07 & 0.74 $\pm$ 0.05 \\
                    \emph{r-i} & 0.22 $\pm$ 0.03 & 0.22 $\pm$ 0.03 \\
                    \emph{r-z} & 0.69 $\pm$ 0.06 & 0.58 $\pm$ 0.03\\
                    \emph{i-z} & 0.45 $\pm$ 0.05 & 0.36 $\pm$ 0.02\\
                \hline
                \end{tabular}
                \vspace{2mm}
        \end{minipage}}
        \hspace{0.8cm}
        \subfloat[table][The 2014 season after the outliers on MJD 56916 are excluded. \label{tab:y2_outlier}]{
        \begin{minipage}{0.46\textwidth}
                \begin{tabular}{C{0.12\textwidth}C{0.37\textwidth}C{0.37\textwidth}}
                \hline
                Colour & Slope with All Epochs & Slope Without Outlier \\
                \hline
                    \emph{g-r} & -0.2 $\pm$ 0.05 & -0.19 $\pm$ 0.05 \\
                    \emph{g-i} & -0.2 $\pm$ 0.09 & -0.14 $\pm$ 0.08 \\
                    \emph{g-z} & 0.03 $\pm$ 0.06 & 0.06 $\pm$ 0.06 \\
                    \emph{r-i} & 0.01 $\pm$ 0.06 & 0.06 $\pm$ 0.03 \\
                    \emph{r-z} & 0.22 $\pm$ 0.03 & 0.23 $\pm$ 0.03 \\
                    \emph{i-z} & 0.21 $\pm$ 0.03 & 0.17 $\pm$ 0.02 \\
                \hline
                \end{tabular}
                \vspace{2mm}
    	\end{minipage}}
        \vfill
        \subfloat[table][The 2015 season after the outliers on MJD 57281 are excluded. \label{tab:y3_outlier}]{
        \begin{minipage}{0.46\textwidth}
                \begin{tabular}{C{0.12\textwidth}C{0.37\textwidth}C{0.37\textwidth}}
                \hline
                Colour & Slope with All Epochs & Slope Without Outliers \\
                \hline
                    \emph{g-r} & -0.31 $\pm$ 0.09 & -0.19 $\pm$ 0.08 \\
                    \emph{g-i} & -0.48 $\pm$ 0.18 & -0.14 $\pm$ 0.13 \\
                    \emph{g-z} & -0.49 $\pm$ 0.20 & 0.12 $\pm$ 0.09 \\
                    \emph{r-i} & -0.14 $\pm$ 0.10 & 0.07 $\pm$ 0.05 \\
                    \emph{r-z} & -0.18 $\pm$ 0.16 & 0.31 $\pm$ 0.05 \\
                    \emph{i-z} & -0.04 $\pm$ 0.11 & 0.24 $\pm$ 0.03 \\
                \hline
                \end{tabular}
        \end{minipage}}
        \hspace{0.8cm}
        \subfloat[table][The 2016 season after the outliers on MJD 57627 are excluded. \label{tab:y4_outlier}]{
        \begin{minipage}{0.46\textwidth}
                \begin{tabular}{C{0.12\textwidth}C{0.37\textwidth}C{0.37\textwidth}}
                \hline
                Colour & Slope with All Epochs & Slope Without Outliers \\
                \hline
                    \emph{g-r} & -0.27 $\pm$ 0.02 & -0.27 $\pm$ 0.02 \\
                    \emph{g-i} & -0.29 $\pm$ 0.03 & -0.29 $\pm$ 0.03 \\
                    \emph{g-z} & -0.14 $\pm$ 0.02 & -0.14 $\pm$ 0.02 \\
                    \emph{r-i} & 0.31 $\pm$ 0.15 & -0.02 $\pm$ 0.02 \\
                    \emph{r-z} & 0.42 $\pm$ 0.09 & 0.13 $\pm$ 0.01 \\
                    \emph{i-z} & 0.11 $\pm$ 0.01 & 0.15 $\pm$ 0.01 \\
                \hline
                \end{tabular}
    	\end{minipage}}
    \end{minipage}
    \begin{minipage}{\textwidth}
            \subfloat[figure][Outlier of the optical colour variability in the 2013 season. \label{fig:yr1_outlier}]{\includegraphics[width=0.49\textwidth]{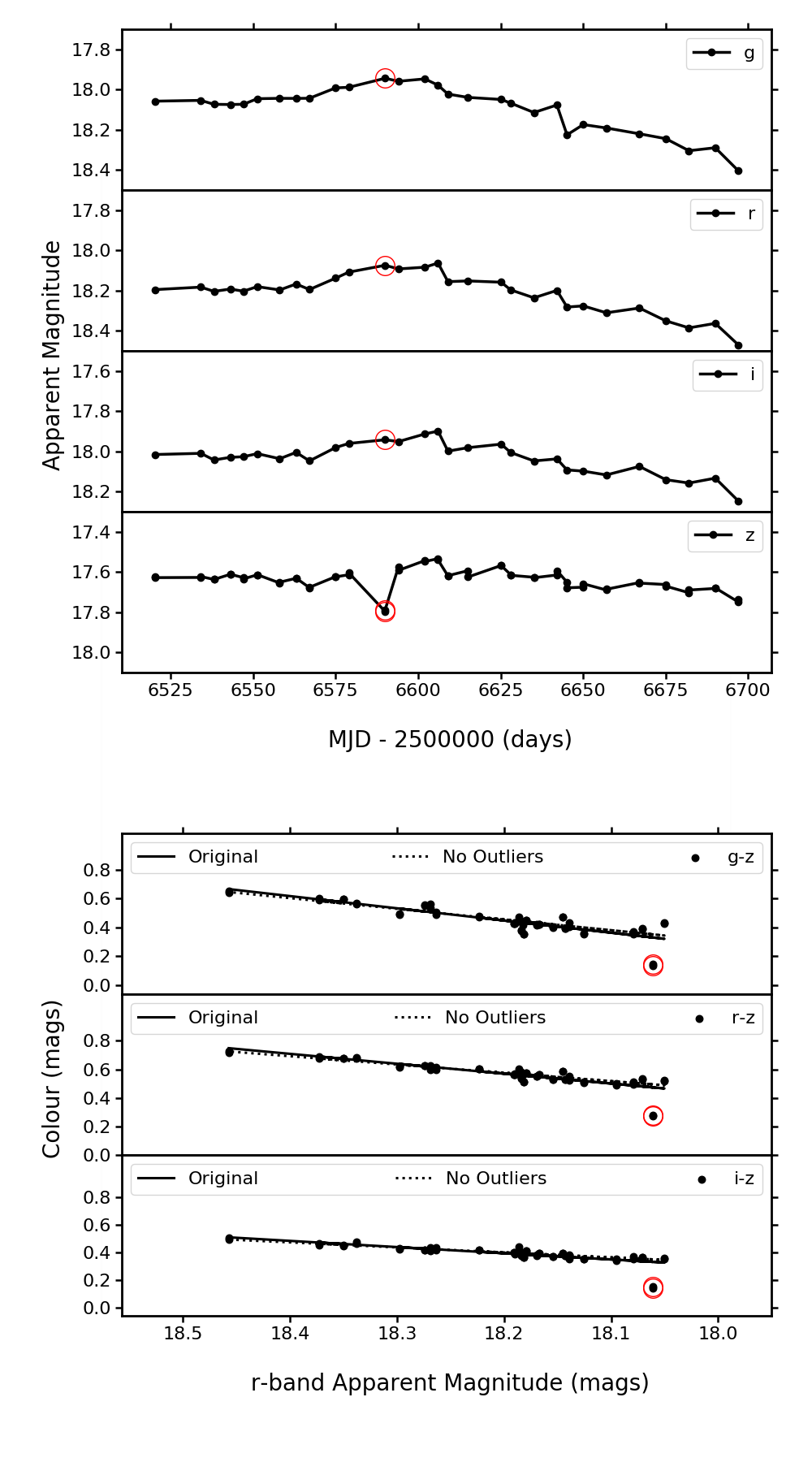}}
            \hfill
            \subfloat[figure][Outlier of the optical colour variability in the 2014 season. \label{fig:yr2_outlier}]{\includegraphics[width=0.49\textwidth]{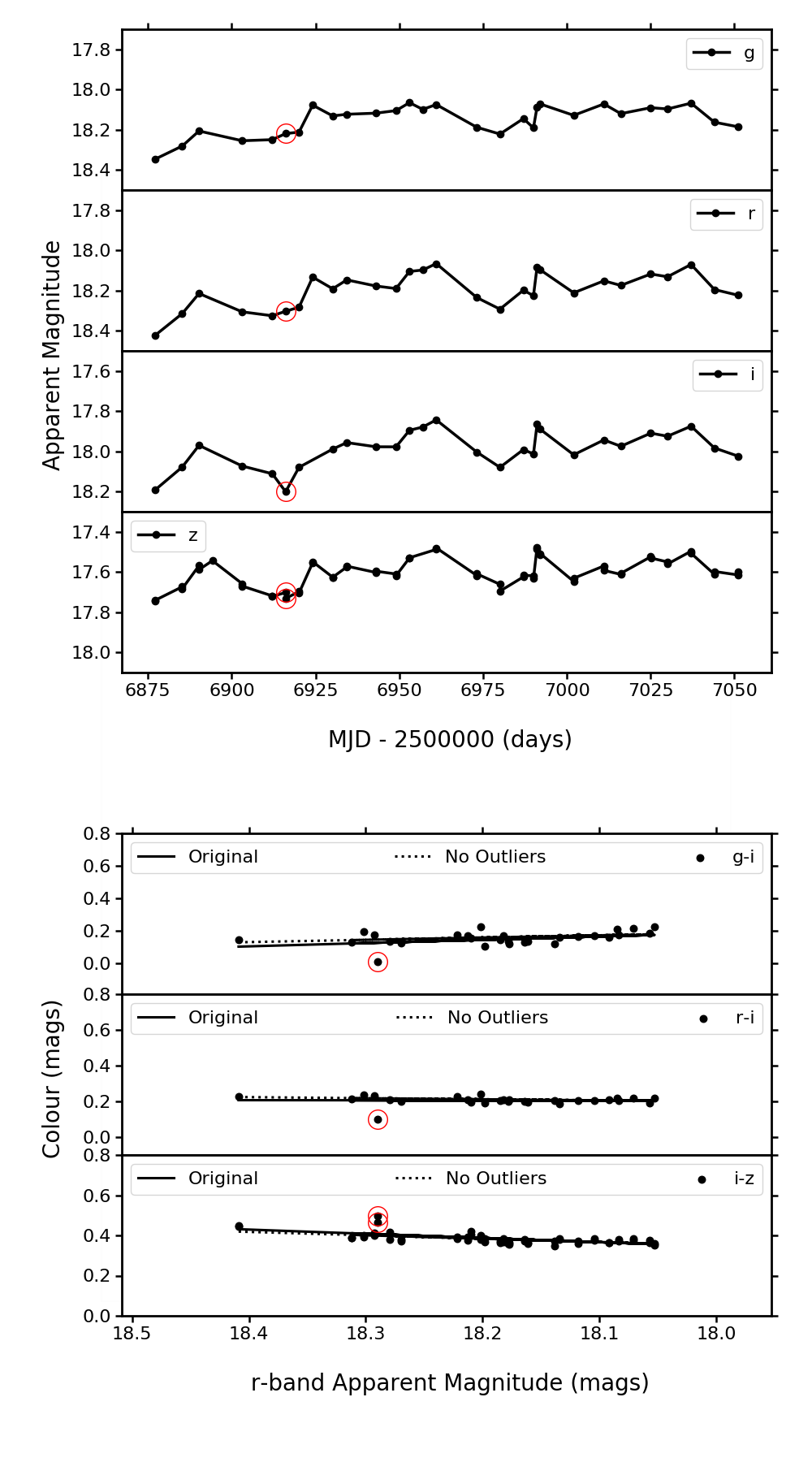}}
            \captionof{figure}{Outliers in the DES light curves of PKS~0027-426 and the optical colour variability. \newline \label{fig:outliers}}
    \end{minipage}
\end{figure*}

\begin{figure*}
    \ContinuedFloat
    \subfloat[figure][Outlier of the optical colour variability in the 2015 season. \label{fig:yr3_outlier}]{\includegraphics[width=0.49\textwidth]{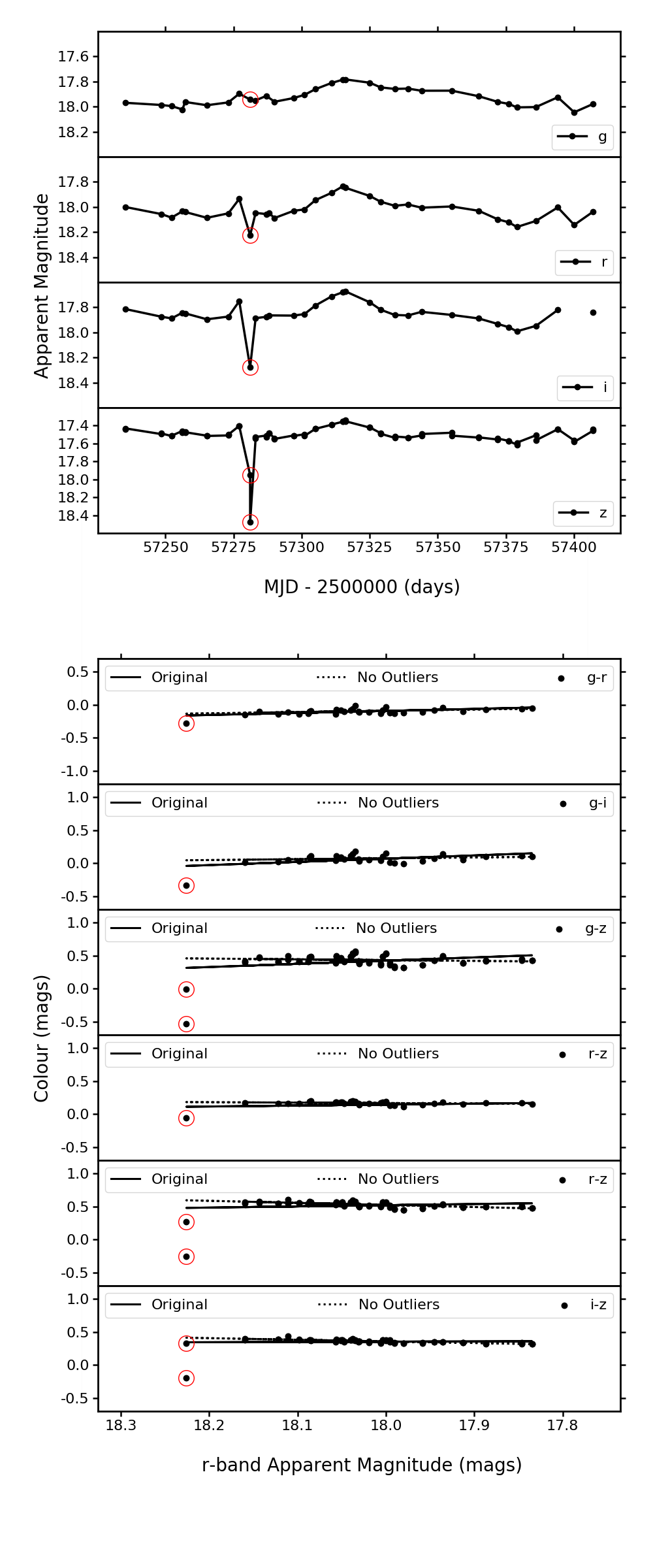}}
    \subfloat[figure][Outlier of the optical colour variability in the 2016 season. \label{fig:yr4_outlier}]{\includegraphics[width=0.49\textwidth]{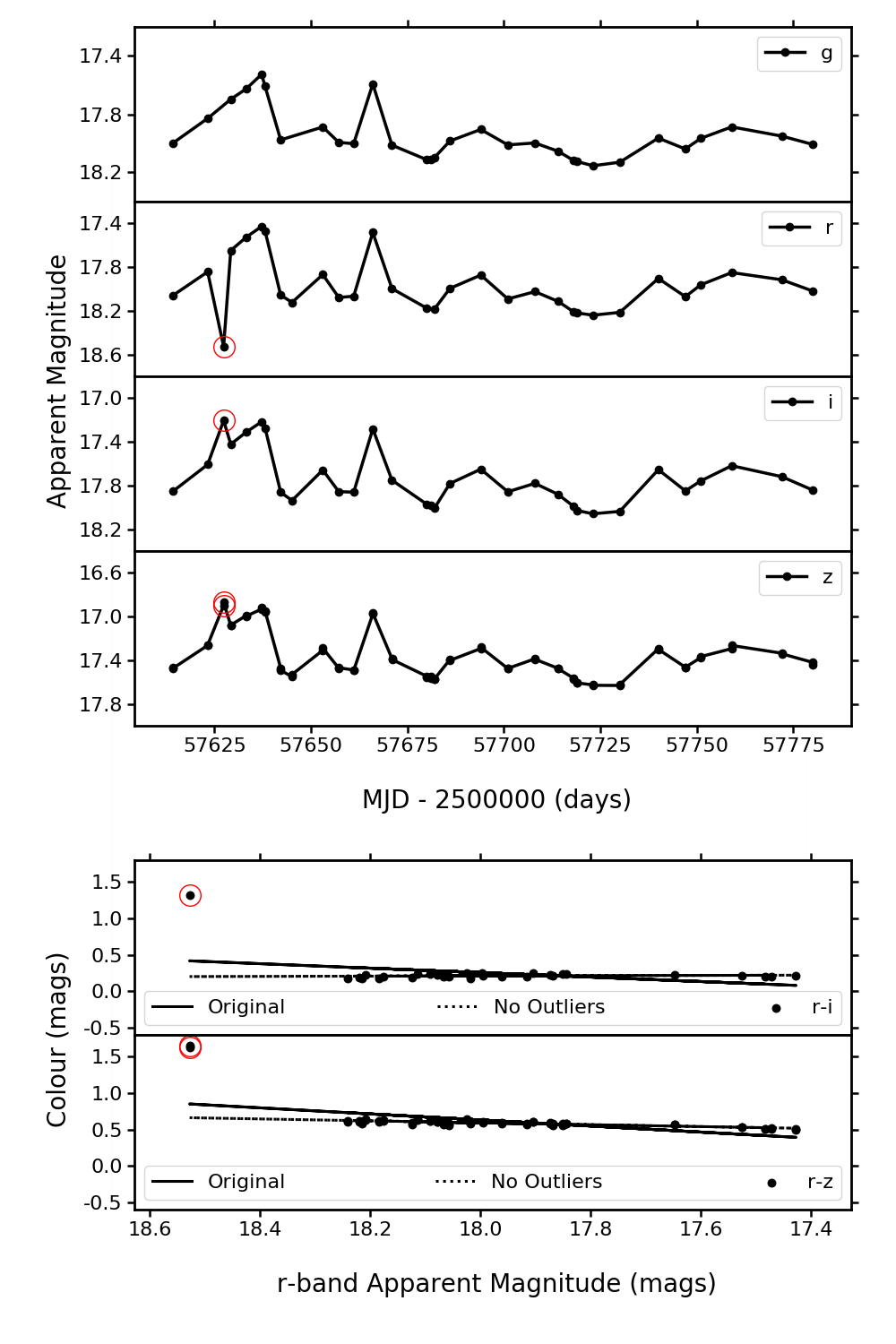}}
    \caption{Continued.}
\end{figure*}

\begin{figure}
    \centering
    \includegraphics[width=\columnwidth]{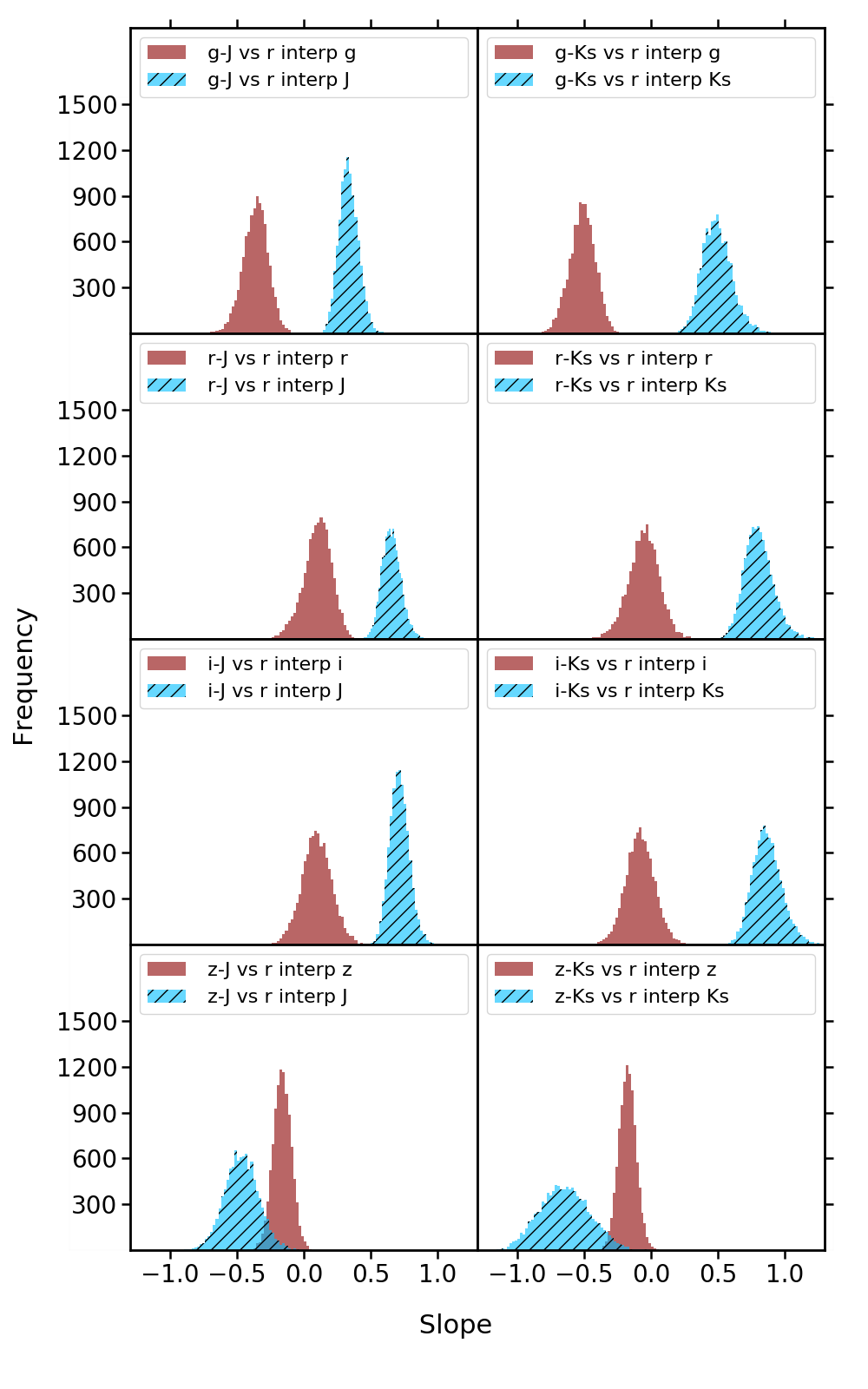}
    \caption{The slopes measured for each combination of the Optical-NIR colour vs r magnitude plots in the 2017 season, for the entire light curve, which includes a $\sim$ month long gap in the NIR light curve between MJD 57993 and 58044.}
    \label{fig:optnir_entirelc_colmag}
\end{figure}

\begin{figure}
    \centering
    \includegraphics[width=\columnwidth]{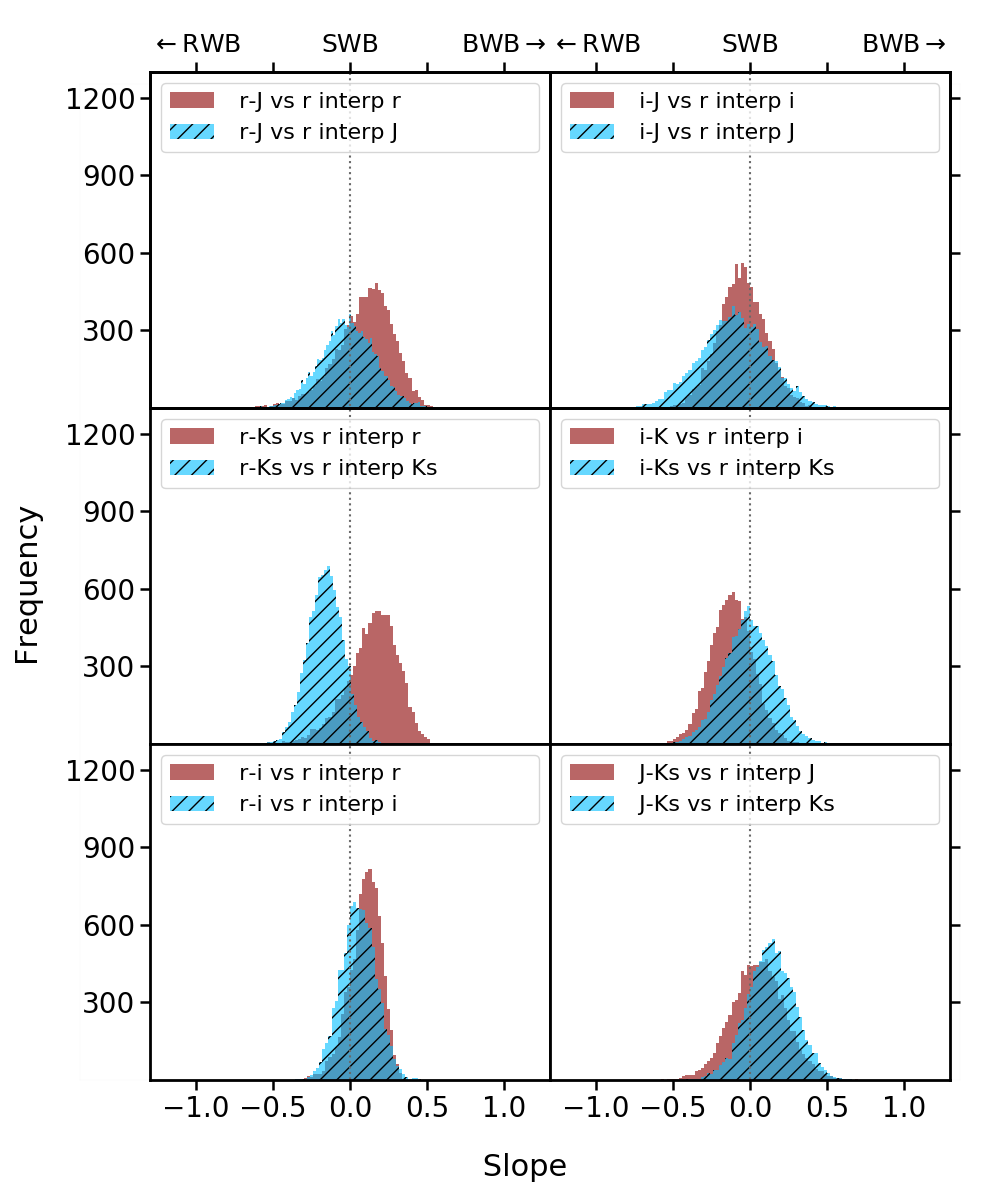}
    \caption{\centering The distributions of the slope of the 2018-19 optical and NIR colours vs \emph{r} magnitude plots returned form 10,000 interpolations of each light curve.}
    \label{fig:nircol201819}
\end{figure}  

\begin{figure}
    \centering
    \includegraphics[width=\columnwidth]{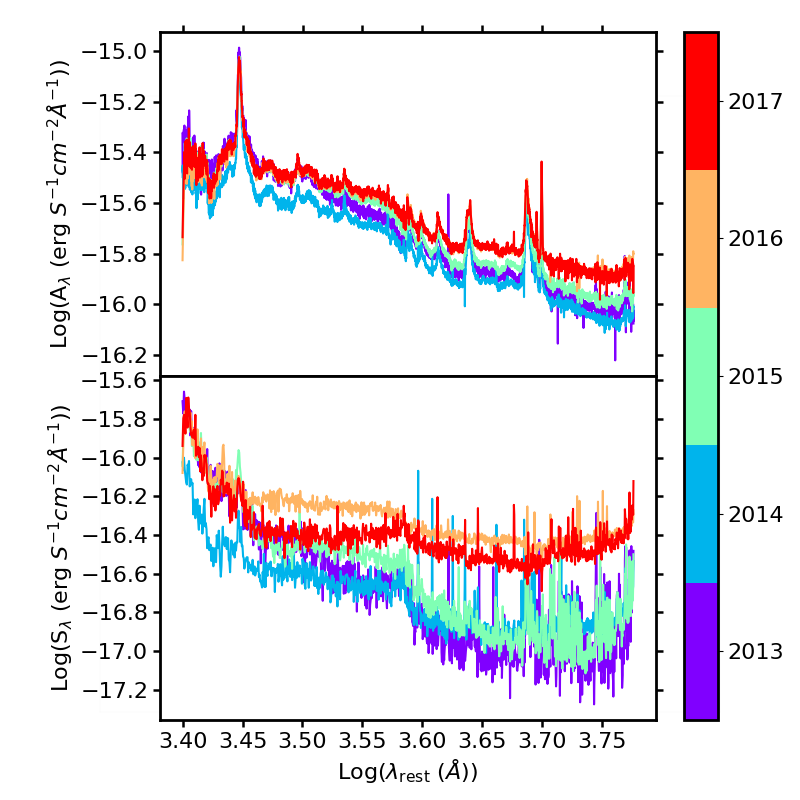}
    \caption{\small Decomposition of the OzDES spectra into the variable (S$_\lambda$) and non-variable (A$_\lambda$) components for each season of DES. The S$_\lambda$ spectra are smoothed in this plot so that they are easier to see. \label{fig:ozdes_comp_decomps} ~ \\}
    \vspace{2mm}   
\end{figure}

\subsection{Optical-NIR Colour - Magnitude Plots from 2017-18, with Gap in Observations}
\label{sec:app_201718_interps}

Figure~\ref{fig:optnir_entirelc_colmag} display histograms of the measured slopes of the 2017-2018 optical-NIR colour vs r magnitude plots, from each interpolated light curve, with the $\sim$ month long break in observations between MJD 57993 and 58044. The measured slope from each light curve are no longer consistent with each other in this scenario, which is assumed to be due the the peak that occurs in the optical light curve within this period. When the optical light curve is interpolated, the results are similar to Figure~\ref{fig:nircolvopt}, however, when the NIR light curve is interpolated, the results are shown to be more positive as the interpolations in the NIR light curve in this gap will not be as drastically variable as the optical light curve within this period. This means that the overall slope will be shifted to be positive as the optical light curve will be a lot more variable in this period. 

\subsection{Optical-NIR Colour Magnitude Plots from 2018-19}
\label{ap:2018colmag}

In the 2018 season, only the optical \emph{r} and \emph{i} bands could be used in the colour-magnitude plots as there were very few epochs observed in the \emph{g} and \emph{z} bands with VOILETTE during this time. Figure~\ref{fig:nircol201819} displays the histograms of the slope from each colour combination of optical, \emph{r} and \emph{i}, and NIR, \emph{J} and \emph{Ks} bands, plotted against the \emph{r} band. It was found that 75\% of the colour vs \emph{r} band slopes for each interpolated filter are consistent with each other within 1$\sigma$ and all are consistent within 1.1$\sigma$.  The slopes for the 2018 season show inconclusive colour behaviours.

\subsection{Decomposition of the OzDES Spectra into Variable and Non-Variable Components}
\label{ap:OzDES_decomp}

In Figure \ref{fig:ozdes_comp_decomps}, the OzDES spectra are decomposed into their variable and non-variable components using Equation \ref{eq:decomp}. The behaviour of the OzDES components is similar to the behaviour seen in Figure \ref{fig:comp_decomps} for the broadband DES spectra, although it is worth noting that the 2017 OzDES spectra were not observed during the brightest epochs of the DES observations. The shape of the A$_\lambda$ component remains similar over all seasons, and is steeper towards the bluer wavelengths, flattening towards the redder wavelengths. The S$_\lambda$ component also behaves similarly to Figure \ref{fig:comp_decomps} however the lack of spectra during the brightest, most variable flare of the DES observations is reflected in the 2017 season. There is a noticeable bump in the log(S$_\lambda$) vs log($\lambda_\text{rest}$) plots especially at log($\lambda_\text{rest}$) $\sim$ 3.58 which corresponds to the split between the red and blue arms of the spectrograph as explained in \cite{Hoormann2019}.

\subsection{Broadband Spectra from DES for each Observation Season}
\label{ap:all_bb_spec}

\begin{figure*}
    \begin{minipage}{\textwidth}
        \centering
        \subfloat[figure][\small Broadband Spectra of the 2013 Season. \label{fig:S1_spec}]{\includegraphics[width=0.49\textwidth]{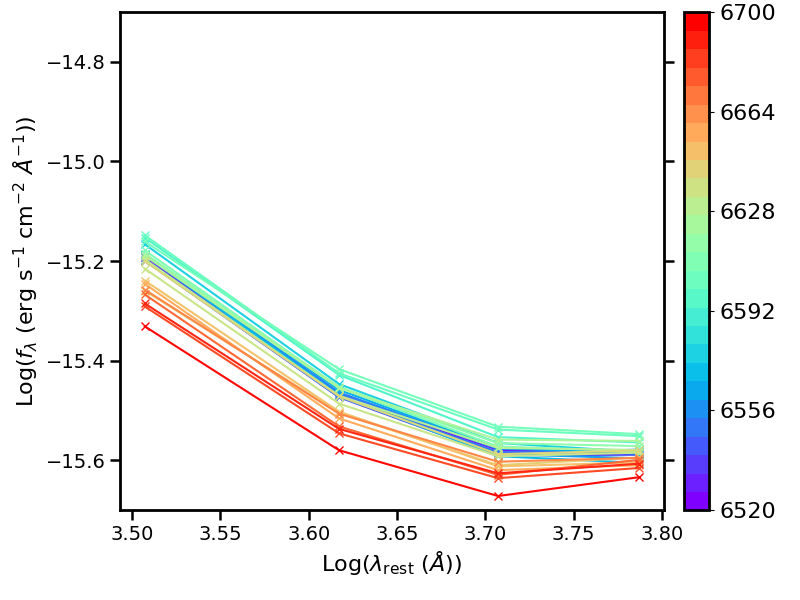}}
        \hfill
        \subfloat[figure][Broadband Spectra of the 2014 Season. \label{fig:S2_spec}]{\includegraphics[width=0.49\textwidth]{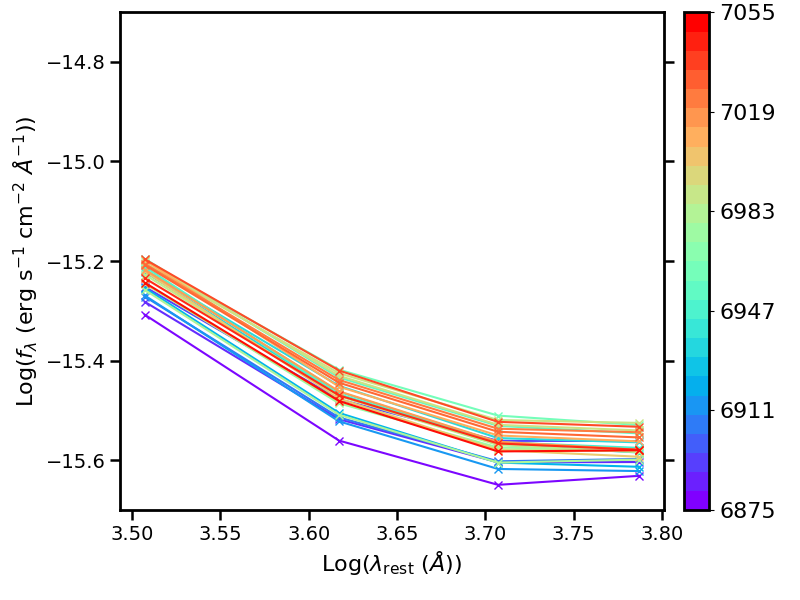}}
        \\
        \subfloat[figure][Broadband Spectra of the 2015 Season. \label{fig:S3_spec}]{\includegraphics[width=0.49\textwidth]{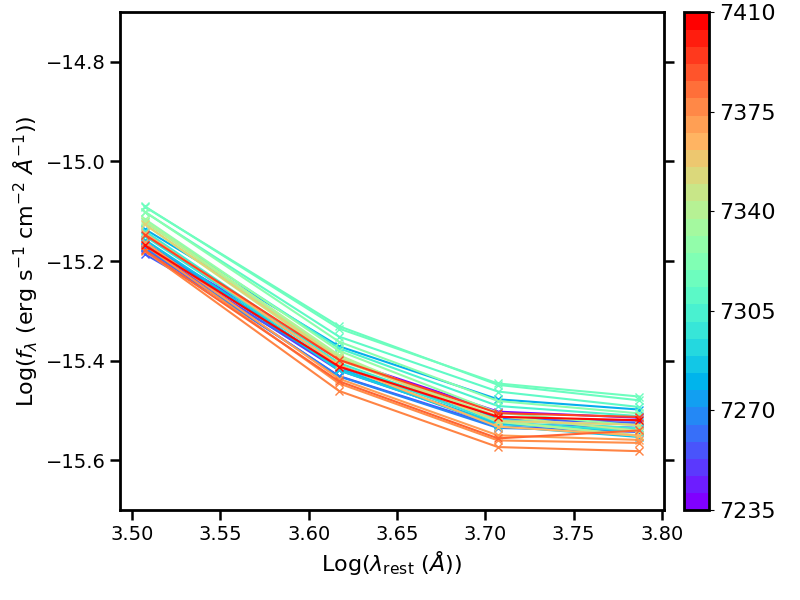}}
        \hfill
        \subfloat[figure][Broadband Spectra of the 2016 Season. \label{fig:S4_spec}]{\includegraphics[width=0.49\textwidth]{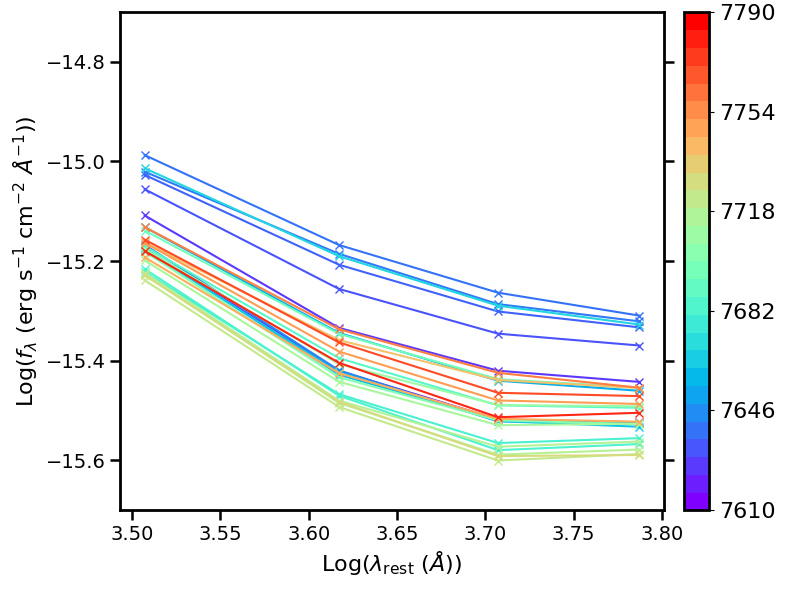}}
        \\
        \subfloat[figure][Broadband Spectra of the 2017 Season. \label{fig:S5_spec}]{\includegraphics[width=0.49\textwidth]{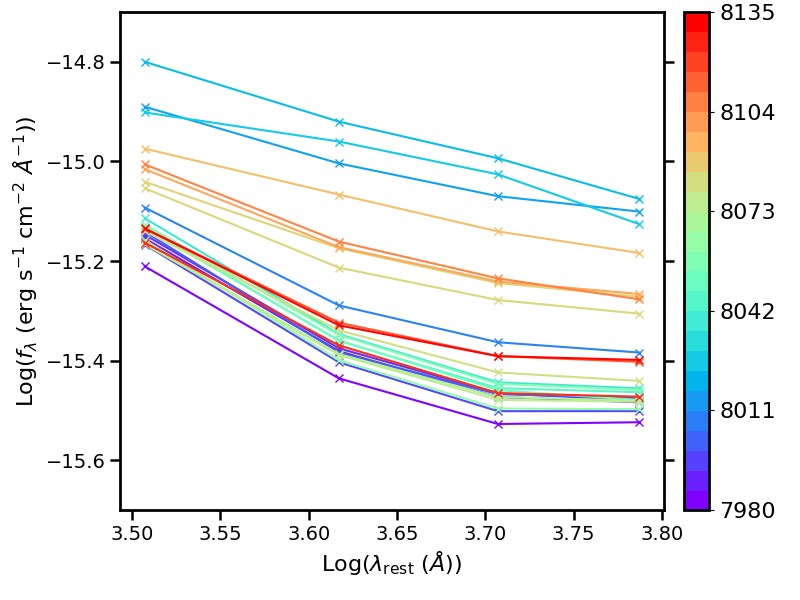}}
        \hspace{8.79cm}
        \caption{All broadband spectra in each observation season of DES, coloured according to observation epoch. \label{fig:season_specs}}
    \end{minipage}
\end{figure*}

In Section~\ref{sec:model_spec}, only the mean brightest and dimmest broadband spectra are shown. Figure \ref{fig:season_specs} displays the broadband spectra for each DES epoch in each observation season, coloured according to the observation epoch, to demonstrate how it changes between the brightest and dimmest states.

\subsection{Additional Models of the Broadband Spectra of DES}
\label{ap:more_models}

\begin{figure*}
    \centering
    \subfloat[figure][ Modelled Spectra of the 2015 Season. \label{fig:S3_fit} ~ \\]{\includegraphics[width=0.49\textwidth]{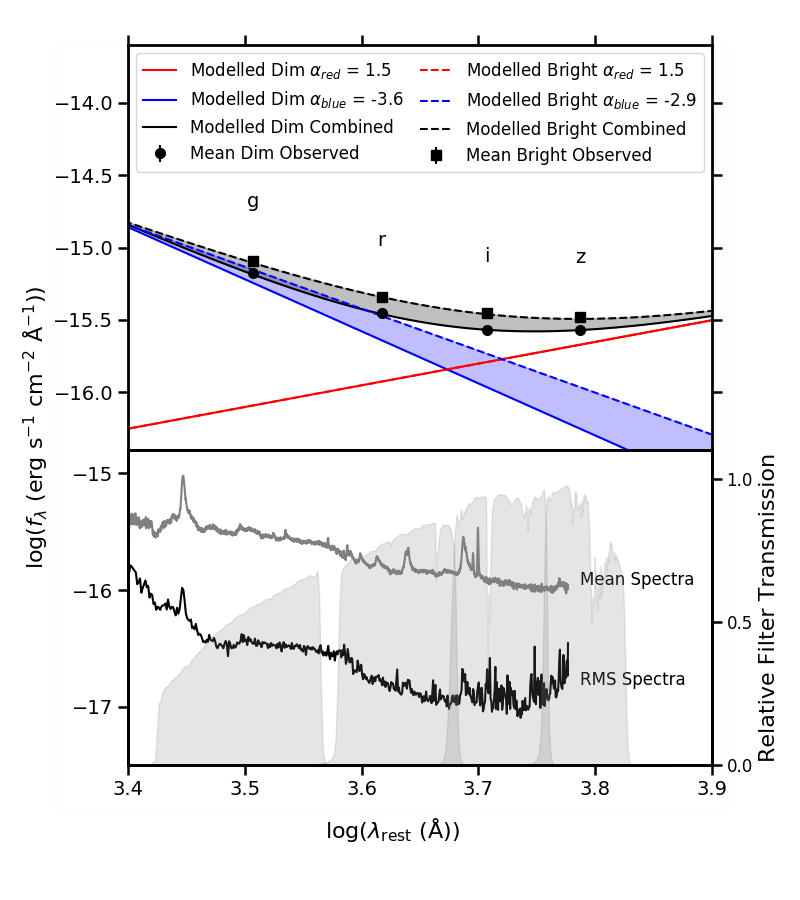}}
    \hfill
    \subfloat[figure][ Modelled Spectra of the Entire DES Observation Period. \label{fig:All_fit} ~ \\]{\includegraphics[width=0.49\textwidth]{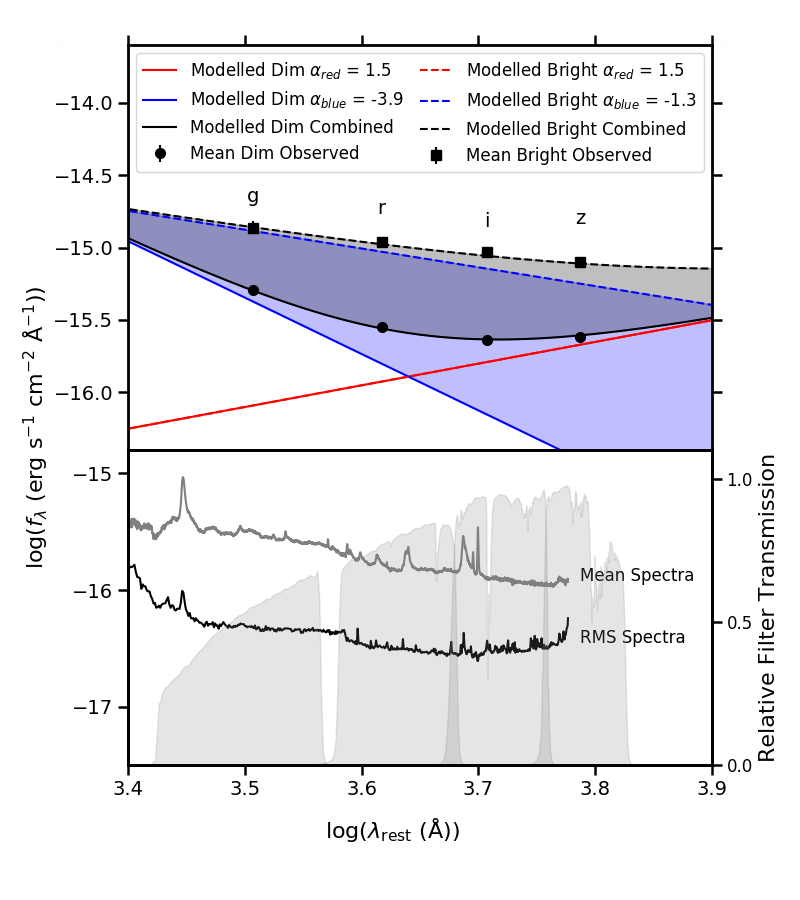}}
    \caption{\textit{Upper panel}: Modelled broadband spectra of the red and blue emission that combine to match the average brightest and dimmest epochs in the 2015 observation season and the entire observational period compared to the observed broadband spectra. The solid lines correspond to the modelled spectra of the dimmest epochs, and the dashed lines correspond to the modelled spectra of the brightest epochs. 
    \textit{Lower panel}: Mean and smoothed RMS OzDES spectra for each season plotted over the DES filter Transmissions.
    \label{fig:Modelled_AD_Sync_More}}
\end{figure*}

Figure~\ref{fig:Modelled_AD_Sync_More} displays the modelled broadband spectra for the 2015 season and for the entire DES observational period. 

\begin{figure*}
    \begin{minipage}{\textwidth}
        \centering
        \subfloat[figure][\label{fig:S2_fit_alt1}~\\]{\includegraphics[width=0.49\textwidth]{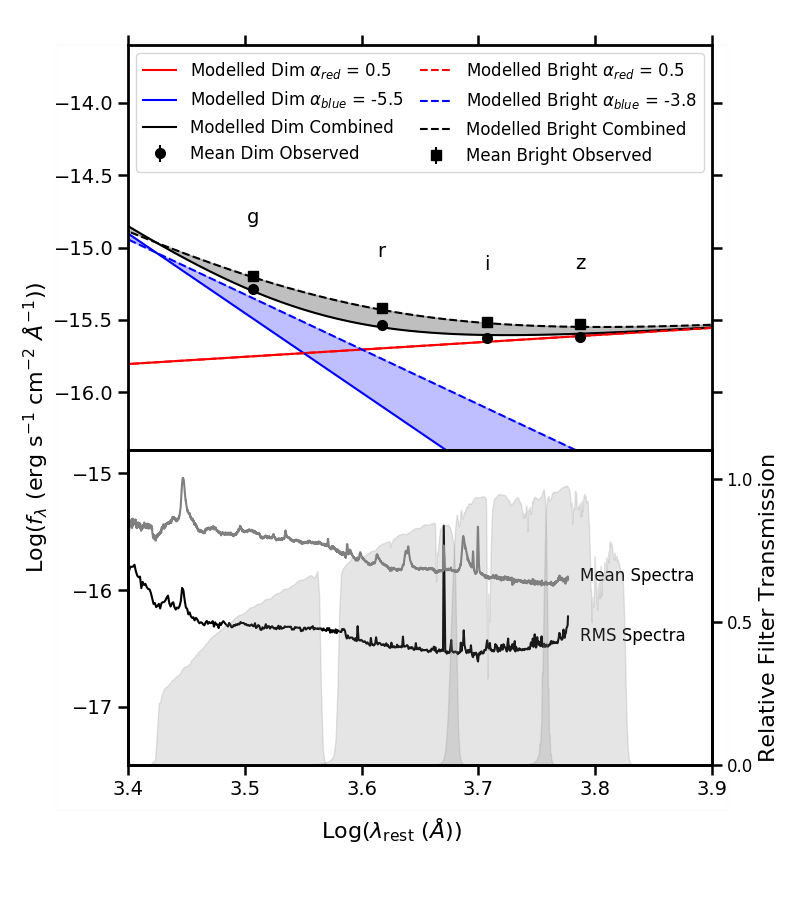}}
        \hfill
        \subfloat[figure][ \label{fig:S2_fit_alt2} ~ \\]{\includegraphics[width=0.49\textwidth]{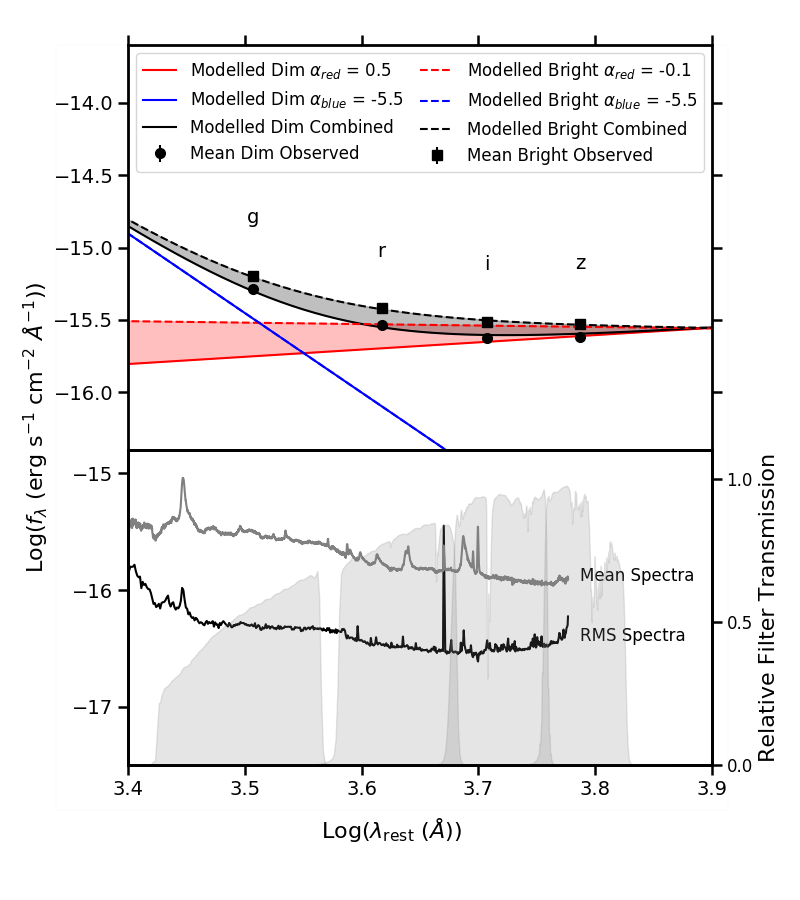}}
        \\
        \subfloat[figure][\label{fig:S3_fit_alt3} ~ \\]{\includegraphics[width=0.49\textwidth]{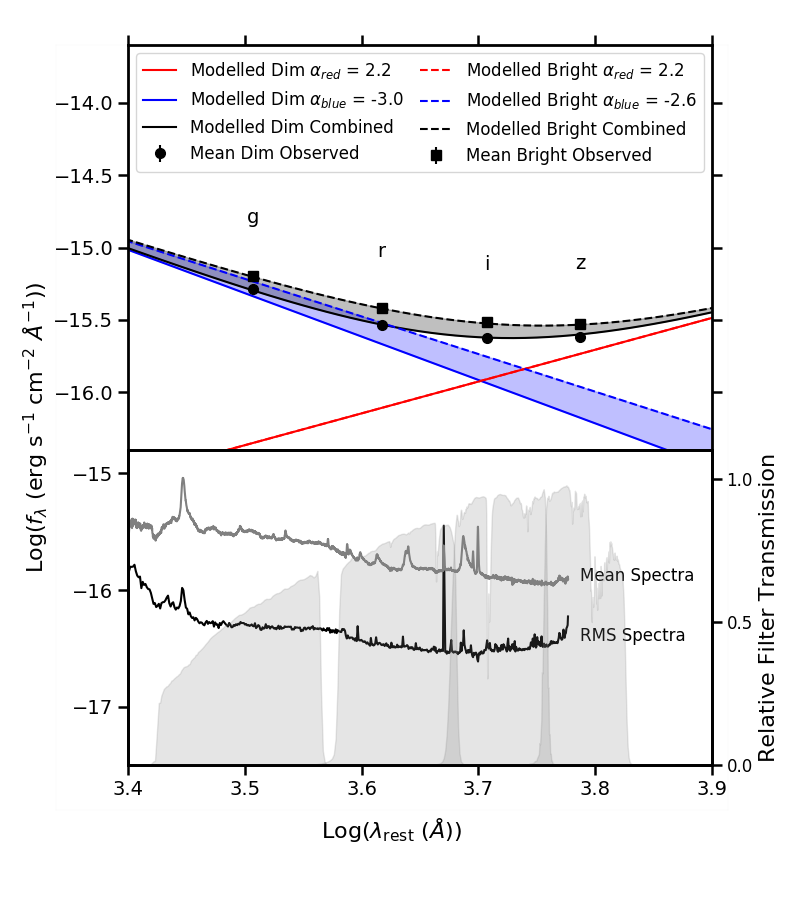}}
        \hfill
        \subfloat[figure][\label{fig:S4_fit_alt4} ~ \\]{\includegraphics[width=0.49\textwidth]{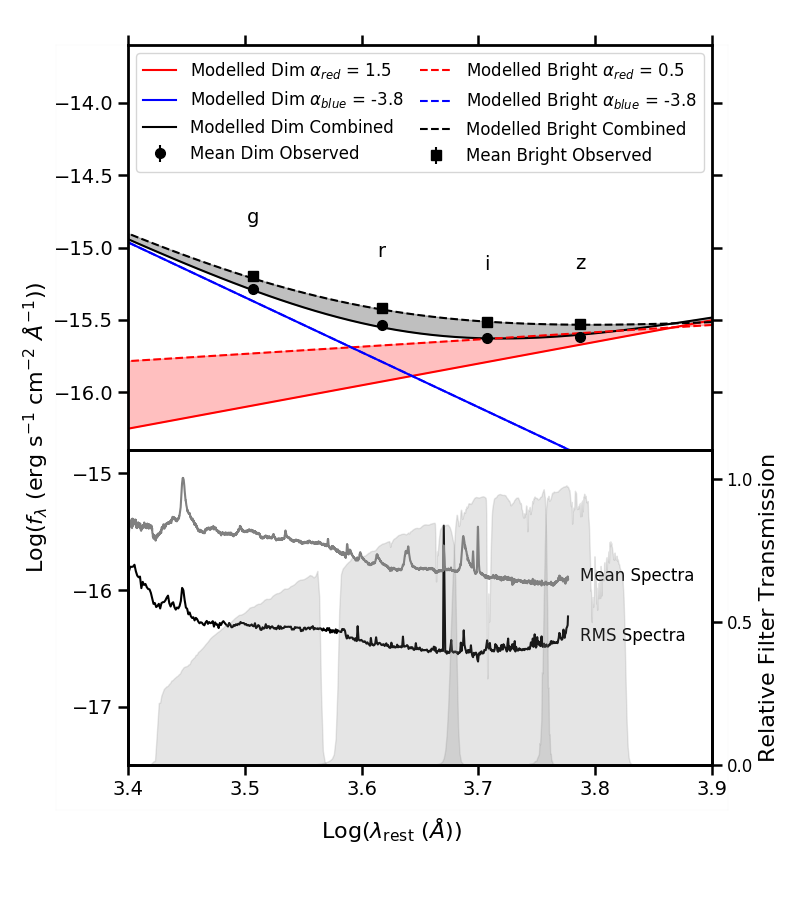}}
        \caption{\textit{Upper panel}: Alternative examples of the modelled broadband spectra of the blue and red emission to match the average observed brightest and dimmest epochs in the 2014 season. The solid lines correspond to the modelled spectra of the dimmest epochs, and the dashed lines correspond to the modelled spectra of the brightest epochs. 
        \textit{Lower panel}: Mean and smoothed RMS OzDES spectra for each season plotted over the DES filter Transmissions.
        \label{fig:More_Modelled_AD_Sync}}
    \end{minipage}
\end{figure*}

In Section~\ref{sec:model_spec}, the broadband spectra are fit using models of blue and red components, however, the models displayed are not unique, and the overall broadband spectra can be fit using a variety of spectral indices for the blue and red emission. In Figure \ref{fig:More_Modelled_AD_Sync}, more examples of the change in spectral slope are displayed for the 2014 season of DES, including models in which the blue component is fixed and the red emission varied.

\bsp	
\label{lastpage}
\end{document}